\documentclass{JHEP3}   
\usepackage{amsmath}
\usepackage{amssymb}
\usepackage{graphicx}

\preprint{arXiv:1006.0470 [hep-ph]\\IFIC/10-06\\FTUV/10-0602}	

\newcounter{figures}
\newcounter{tables}
\newcommand{\mytab}[4][abcd\thetables]{\refstepcounter{tables}%
                       \begin{table}[htb]%
		       \begin{center}%
		       \begin{tabular}{#2}%
		       #3%
		       \end{tabular}%
		       \end{center}%
		       \caption{\label{#1}\small #4}
		       \end{table}}

\newcommand{\athdm}[0]{A2HDM }
\newcommand{\athdmws}[0]{A2HDM}
\newcommand{\thdm}[0]{2HDM }
\newcommand{\thdmws}[0]{2HDM}
\newcommand{\e}{\mathrm{e}}
\newcommand{\cR}{\mathcal{R}}
\newcommand{\Br}{\mathrm{Br}}
\newcommand{\cO}{\mathcal{O}}
\newcommand{\cP}{\mathcal{P}}

\author{Martin Jung, Antonio Pich, Paula Tuz\'on\\
Instituto de F\'{\i}sica Corpuscular (IFIC),CSIC-Universitat de Val\`encia\\Apartado de Correos 22085,E-46071 Valencia, Spain\\
E-Mail: \email{jung@ific.uv.es}\\
E-Mail: \email{pich@ific.uv.es}\\
E-Mail: \email{tuzon@ific.uv.es}
}

\title{Charged-Higgs phenomenology in the Aligned two-Higgs-doublet model}
\date{28 May 2010}
\abstract{
The alignment in flavour space of the Yukawa matrices of a general two-Higgs-doublet model results in the absence of tree-level flavour-changing neutral currents. In addition to the usual fermion masses and mixings, the aligned Yukawa structure only contains three complex parameters $\varsigma_f$, which are potential new sources of $CP$ violation \cite{Pich:2009sp}. For particular values of these three parameters
all known specific implementations of the model based on discrete $\mathcal Z_2$ symmetries
are recovered.
One of the most distinctive features of the two-Higgs-doublet model is the presence of a charged scalar $H^{\pm}$.
In this work, we discuss its main phenomenological consequences in flavour-changing processes at low energies
and derive the corresponding constraints on the parameters of the aligned two-Higgs-doublet model.
}

\begin{document}

\setcounter{page}{1}

\section{Introduction}
The simplicity of the idea of including one additional Higgs doublet to the Standard Model (SM) and the versatility of the resulting phenomenology are the main ingredients that have made the two-Higgs-doublet model
(\thdmws, see e.g. \cite{guide, branco} and references therein) so interesting.
In the most general version of the model, the fermionic couplings of the neutral scalars are non-diagonal in flavour and, therefore, generate unwanted flavour-changing neutral-current (FCNC) phenomena. Different ways
to suppress FCNCs have been developed, giving rise to a variety of specific implementations of the \thdmws.
The simplest and most common approach is to impose a $\mathcal{Z}_2$ symmetry forbidding all non-diagonal terms in the Lagrangian
\cite{Glashow:1976nt}. Depending on the charge assignments under this symmetry, the model is called type I \cite{Haber:1978jt,Hall:1981bc}, II \cite{Hall:1981bc,Donoghue:1978cj}, X and Y \cite{Barger:1989fj,Savage:1991qh,Grossman:1994jb,Akeroyd:1998ui,Akeroyd:1996di,Akeroyd:1994ga,Aoki:2009ha} or \emph{inert} \cite{Ma:2008uza,Ma:2006km, Barbieri:2006dq,LopezHonorez:2006gr}.
In these types of models with \emph{natural flavour conservation} the Cabibbo-Kobayashi-Maskawa (CKM) quark mixing matrix \cite{Cabibbo:1963yz,Kobayashi:1973fv} is the only possible source of $CP$ violation.
Another possibility is to assume particular Yukawa textures which force the non-diagonal Yukawa couplings to be proportional to the geometric mean of the two fermion masses, $g_{ij}\propto \sqrt{m_i m_j}$,
the so-called type III \thdm \cite{Cheng:1987rs,Atwood:1996vj,DiazCruz:2004pj,DiazCruz:2009ek}.

Our work focuses on the recent suggestion \cite{Pich:2009sp} to enforce the alignment in flavour space of the Yukawa couplings of the two scalar doublets, which guarantees the absence of tree-level FCNC interactions. The Yukawa structure of the resulting aligned two-Higgs-doublet model (\athdmws) is fully characterized by the fermion masses, the CKM quark mixing matrix
and three complex parameters $\varsigma_f$ ($f=u,d,l$), whose phases are potential new sources of $CP$ violation \cite{Pich:2009sp}. The usual models based on $\mathcal{Z}_2$ symmetries are recovered for particular (real) values of these three parameters. The \athdmws\ provides a more general setting to discuss the phenomenology of \thdmws s without tree-level FCNCs, leaving open the possibility of having additional $CP$-violating phases in the Yukawa sector beyond the CKM-matrix one.

The presence of a charged scalar $H^{\pm}$ is one of the most distinctive features of an extended scalar sector. In the following we analyze its phenomenological impact in low-energy flavour-changing processes within the \athdmws, and constrain the three complex parameters $\varsigma_f$ with present data on different leptonic, semileptonic and hadronic decays.
We proceed as follows: the formulation of the general \thdm is recalled in section~\ref{model}, where the \emph{aligned} condition is implemented and the resulting Yukawa structure discussed.
Section~\ref{sec:inputs} explains our statistical treatment of theoretical uncertainties and compiles the inputs used in our analysis.
The phenomenological consequences of having a charged scalar field are analyzed next, process by process, extracting the corresponding constraints on the new-physics parameters $\varsigma_f$.
In section~\ref{sec:tree} we discuss the constraints derived from tree-level leptonic and semileptonic decays, while section \ref{sec:loops} describes the information obtained from loop-induced processes. Finally, we give our conclusions in section \ref{summary}. Some technical aspects related to $\Delta F=2$ transitions have been relegated to the appendix.

\section{Aligned two-Higgs-doublet model}\label{model}

The \thdm extends the SM with a second Higgs doublet of hypercharge $Y=\frac{1}{2}$.
The neutral components of the scalar doublets $\phi_a(x)$ ($a=1,2$) acquire vacuum expectation values (VEVs) that are, in general, complex: $\langle 0|\phi_a^T(x)|0\rangle =\frac{1}{\sqrt{2}}\, (0,v_a\, \e^{i\theta_a})$. Through an appropriate $U(1)_Y$ transformation we can enforce $\theta_1=0$, since only the relative phase $\theta \equiv \theta_2 - \theta_1$
is observable. The combination $v\equiv \sqrt{v_1^2+v_2^2}\simeq (\sqrt{2}\, G_F)^{-1/2} = 246~\mathrm{GeV}$ plays the role of the SM VEV when generating the gauge boson masses.

A global SU(2) transformation in the scalar space $(\phi_1,\phi_2)$ takes us to the so-called Higgs basis
$(\Phi_1,\Phi_2)$, where only one doublet acquires a VEV:
\begin{equation}
\left( \begin{array}{c} \Phi_1 \\ -\Phi_2  \end{array} \right) \equiv \; \frac{1}{v} \left[ \begin{array}{cc} v_1 & v_2 \\ v_2 & -v_1 \end{array} \right] \; \left( \begin{array}{c} \phi_1 \\ e^{-i\theta}\phi_2  \end{array} \right) \; .
\end{equation}
In this basis, the two doublets are parametrized as
\begin{equation}
\Phi_1=\left[ \begin{array}{c} G^+ \\ \frac{1}{\sqrt{2}}\, (v+S_1+iG^0) \end{array} \right] \; , \qquad \Phi_2 = \left[ \begin{array}{c} H^+ \\ \frac{1}{\sqrt{2}}\, (S_2+iS_3)   \end{array}\right] \; ,
\end{equation}
where $G^\pm$ and $G^0$ denote the Goldstone fields
and $\langle H^+\rangle=\langle G^+\rangle=\langle G^0\rangle=\langle S_i\rangle=0$.
The five physical scalars are given by two charged fields $H^{\pm}(x)$ and three neutral ones $\varphi_i^0(x)=\{h(x),H(x),A(x)\}$, which are related to the $S_i$ fields through an orthogonal transformation $\varphi^0_i(x)=\mathcal R_{ij}S_j(x)$. The form of $\mathcal R_{ij}$ depends on the scalar potential, which could violate $CP$ in its most general version; in that case the resulting mass eigenstates do not have a definite $CP$ parity.

The most general Yukawa Lagrangian of the \thdm is given by
\begin{equation}
\mathcal L_Y \; = \; -\left\{
\bar{Q}_L' (\Gamma_1 \phi_1 +\Gamma_2 \phi_2)\, d_R' + \bar{Q}_L' (\Delta_1 \widetilde{\phi}_1 +\Delta_2 \widetilde{\phi}_2)\, u_R'  + \bar{L}_L' (\Pi_1 \phi_1 + \Pi_2 \phi_2)\, l_R'\right\}
\, +\, \mathrm{h.c.}  \; ,
\end{equation}
where $ \bar{Q}_L'$ and $\bar{L}_L' $ are the left-handed quark and lepton doublets, respectively, and $\tilde{\phi}_a(x)\equiv i \tau_2\phi_a^*(x)$ the charge-conjugated scalar doublets with $Y=-\frac{1}{2}$. All fermionic fields are written as $N_G$-dimensional vectors and the couplings $\Gamma_a$, $\Delta_a$ and $\Pi_a$ are $N_G\times N_G$ complex matrices in flavour space, $N_G$ being the number of fermion generations. Moving to the Higgs basis, the Lagrangian reads
\begin{eqnarray}
\mathcal L_Y \; &=&   -\frac{\sqrt{2}}{v}  \left\{  \bar{Q}_L'  (M_d'  \Phi_1 + Y_d' \Phi_2)\, d_R' + \bar{Q}_L'  (M_u' \tilde{\Phi}_1  + Y_u' \tilde{\Phi}_2)\, u_R'  + \bar{L}_L'  (M_l' \Phi_1  + Y_l' \Phi_2)\, l_R'  \right\}+\nonumber\\ &&+\,\mathrm{h.c.}   \,,
\end{eqnarray}
with
\begin{align}
M_d'&=\frac{1}{\sqrt{2}}\,\left( v_1\Gamma_1+v_2\Gamma_2\e^{i\theta}\right)\, ,
\qquad &
Y_d'&=\frac{1}{\sqrt{2}}\,\left( v_1\Gamma_2\e^{i\theta} - v_2\Gamma_1\right)\, ,
\\
M_u'&=\frac{1}{\sqrt{2}}\,\left( v_1\Delta_1+v_2\Delta_2\e^{-i\theta}\right)\,,
\qquad &
Y_u'&=\frac{1}{\sqrt{2}}\,\left( v_1\Delta_2\e^{-i\theta} - v_2\Delta_1\right) \, ,
\\
M_l'&=\frac{1}{\sqrt{2}}\,\left( v_1\Pi_1+v_2\Pi_2\e^{i\theta}\right)\, , 
\qquad &
Y_l'&=\frac{1}{\sqrt{2}}\,\left( v_1\Pi_2\e^{i\theta} - v_2\Pi_1\right)\, .
\end{align}
In general, the complex matrices $M_f'$ and $Y_f'$ ($f=d,u,l$) cannot be simultaneously diagonalized.
Thus, in the fermion mass-eigenstate basis, with diagonal mass matrices $M_f$, the Yukawa-coupling matrices
$Y_f$ remain non-diagonal giving rise to FCNC interactions.

The unwanted non-diagonal neutral couplings can be eliminated requiring the alignment in flavour space of the Yukawa matrices  \cite{Pich:2009sp}. It is convenient to implement this condition in the form:
\begin{eqnarray}
\Gamma_2 = \xi_d\, \e^{-i\theta} \,\Gamma_1 \; , \qquad \Delta_2=\xi_u^*\, \e^{i\theta}\Delta_1\; ,  \qquad \Pi_2=\xi_l\, \e^{-i\theta} \,\Pi_1 \; ,
\end{eqnarray}
where $\xi_f$ are arbitrary complex parameters. The proportionality of the matrices $Y'_f$ and $M'_f$ guarantees that all FCNC couplings vanish at tree level:
\begin{eqnarray}
Y_{d,l} = \varsigma_{d,l} M_{d,l}\; , \qquad Y_u = \varsigma_u^* M_u \; ,  \qquad \varsigma_f \equiv \frac{\xi_f-\tan{\beta}}{1+\xi_f\tan{\beta}}
\; ,  \qquad\tan{\beta}\equiv v_2/v_1 \; .
\end{eqnarray}
In the \athdm the mass-eigenstate Yukawa Lagrangian reads \cite{Pich:2009sp}
\begin{eqnarray}\label{lagrangian}
\mathcal L_Y &\! =&\! - \frac{\sqrt{2}}{v}\, H^+(x) \left\{  \bar{u}(x)   \left[  \varsigma_d\, V M_d \mathcal P_R - \varsigma_u\, M_u^\dagger V \mathcal P_L  \right]   d(x)\, +\, \varsigma_l\; \bar{\nu}(x) M_l \mathcal P_R l(x)\right\}-\nonumber \\ 
& & -\,\frac{1}{v}\; \sum_{\varphi, f}\, y^{\varphi^0_i}_f\, \varphi^0_i(x)  \; \bar{f}(x)\,  M_f \mathcal P_R  f(x)\;  + \;\mathrm{h.c.} \, ,
\end{eqnarray}
where $V$ denotes the CKM matrix, $\mathcal P_{R,L}\equiv \frac{1\pm \gamma_5}{2}$ are the right-handed and left-handed projectors and the  couplings of the neutral scalar fields are given by:
\begin{equation}
y_{d,l}^{\varphi^0_i} = \cR_{i1} + (\cR_{i2} + i\,\cR_{i3})\,\varsigma_{d,l}  \, ,
\qquad\qquad
y_u^{\varphi^0_i} = \cR_{i1} + (\cR_{i2} -i\,\cR_{i3}) \,\varsigma_{u}^* \, .
\end{equation}

Some conclusions can be drawn from \eqref{lagrangian}. In the \athdm all fermionic couplings to scalars are proportional to the corresponding fermion masses and the neutral-current interactions are diagonal in flavour. The only source of flavour-changing interactions is the CKM matrix in the quark charged current, while all leptonic couplings are diagonal in flavour because of the absence of right-handed neutrinos in our framework, which could however easily be included.
There are only three new parameters $\varsigma_f$, which encode all possible freedom allowed by the alignment conditions; these couplings satisfy universality among the different generations, i.e. all fermions with a given electric charge have the same universal coupling $\varsigma_f$. The three parameters $\varsigma_f$ are also invariant under global SU(2) transformations of the scalar fields $\phi_a \rightarrow \phi_a'=U_{ab}\phi_b$ \cite{Davidson:2005cw}, i.e. they are scalar-basis independent. Taking the particular values shown in table 1, the different models based on $\mathcal Z_2$ symmetries are recovered, with a single scalar doublet coupling to each type of right-handed fermions \cite{Glashow:1976nt}. Finally, it should be pointed out again that $\varsigma_f$ are arbitrary complex numbers, opening the possibility of having new sources of $CP$ violation without tree-level FCNCs.

\begin{table}\begin{center}
\begin{tabular}{|c|c|c|c|c|}
\hline
Model & $(\xi_d,\xi_u,\xi_l)$ & $\varsigma_d$ & $\varsigma_u$ & $\varsigma_l$  \\
\hline
Type I & $(\infty,\infty,\infty)$& $\cot{\beta}$ &$\cot{\beta}$ & $\cot{\beta}$ \\
Type II & $(0,\infty,0)$& $-\tan{\beta}$ & $\cot{\beta}$ & $-\tan{\beta}$ \\
Type X & $(\infty,\infty,0)$& $\cot{\beta}$ & $\cot{\beta}$ & $-\tan{\beta}$ \\
Type Y & $(0,\infty,\infty)$& $-\tan{\beta}$ & $\cot{\beta}$ & $\cot{\beta}$ \\
Inert & $(\tan{\beta},\tan{\beta},\tan{\beta})$ & 0 & 0 & 0 \\
\hline
\end{tabular}
\caption{Limits on $\xi_f$ that recover the different $\mathcal Z_2$ models and corresponding $\varsigma_f$ values.}\label{tab:models}
\end{center}\end{table}

\subsection{Quantum corrections}

Quantum corrections induce some misalignment of the Yukawa coupling matrices, generating small FCNC effects suppressed by the corresponding loop factors. However, the special structure of the \athdm strongly constrains the possible FCNC interactions \cite{Pich:2009sp}. Obviously, the alignment condition remains stable under renormalization when it is protected by a $\mathcal{Z}_2$ symmetry \cite{Ferreira:2010xe}, i.e. for the particular cases indicated in table~\ref{tab:models}.
In the most general case loop corrections do generate some FCNC effects, but the resulting structures are enforced to satisfy the flavour symmetries of the model.
The Lagrangian of the \athdm is invariant under flavour-dependent phase transformations of the fermion mass eigenstates ($f=d,u,l,\nu$, $X=L,R$, $\alpha^{\nu,L}_i = \alpha^{l,L}_i$):
\begin{equation}
f_X^i(x)\,\to\,\e^{i\alpha^{f,X}_i}\, f_X^i(x)\, , \quad
V_{ij}\,\to\,\e^{i\alpha^{u,L}_i} V_{ij}\, \e^{-i\alpha^{d,L}_j}\, , \quad
M_{f,ij}\,\to\,\e^{i\alpha^{f,L}_i} M_{f,ij}\, \e^{-i\alpha^{f,R}_j}\, .
\end{equation}
Owing to this symmetry, lepton-flavour-violating neutral couplings are identically zero to all orders in perturbation theory, while in the quark sector the CKM mixing matrix remains the only possible source of flavour-changing transitions. The only allowed local FCNC structures are of the type $\bar u_L V (M_d^{\phantom{\dagger}} M_d^\dagger)^n V^\dagger (M_u^{\phantom{\dagger}} M_u^\dagger)^m M_u^{\phantom{\dagger}} u_R$,\
$\bar d_L V^\dagger (M_u^{\phantom{\dagger}} M_u^\dagger)^n V (M_d^{\phantom{\dagger}} M_d^\dagger)^m M_d^{\phantom{\dagger}} d_R$,\ or similar structures with additional factors of $V$, $V^\dagger$ and
quark mass matrices \cite{Pich:2009sp}. Therefore, at the quantum level the \athdm provides an explicit implementation
of the popular Minimal Flavour Violation scenarios \cite{D'Ambrosio:2002ex,Chivukula:1987py,Hall:1990ac,Buras:2000dm,Cirigliano:2005ck,Kagan:2009bn}, but allowing at the same time for new $CP$-violating phases.\footnote{Minimal flavour violation within the context of the Type II \thdm  has been discussed in \cite{D'Ambrosio:2002ex}. This reference didn't consider the possibility of incorporating new $CP$-violating phases.}
Structures of this type have been recently discussed in \cite{Botella:2009pq}.

Using the renormalization-group equations \cite{Ferreira:2010xe,Cvetic:1998uw}, one can easily check that the one-loop gauge corrections preserve the alignment while the only FCNC structures induced by the scalar contributions take the form \cite{JPT:2010}:
\begin{eqnarray}
\mathcal L_{\mathrm{FCNC}} & =& \frac{C(\mu)}{4\pi^2 v^3}\; (1+\varsigma_u^*\varsigma_d^{\phantom{*}})\;\times\nonumber\\
&& \times\sum_i\, \varphi^0_i(x)\;\left\{
(\cR_{i2} + i\,\cR_{i3})\, (\varsigma_d^{\phantom{*}}-\varsigma_u^{\phantom{*}})\;
\left[\bar d_L\, V^\dagger M_u^{\phantom{\dagger}} M_u^\dagger\, V M_d^{\phantom{\dagger}}\, d_R\right]-\right.\label{eq:FCNCop}
\\ 
&&\hskip 2.5cm \left.
-\, (\cR_{i2} - i\,\cR_{i3})\, (\varsigma_d^*-\varsigma_u^*)\;
\left[\bar u_L\, V M_d^{\phantom{\dagger}} M_d^\dagger\, V^\dagger M_u^{\phantom{\dagger}}\, u_R\right] \right\}
\; +\; \mathrm{h.c.}\nonumber
\end{eqnarray}
As they should, these FCNC effects vanish identically when $\varsigma_d =\varsigma_u$ ($\mathcal{Z}_2$ models of type I, X and Inert) or $\varsigma_d =-1/\varsigma_u^*$ (types II and Y).
The leptonic coupling $\varsigma_l$ does not induce any FCNC interaction, independently of its value; the usually adopted $\mathcal{Z}_2$ symmetries are unnecessary in the lepton sector.
Assuming the alignment to be exact at some scale $\mu_0$, i.e. $C(\mu_0)=0$, a non-zero value for the FCNC coupling, $C(\mu)=-\log{(\mu/\mu_0)}$, is generated when running to a different scale. 

The numerical effect of these contributions is, in any case, suppressed by  $m_{q}m_{q'}^2/v^3$ and quark-mixing factors. This implies an interesting hierarchy of FCNC effects, avoiding the stringent experimental constraints for light-quark systems, while allowing at the same time for potential interesting signals in heavy-quark transitions. Obviously, the most relevant terms in (\ref{eq:FCNCop}) are the $\bar s_L b_R$ and $\bar c_L t_R$ operators. The $\bar s_L b_R$ term induces a calculable contribution to $B^0_s$--$\bar B^0_s$ mixing through $\varphi^0_i$ exchanges, which modifies the mixing phase and could explain the like-sign dimuon charge asymmetry recently observed by D0 \cite{Abazov:2010hv}. Tree-level scalar exchanges from FCNC vertices have been already suggested as a possible explanation of the D0 measurement \cite{Dobrescu:2010rh}. We defer the phenomenological analysis of the FCNC operator (\ref{eq:FCNCop}) to a future publication \cite{JPT:2010}, where the neutral sector of the \athdm will be studied in detail. In the present paper we will concentrate in the phenomenology of the charged-scalar Yukawa Lagrangian (\ref{lagrangian}).

\section{Inputs and statistical treatment}\label{sec:inputs}

In the following sections we will analyze the most important flavour-changing processes that are sensitive to charged-scalar exchange and will try to constrain from them the new-physics parameters $\varsigma_f$. Most of these observables have been discussed in recent phenomenological analyses, usually in the framework of the type II \thdm \cite{WahabElKaffas:2007xd,Deschamps:2009rh,Flacher:2008zq,Bona:2009cj}, but also in the type III \thdm \cite{Mahmoudi:2009zx}.

For that purpose, a good control of the hadronic decay parameters is necessary. These usually involve large theoretical uncertainties whose treatment is not well defined. In our work we use the statistical approach
RFit \cite{Hocker:2001xe}, which has been implemented in the CKMfitter package \cite{CKMfitterWeb}.
The new-physics parameter space is explored, assigning to each point the maximal relative likelihood under variation of the theoretical parameters which are not shown. Theoretical uncertainties are treated by defining allowed ranges within which the contribution of the corresponding theoretical quantity to the $\Delta\chi^2$ is set to zero, while it is set to infinity outside. This treatment implies that uncertainties of this kind should be chosen conservatively and added linearly.

Another related problem is the combination of different theoretical determinations of a hadronic quantity, which is even less well defined. We follow the prescription given in \cite{CKMfitterWeb}. 
However, unless commented explicitly, we only take lattice results coming from numerical
simulations with $2+1$ flavours. For quantities concerning the light hadrons, we consider the determinations recommended by the Flavour Lattice Averaging Group (FLAG) \cite{FLAG,Lubicz:2010nx}. The obtained values are collected in table~\ref{tab::hadronic}.

For $f_+^{K\pi}(0)$ the only published value with 2+1 dynamical quarks is the one from RBC/UKQCD \cite{Boyle:2010bh,Boyle:2007qe}, which however fails to fulfill the FLAG standards. On the other hand, there is one 2-flavour result, which fulfills the FLAG criteria \cite{Lubicz:2009ht}.
Although consistent with the old Leutwyler-Roos estimate \cite{Leutwyler:1984je}, based on $O(p^4)$ Chiral Perturbation Theory ($\chi$PT), these lattice determinations are somewhat smaller than the $O(p^6)$ analytical calculations \cite{Bijnens:2003uy,Jamin:2004re,Cirigliano:2005xn,Kastner:2008ch}.
We take this into account and adopt the conservative range $f_+^{K\pi}(0)=0.965\pm0.010$.

\begin{table}[tbh!]
\begin{tabular}{p{2.7cm} c p{6.cm}}\hline
Parameter                                            & Value                                 & Comment\\\hline
$f_{B_s}$                                            & $(0.242\pm0.003\pm0.022)$~GeV         & Our average \cite{Wingate:2003gm,Gamiz:2009ku,Bernard:2009wr}\\
$f_{B_s}/f_{B_d}$                                    & $1.232\pm0.016\pm0.033$               & Our average \cite{Gamiz:2009ku,Bernard:2009wr}\\
$f_{D_s}$                                            & $(0.2417\pm0.0012\pm0.0053)$~GeV      & Our average
 \cite{Wingate:2003gm,Follana:2007uv,Bernard:2009wr}\\
$f_{D_s}/f_{D_d}$                                    & $1.171\pm0.005\pm0.02$                & Our average  \cite{Follana:2007uv,Bernard:2009wr}\\
$f_{K}/f_\pi$                                        & $1.192\pm 0.002\pm0.013$            & Our average
\cite{Follana:2007uv,Bernard:2007ps,Durr:2010hr}\\
$f_{B_s} \sqrt{\hat B_{B^0_s}}$ & $(0.266\pm0.007\pm0.032)$~GeV & \cite{Gamiz:2009ku}\\
$f_{B_d}\sqrt{\hat B_{B^0_s}}/(f_{B_s}\sqrt{\hat B_{B^0_s}})$ & $1.258\pm 0.025\pm 0.043$ & \cite{Gamiz:2009ku}\\
$\hat B_K$ & $0.732\pm 0.006\pm 0.043$ & \cite{Aubin:2009jh,Kelly:2009fp}
\\\hline
$|V_{ud}|$     & $0.97425\pm 0.00022$  &   \cite{Hardy:2008gy}\\
$\lambda$   & $0.2255\pm0.0010$    & $\left(1-|V_{ud}|^2\right)^{1/2}$ \\
$|V_{ub}|$                                  & $(3.8\pm 0.1\pm 0.4) \cdot 10^{-3}$         & $b\to u l\nu$ (excl. + incl.) \cite{Antonelli:2009ws,Barberio:2008fa}\\
$A$                                       & $0.80\pm 0.01\pm 0.01$          & $b\to c l\nu$ (excl. + incl.) \cite{Antonelli:2009ws,Barberio:2008fa}\\
$\bar\rho$     & $0.15\pm 0.02\pm 0.05$ & Our fit\\
$\bar\eta$     & $0.38\pm 0.01\pm 0.06$ & Our fit
\\\hline
$\bar{m}_u(2~{\rm GeV})$                             & $(0.00255\, {}^{+\, 0.00075}_{-\, 0.00105})$~GeV & \cite{Amsler:2008zzb}\\
$\bar{m}_d(2~{\rm GeV})$                             & $(0.00504\, {}^{+\, 0.00096}_{-\, 0.00154})$~GeV & \cite{Amsler:2008zzb}\\
$\bar{m}_s(2~{\rm GeV})$                             & $(0.105\, {}^{+\, 0.025}_{-\, 0.035})$~GeV       & \cite{Amsler:2008zzb}\\
$\bar{m}_c(2~{\rm GeV})$                             & $(1.27\, {}^{+\, 0.07}_{-\, 0.11})$~GeV         & \cite{Amsler:2008zzb}\\
$\bar{m}_b(m_b)$                                     & $(4.20\, {}^{+\, 0.17}_{-\, 0.07})$~GeV         & \cite{Amsler:2008zzb}\\
$\bar{m}_t(m_t)$            & $(165.1\pm 0.6\pm 2.1)$~GeV  & \cite{:2009ec}
\\\hline\\[-13pt]
$\delta_{\mathrm{em}}^{K\ell2/\pi\ell2}$                      & $-0.0070\pm0.0018$                    & \cite{Marciano:2004uf,Cirigliano:2007ga,Cirigliano:2007xi,Antonelli:2010yf}\\
$\delta_{\mathrm{em}}^{\tau K2/K\ell2}$                      & $0.0090\pm0.0022$                    &
\cite{Decker:1994ea,Decker:1994kw,Marciano:1993sh}\\
$\delta_{\mathrm{em}}^{\tau \pi2/\pi\ell2}$                      & $0.0016\pm0.0014$                    &
\cite{Decker:1994ea,Decker:1994kw,Marciano:1993sh}\\ \hline
$\rho^2|_{B\to D l\nu}$                              & $1.18\pm0.04\pm0.04$                  & \cite{Barberio:2008fa}\\
$\Delta|_{B\to D l\nu}$                               & $0.46\pm0.02$                         & \cite{deDivitiis:2007uk} \\
$f_+^{K\pi}(0)$                 &    $0.965\pm0.010$  &
\cite{Boyle:2010bh,Boyle:2007qe,Lubicz:2009ht,Bijnens:2003uy,Jamin:2004re,Cirigliano:2005xn,Kastner:2008ch}\\
$\bar{g}_{b,SM}^L$                         &  $-0.42112\, {}_{-\, 0.00018}^{+\, 0.00035}$  & \cite{LEPZresonance,Degrassi:2010ne}\\
$\kappa_\epsilon$ & $0.94\pm 0.02$ & \cite{Buras:2010pz} \\
$\bar{g}_{b,SM}^R$                         & $0.07744\, {}^{+\, 0.00006}_{-\, 0.00008}$    & \cite{LEPZresonance,Degrassi:2010ne}\\
\hline
\end{tabular}
\caption{\it \label{tab::hadronic} Input values for the hadronic parameters, obtained as described in the text. The first error denotes statistical uncertainty, the second systematic/theoretical.}
\end{table}

To fix the values of the relevant CKM entries we only use determinations \cite{Antonelli:2009ws,Bona:2005vz,Charles:2004jd} which are not sensitive to the new-physics contributions. Thus, we use the $V_{ud}$ value extracted from superallowed ($0^+\to 0^+$) nuclear $\beta$ decays and CKM unitarity to determine $V_{us} \equiv \lambda$. The values of $V_{ub}$ and $V_{cb} = A\lambda^2$ are determined from exclusive and inclusive $b\to u l\bar\nu_l$ and $b\to c l\bar\nu_l$ transitions, respectively, with $l=e,\mu$. The apex $(\bar\rho,\bar\eta)$ of the unitarity triangle has been determined from $|V_{ub}/V_{cb}|$, $\lambda$ and the ratio $\Delta m_{B^0_s}/\Delta m_{B^0_d}$ (see section \ref{subsec:Bmixing}).
For the top quark mass we have adopted the usual assumption that the Tevatron value \cite{:2009ec} corresponds to the pole mass, but increasing its systematic error by 1 GeV to account for the intrinsic ambiguity in the $m_t$ definition; i.e. we have taken
$m_t^{\mathrm{pole}} = (173.1\pm 0.6\pm 2.1)~\mathrm{GeV}$ and have converted this value into the running $\overline{\mathrm{MS}}$ mass.
The measurements used in our analysis are listed in table~\ref{tab::measurements}.

Concerning the charged-scalar mass, we will use the LEP lower bound $M_{H^\pm}> 78.6$~GeV (95\% CL), which does not refer to any specific Yukawa structure \cite{Pich:2009sp,:2001xy}. This limit assumes only that $H^+$ decays dominantly into $u_i\bar d_j$ and $l^+\nu_l$. Obviously, the bound is avoided by a fermiophobic (inert) \athdm with $\varsigma_f\ll 1$, but all our constraints would also disappear in this case. The charged scalar could still be detected through the decay mode $H^\pm\to W^\pm A$, provided it is kinematically allowed. Assuming a $CP$-conserving scalar potential, OPAL finds the 95\% CL constraints $M_{H^\pm}> 56.5\; (64.8)$~GeV,
for $12\; (15)~\mathrm{GeV} < M_A < M_{H^\pm}-M_{W^\pm}$ \cite{:2008be}.

\begin{table}[tb]
\begin{center}
\begin{tabular}{p{5.5cm} c p{5cm}}\hline
Observable                                   & Value                               & Comment\\\hline
$|g_{RR}^S|_{\tau\to \mu}$ & $ < 0.72$ \ (95\% CL) & \cite{Amsler:2008zzb}\\
$\Br (\tau\to\mu\nu_\tau\bar{\nu}_\mu)$        & $(17.36\pm0.05)\times10^{-2}$                  & \cite{Amsler:2008zzb}\\
$\Br (\tau\to e\nu_\tau\bar{\nu}_e)$           & $(17.85\pm0.05)\times10^{-2}$                  & \cite{Amsler:2008zzb}\\
$\Br (\tau\to \mu \nu_{\tau}\bar{\nu}_{\mu})/\Br (\tau\to e \nu_{\tau}\bar{\nu}_e)$&$0.9796\pm0.0039$& \cite{Aubert:2009qj}\\
$\Br (B\to\tau\nu)$                            & $(1.73\pm0.35)\times10^{-4}$        & \cite{Charles:2004jd}\\
$\Br (D\to\mu\nu)$                             & $(3.82\pm0.33)\times10^{-4}$        & \cite{:2008sq}\\
$\Br (D\to\tau\nu)$                            & $\leq 1.3\times 10^{-3}$ \ (95\% CL)  & \cite{:2008sq}\\
$\Br (D_s\to\tau\nu)$                          & $(5.58\pm0.35)\times10^{-2}$        &  \cite{Alexander:2009ux,Onyisi:2009th,Naik:2009tk,:2010qj,Rosner:2010ak}\\
$\Br (D_s\to\mu\nu)$                           & $(5.80\pm0.43)\times10^{-3}$        & \cite{Alexander:2009ux,Rosner:2010ak,:2007ws} \\
$\Gamma(K\to\mu\nu)/\Gamma(\pi\to\mu\nu)$    & $1.334\pm0.004$               & \cite{Antonelli:2010yf} \\
$\Gamma(\tau\to K\nu)/\Gamma(\tau\to\pi\nu)$ & $(6.50\pm 0.10)\times10^{-2}$        & \cite{Amsler:2008zzb,Aubert:2009qj}\\
$\log C$                                     & $0.194\pm0.011$                     & \cite{:2007yza,Abouzaid:2009ry} \\
$\Br (B\to D\tau\nu)/BR(B\to D\ell\nu)$        & $0.392\pm0.079$                     & \cite{:2009xy,Bozek:2010xy,Adachi:2009qg}\\
$\Gamma(Z\to b\bar{b})/\Gamma(Z\to\mbox{hadrons})$ & $0.21629\pm0.00066$            & \cite{Alcaraz:2009jr}\\
$\Br (\bar B\to X_s\gamma)_{E_\gamma>1.6\mathrm{GeV}}$       & $(3.55\pm0.26)\times 10^{-4} $   
   &\cite{Barberio:2008fa}\\
$\Br (\bar B\to X_c e\bar\nu_e)$     & $(10.74\pm 0.16)\times10^{-2}$   & \cite{Barberio:2008fa}\\
$\Delta m_{B^0_d}$                                 & $(0.507\pm0.005)~{\rm ps}^{-1}$       & \cite{Barberio:2008fa}\\
$\Delta m_{B^0_s}$                                 & $(17.77\pm0.12)~{\rm ps}^{-1}$       & \cite{Barberio:2008fa}\\
$|\epsilon_K|$ & $(2.228 \pm 0.011)\times 10^{-3}$ & \cite{Amsler:2008zzb}\\
\hline
\end{tabular}
\caption{\it \label{tab::measurements} Measurements used in the analysis. Masses and lifetimes are taken from the PDG \cite{Amsler:2008zzb}.}
\end{center}
\end{table}

\section{Tree-level decays}\label{sec:tree}

\subsection{Lepton decays}

The pure leptonic decays $l\to l'\bar\nu_{l'}\nu_l$ provide accurate tests of the universality of the
leptonic $W$ couplings and of their left-handed current structure
\cite{Amsler:2008zzb,Pich:2008ni,Pich:1997ym}. The exchange of a charged scalar induces an additional amplitude mediating the decay of a right-handed initial lepton into a right-handed final charged lepton; in standard notation \cite{Pich:1997ym,Pich:1995vj}, this scalar contribution gets parametrized
through the effective low-energy coupling
$g_{RR}^S = -\frac{m_{l} m_{l'}}{M_{H^\pm}^2}\, |\varsigma_l|^2$.
Its phenomenological effects can be isolated through the Michel parameters governing the decay distribution,
\begin{equation}
\rho-\frac{3}{4} = 0\, , \quad
\eta =\frac{1}{2N}\,\mathrm{Re}(g_{RR}^S)\, ,\quad
\xi-1 = -\frac{1}{2N}\, |g_{RR}^S|^2\, ,\quad
\xi\delta - \frac{3}{4} =-\frac{3}{8N}\, |g_{RR}^S|^2\, ,
\end{equation}
and in the total decay width
\begin{eqnarray}
\Gamma(l\rightarrow l' \; \bar{\nu}_{l'} \; \nu_{l}) = \frac{G_F^2}{192\pi^3}\; m_l^5 \; N\;  \; \left[ f\left( \frac{m_{l'}^2}{m_l^2} \right)+4\; \eta \; \frac{m_{l'}}{m_l} \; g\left( \frac{m_{l'}^2}{m_l^2} \right) \right] \;
r_{\mathrm{RC}}\; ,
\end{eqnarray}
where
$f(x) = 1 - 8x + 8 x^3 - x^4 - 12 x^2 \log{x}$,
$g(x) = 1 + 9x - 9 x^2 - x^3 + 6 x (1+ x) \log{x}$,
$N = 1 + \frac{1}{4}\, |g_{RR}^S|^2$
and \cite{Marciano:1988vm}
\begin{equation}
r_{\mathrm{RC}} =\left[ 1 +\frac{\alpha(m_{l})}{2\pi}\,\left(\frac{25}{4}-\pi^2\right)\right]\,
\left[ 1 + \frac{3}{5}\frac{m_{l}^2}{M_W^2} - 2\, \frac{m_{l'}^2}{M_W^2}\right]\, .
\end{equation}

Since the scalar couplings are proportional to lepton masses, the decay
$\tau\to\mu\bar\nu_\mu\nu_\tau$ is the most sensitive one to the scalar-exchange contribution.
The present bound $|g_{RR}^S|_{\tau\to \mu} < 0.72$ (95\% CL) \cite{Amsler:2008zzb} translates into
$|\varsigma_l|/M_{H^\pm}\le 1.96~\mbox{GeV}^{-1}$ (95\% CL). A better limit can be obtained from the ratio of the total $\tau$ decay widths into the muon and electron modes. The universality test
$|g_\mu/g_e|^2\equiv |\mathrm{Br}(\tau\to\mu)/\mathrm{Br}(\tau\to e)| |f(m_e^2/m_\tau^2)/f(m_\mu^2/m_\tau^2)| = 1.0036\pm 0.0029$ \cite{Amsler:2008zzb,Aubert:2009qj} implies:
\begin{equation}\label{eq:sigma_l}
\frac{|\varsigma_l|}{M_{H^\pm}}\;\le\; 0.40~\mbox{GeV}^{-1}
\qquad (95\%\, \mathrm{CL}).
\end{equation}

\vspace{1mm}

\subsection{Leptonic decays of pseudoscalar mesons}
Information about new-physics parameters can be also extracted from leptonic
decays of pseudoscalar mesons, $P^+\rightarrow l^+ \nu_l$, which are very sensitive to $H^+$ exchange due to the helicity suppression of the SM amplitude. The total decay width is given 
by\footnote{The normalization of the meson decay constant corresponds to $f_\pi = \sqrt{2} F_\pi = 131 ~\mathrm{MeV}$.}
\begin{eqnarray}\label{eq:Gamma_Plnu}
\Gamma(P^+_{ij}\rightarrow l^+ \nu_l)\, =\, G_F^2m_l^2f_P^2|V_{ij}|^2 \,\frac{m_{P^+_{ij}}}{8\pi} \left( 1- \frac{m_l^2}{m_{P^+_{ij}}^2} \right)^2  (1+\delta_{\mathrm{em}}^{M\ell2})\; |1-\Delta_{ij}|^2 \; ,
\end{eqnarray}	
where $i,j$ represent the valence quarks of the meson under consideration. The correction
\begin{eqnarray}\label{eq::Delta}
\Delta_{ij}\, =\,\left( \frac{m_{P^{\pm}_{ij}}}{M_{H^{\pm}}} \right)^2\varsigma_l^*\,\frac{\varsigma_um_{u_i}+\varsigma_dm_{d_j}}{m_{u_i}+m_{d_j}}
\end{eqnarray}
encodes the new-physics information and $\delta_{\mathrm{em}}^{M\ell2}$ denotes the electromagnetic radiative contributions. These corrections are relevant because the additional photon lifts the helicity suppression of the two-body decay, thereby compensating in part for the additional electromagnetic coupling, and the two processes are not distinguishable experimentally for low photon energies. Their relative importance therefore increases for decreasing lepton masses.

The correction $\Delta_{ij}$ is predicted to be positive in model I, negative in model X and
can have either sign in the models II and Y, depending on the decaying meson, while
it is of course absent in the {\it inert}\/ scenario. In the more general A2HDM it is a complex number with a real part of either sign. To determine its size one needs to know $|V_{ij}|$ and a theoretical determination of the meson decay constant.

The SM as well as the 2HDM contribution to this class of decays start at tree level. Therefore they can be assumed to remain the dominant contributions, relatively independent of a possible high-energy completion of the theory. Electroweak loop corrections are of course expected and they could be sizeable in some cases,
for example in supersymmetry at large values of $\tan \beta$ \cite{Buras:2002vd,Akeroyd:2003zr}.

\subsubsection{Heavy pseudoscalar mesons}

\begin{figure}[tbh]
\begin{center}
\begin{tabular}{cc}
\includegraphics[width=7cm]{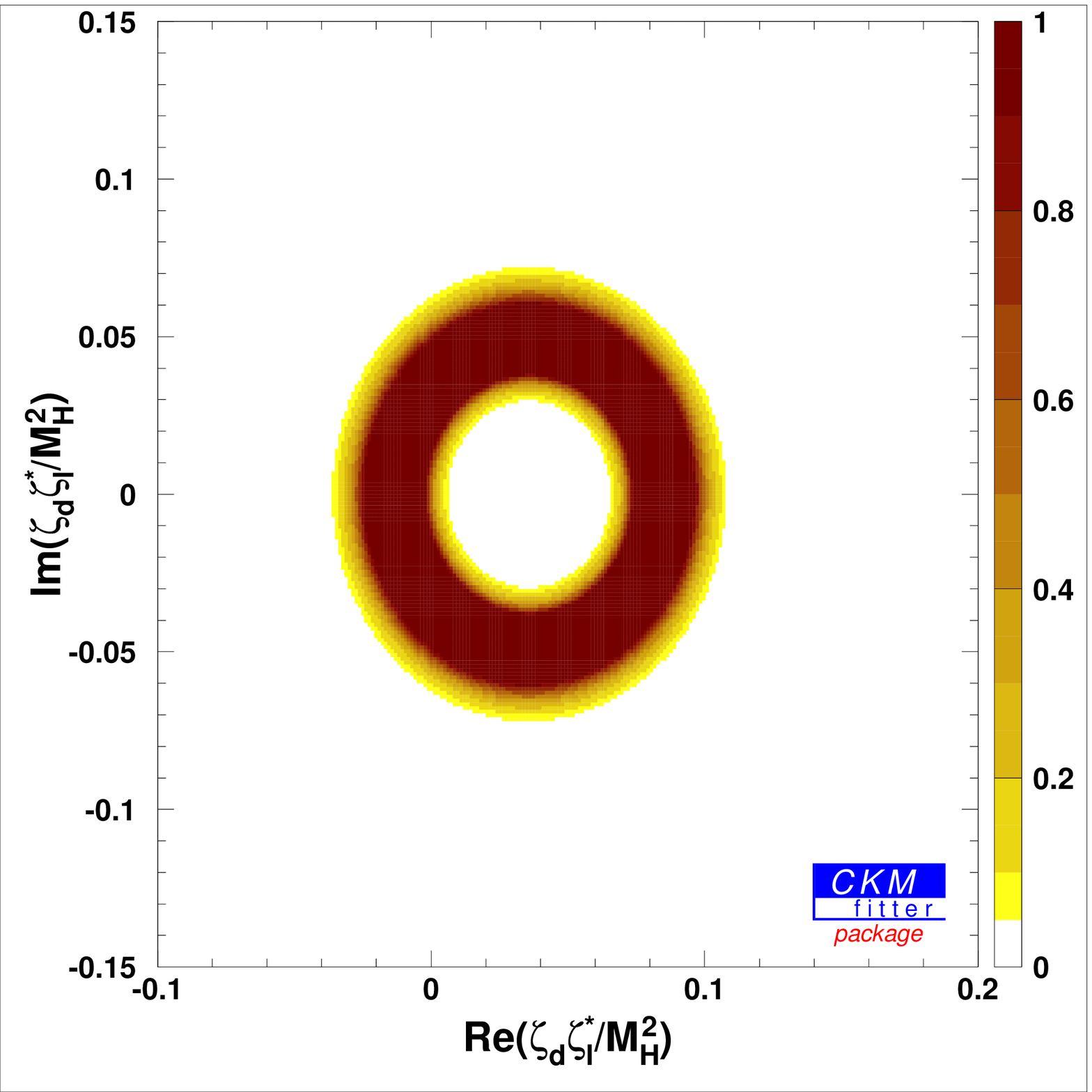} &
\includegraphics[width=7cm]{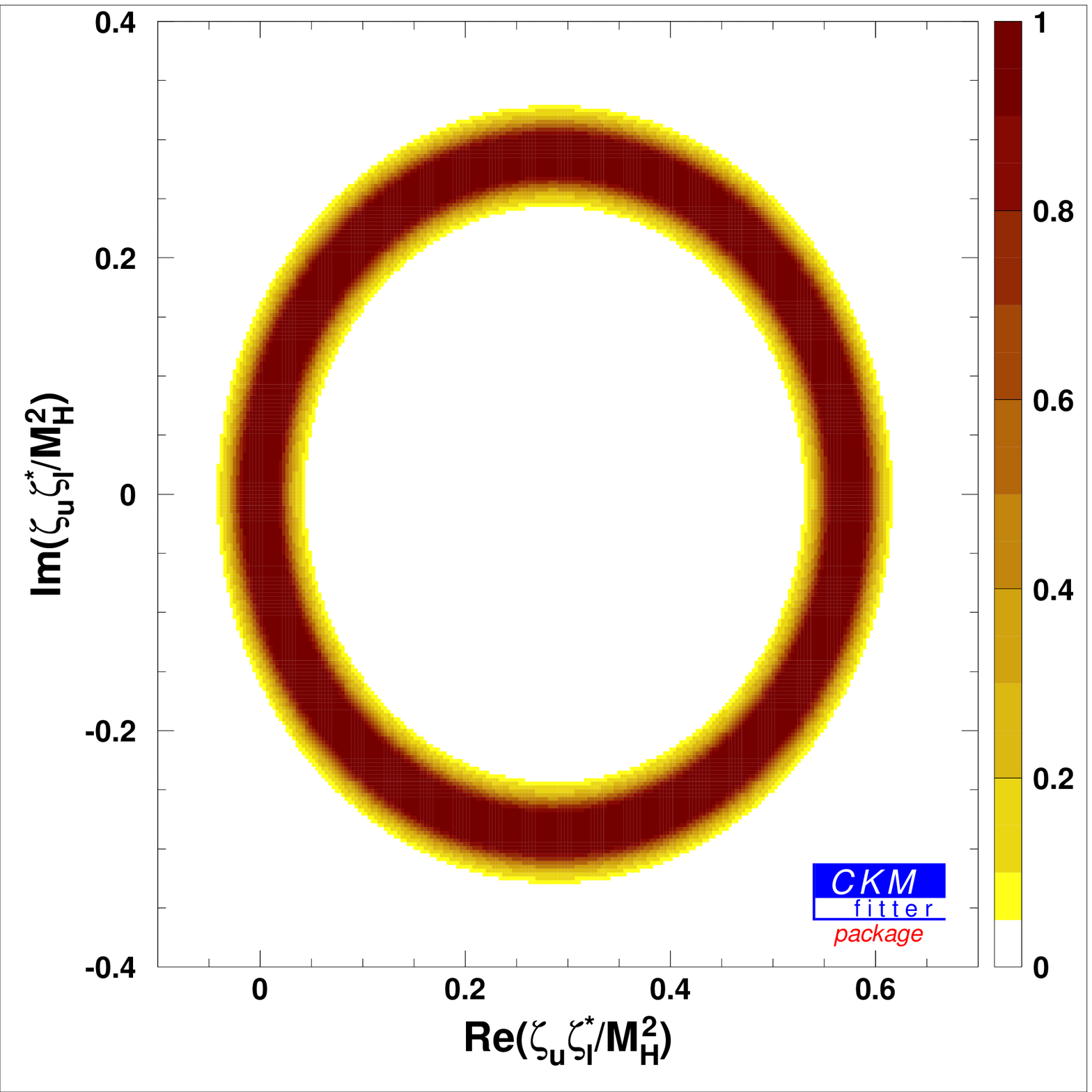}
\end{tabular}
\caption{\it Constraints in the complex $\varsigma_l^*\varsigma_{u,d}/M_{H^{\pm}}^2$ planes from $B\to\tau\nu$ (left) and $D\to\mu\nu$ (right), in units of $GeV^{-2}$. The colour code indicates confidence levels ($1-CL$). \label{fig::btaudmu}}
\end{center}
\end{figure}

The leptonic decays of heavy pseudoscalar mesons that have been measured up to now are $B \rightarrow \tau \nu$, $D_s \rightarrow \mu \nu$, $D_s \rightarrow \tau \nu$  and $D\rightarrow \mu \nu$. The radiative corrections for the leptonic decays of heavy mesons have been estimated in \cite{Burdman:1994ip}, and are already taken into account in the experimental values given in table~\ref{tab::measurements}; therefore the electromagnetic correction is set to zero in Eq.~(\ref{eq:Gamma_Plnu}).

In $B$ and $D$ decays the function $\Delta_{ij}$ can be approximated by neglecting the contribution proportional to the light quark mass, because $m_u/m_b\lesssim m_d/m_c\sim\mathcal O(10^{-3})$. Therefore the relations
\begin{eqnarray}
\Delta_{ub} \approx \frac{m_B^2}{M_{H^{\pm}}^2}\,\varsigma_l^*\varsigma_d \; , \qquad\qquad \Delta_{cd} \approx \frac{m_D^2}{M_{H^{\pm}}^2}\,\varsigma_l^*\varsigma_u \; 
\end{eqnarray}
hold, leading to a direct constraint on these combinations.
While for $D_{(s)}\to\tau\nu$ the helicity suppression is absent, the corresponding phase space is small
and there are two neutrinos in the final state, which is why $D\to\tau\nu$ has not been measured up to now. Nevertheless, the upper limit set by CLEO \cite{:2008sq} starts to become relevant in constraining our parameters: $|1-\Delta_{cd}| < 1.19$ (95\% CL).
The present experimental limit on $B\to\mu\nu$ gives
$|1-\Delta_{ub}| < 2.04$ (95\% CL). The information obtained from the decays $B\to\tau\nu$ and $D\to\mu\nu$ is shown in figure~\ref{fig::btaudmu}. The broad dark red (black) ring in the middle reflects the fact, that the systematic error is dominant in these constraints, leading to a large amount of degeneracy for the `best fit value'. To infer a limit at a certain confidence level, the corresponding number of rings has to be included, for example for $95\%$ up to the yellow (light grey) corresponding to $1-CL=0.05$. The resulting 95\% CL constraints,
$|1-\Delta_{ub}|\in[0.8,2.0]$ \ and \ $|1-\Delta_{cd}|\in[0.87,1.12]$, translate into allowed circular bands in the $\varsigma_l^*\varsigma_{u,d}/M_{H^{\pm}}^2$ complex planes.
For real Yukawa couplings there is a two-fold sign ambiguity generating two possible solutions,
the expected one around $\Delta_{ij}=0$ (the SM amplitude dominates) and its mirror around
$\Delta_{ij}=2$, corresponding to a new-physics contribution twice as large as the SM one and of opposite sign. The real solutions are
$\varsigma_l^*\varsigma_d/M_{H^{\pm}}^2 \in[-0.036,0.008]~\mathrm{GeV}^{-2}$ and $[0.064,0.108]~\mathrm{GeV}^{-2}$, and
$\varsigma_l^*\varsigma_u/M_{H^{\pm}}^2 \in [-0.037,0.037]~\mathrm{GeV}^{-2}$ and $[0.535,0.609]~\mathrm{GeV}^{-2}$.

\begin{figure}[tb]
\begin{center}
\begin{tabular}{cc}
\includegraphics[width=7cm]{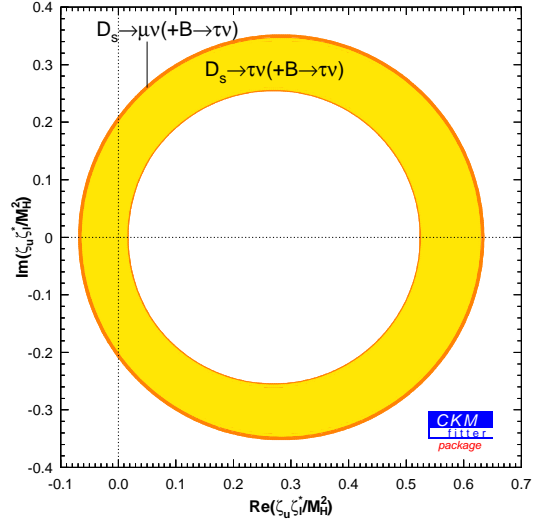} &
\end{tabular}
\caption{\it 95\% CL constraints in the complex $\varsigma_l^*\varsigma_{u}/M_{H^{\pm}}^2$ plane
 from $D_s\to (\tau,\mu)\nu$, in units of $GeV^{-2}$, using $B\to\tau\nu$ to constrain $\varsigma_l^*\varsigma_d/M_{H^{\pm}}^2$.
 \label{fig::dslnu}}
\end{center}
\end{figure}

\begin{figure}[tb]
\begin{center}
\begin{tabular}{cc}
\includegraphics[width=7cm]{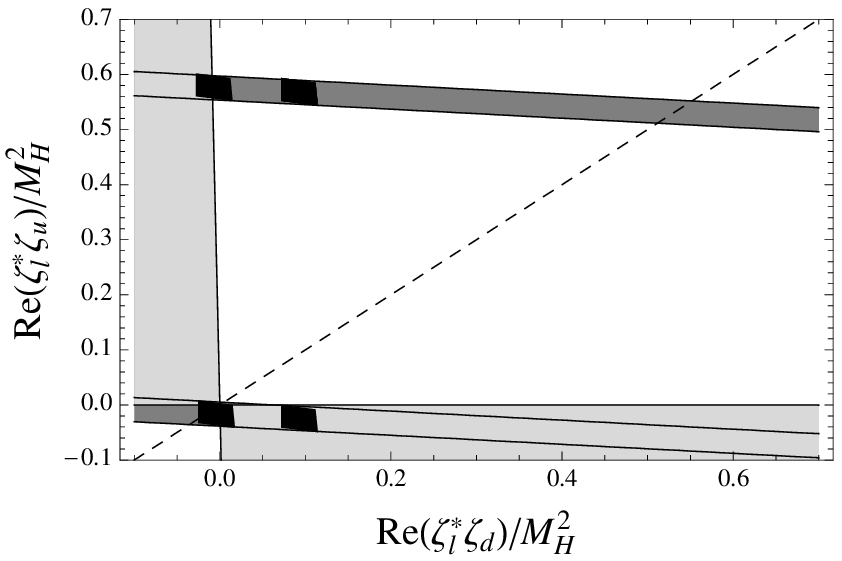} &
\includegraphics[width=7cm]{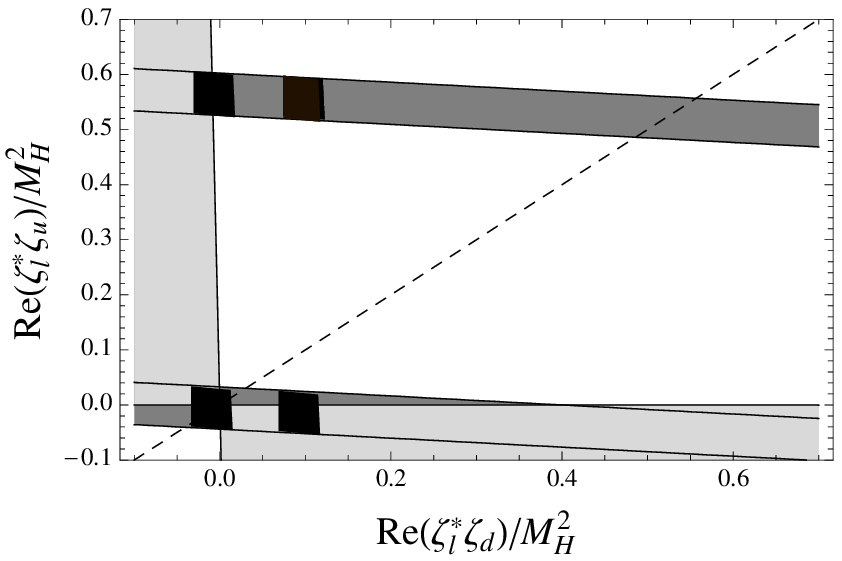}
\end{tabular}
\caption{\it
Constraints from
$D_s \rightarrow \tau \nu_{\tau}$ (left) and $D_s \rightarrow \mu \nu_{\mu}$ (right), in units of $GeV^{-2}$,
under the assumption of real parameters $\varsigma_f$. The grey bands correspond to $95\%$ CL.
Also shown are the cuts for the
\thdm of type I/X (dashed line) and II (lighter grey area, $\tan{\beta}\in [0.1,60]$). Finally, the four black regions are the possible allowed areas considering the information coming from $B\rightarrow \tau \nu_{\tau}$.\label{fig::Dsreal}}
\end{center}
\end{figure}

In $\mathrm{D}_s$ decays we get $|1-\Delta_{cs}|\in[0.97,1.18]$ from $D_s\to\mu\nu$ and $|1-\Delta_{cs}|\in[0.98,1.16]$ from $D_s\to\tau\nu$. Here
the situation is a bit more complex, because $m_s/m_c\approx 10\%$ and the light-quark term in the $\Delta_{cs}$ function cannot be neglected since this suppression could be compensated by the different $\varsigma_f$. Therefore there is no direct constraint, neither on $\varsigma_l^*\varsigma_u/M_{H^{\pm}}^2$ nor on $\varsigma_l^*\varsigma_d/M_{H^{\pm}}^2$,
only a correlation among them.
For that reason, we use the additional information from $B\to\tau\nu$
to constrain the parameters which are not shown.
This suffices to render the influence of the mass-suppressed term subdominant.

If $CP$ symmetry were only broken by the CKM phase, the parameters $\varsigma_f$ would be real.
In this case, the constraints from $D_s \rightarrow \tau \nu_{\tau}$ and $D_s \rightarrow \mu \nu_{\mu}$ can be visualized as shown in figure~\ref{fig::Dsreal}, plotting the correlation between the two real parameters. The two grey bands are associated with the two possible solutions around
$\Delta_{cs} = 0$ and  $\Delta_{cs} = 2$.
The different models with $\mathcal{Z}_2$-symmetry correspond to cuts in these plots. The plots show the small influence of the term proportional to the strange quark mass, as long as the couplings are of the same order.
Using the constraints on $\varsigma_l^*\varsigma_d/M_{H^{\pm}}^2$ from $B\to\tau\nu$, one finds
for the other coupling combination the two real solutions
$\varsigma_l^*\varsigma_u/M_{H^{\pm}}^2  \in[-0.005,0.041]~\mathrm{GeV}^{-2}$ and $[0.511,0.557]~\mathrm{GeV}^{-2}$, at 95\% CL, which
agree with the corresponding constraints from $D\to\mu\nu$. Putting together all the information
from leptonic B, D and $\mathrm{D}_s$ decays, the real solutions are:
\begin{equation}
\frac{\varsigma_l^*\varsigma_d}{M_{H^{\pm}}^2}\; \in\; \left\{
\begin{array}{l}
[-0.036,0.008]~\mathrm{GeV}^{-2}\, , \\[5pt] [0.064,0.108]~\mathrm{GeV}^{-2}\, ,
\end{array}\right.\qquad\qquad
\frac{\varsigma_l^*\varsigma_u}{M_{H^{\pm}}^2}\; \in\; \left\{
\begin{array}{l}
[-0.006,0.037]~\mathrm{GeV}^{-2}\, , \\[5pt] [0.511,0.535]~\mathrm{GeV}^{-2}\, .
\end{array}\right.
\end{equation}

\subsubsection{Light pseudoscalar mesons}

\begin{figure}[tbh]
\begin{center}
\begin{tabular}{cc}
\parbox{7cm}{\includegraphics[width=7cm]{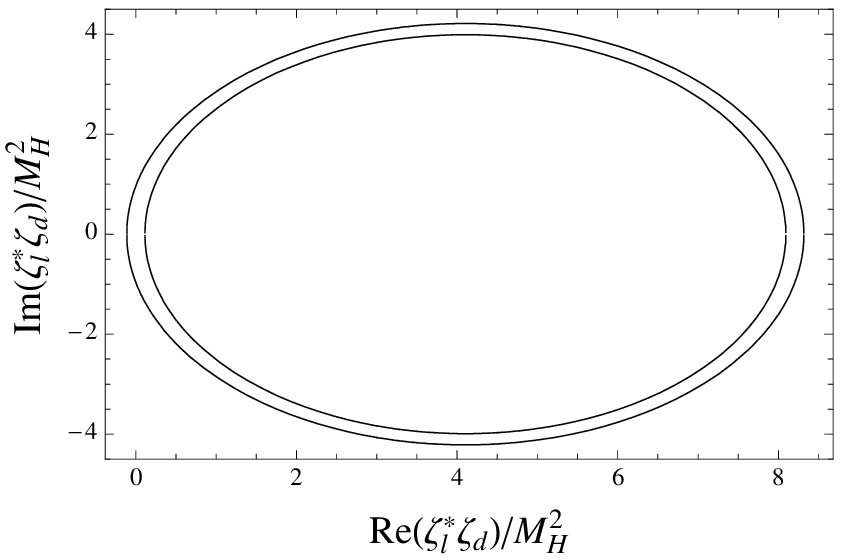}\\[10pt]\includegraphics[width=7cm]{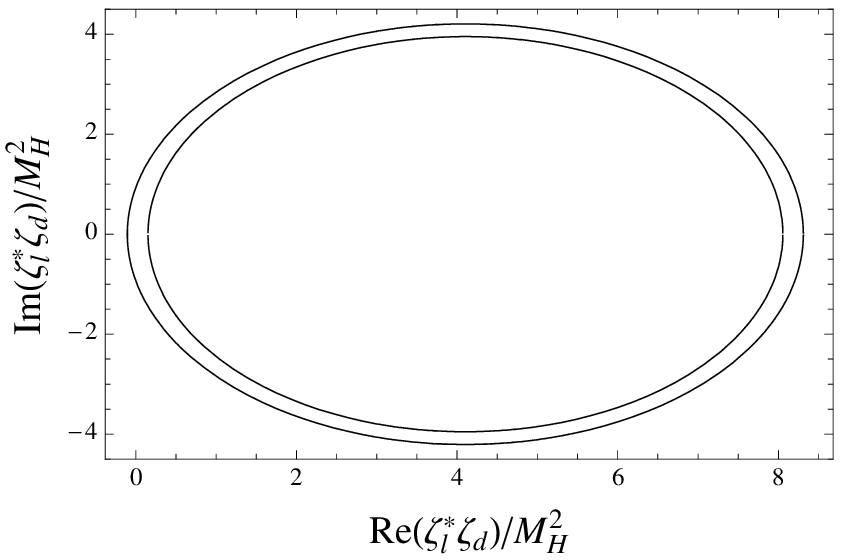}} &\quad\raisebox{-3.3cm}{\includegraphics[width=7cm]{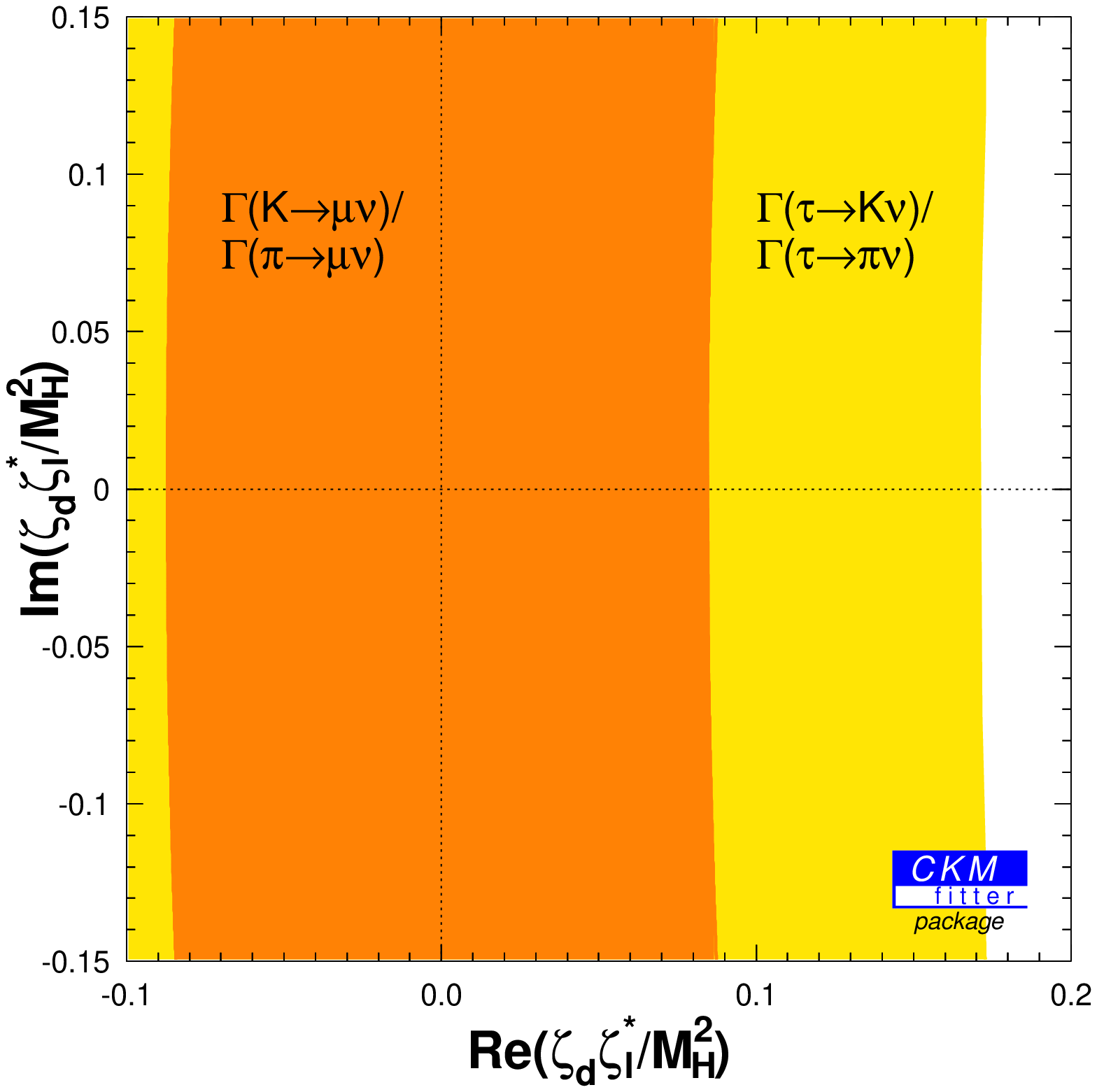}}
\end{tabular}
\caption{\it Constraints in the complex plane
$(\varsigma_l^*\varsigma_{d})/M_{H^{\pm}}^2$, in units of $GeV^{-2}$. Left: Full regions allowed at $95\%~CL$ for $K/\pi\to\mu\nu$ (upper plot) and $\tau\to K/\pi\nu$ (lower plot). Right: 95\% CL constraints in the interesting region (from the global fit) for both constraints, using $D\to\mu\nu$ to constrain $\varsigma_l^*\varsigma_{u}/M_{H^{\pm}}^2$.
\label{fig::kopilnu}}
\end{center}
\end{figure}

Due to the cancellation of common uncertainties, lattice calculations of the ratio $f_K/f_\pi$ are
more precise than the determinations of the individual decay constants. This ratio can be extracted
experimentally from two different ratios of decay widths:
\begin{eqnarray}
\frac{\Gamma(K\to\mu\nu)}{\Gamma(\pi\to\mu\nu)}&=&\frac{m_K}{m_\pi}\left(\frac{1-m_\mu^2/m_K^2}{1-m_\mu^2/m_\pi^2}\right)^2
\left|\frac{V_{us}}{V_{ud}}\right|^2\left(\frac{f_K}{f_\pi}\right)^2(1+\delta_{\mathrm{em}}^{Kl2/\pi l2})
\left|\frac{1-\Delta_{us}}{1-\Delta_{ud}}\right|^2 \; ,\\
\frac{\Gamma(\tau\to K\nu)}{\Gamma(\tau\to\pi\nu)}&=&\left(\frac{1-m_K^2/m_\tau^2}{1-m_\pi^2/m_\tau^2}\right)^2\left|\frac{V_{us}}{V_{ud}}\right|^2
\left(\frac{f_K}{f_\pi}\right)^2(1+\delta_{\mathrm{em}}^{\tau K2/\tau\pi 2})\left|\frac{1-\Delta_{us}}{1-\Delta_{ud}}\right|^2\; ,
\end{eqnarray}
where $\delta_{\mathrm{em}}^{Kl2/\pi l2}$ is given in table~\ref{tab::hadronic} and
$\delta_{\mathrm{em}}^{\tau K2/\tau\pi2}=\delta_{\mathrm{em}}^{(\tau K2/K\ell2)} +\delta_{\mathrm{em}}^{K\ell2/\pi\ell2}-\delta_{\mathrm{em}}^{\tau \pi2/\pi\ell2} = 0.0004\pm 0.0054$.

The new-physics corrections are dominated by $\Delta_{us}\simeq \varsigma_l^*\varsigma_d m_K^2/M_{H^\pm}^2$.
As $m_K^2/m_B^2\sim1\%$, the scalar contributions to these decays are much smaller than for the heavy mesons. However, the good experimental precision achieved provides interesting constraints, as shown in figure~\ref{fig::kopilnu}, which are dominated by the $K_{\mu 2}/\pi_{\mu 2}$ ratio.
At 95\% CL, one finds $|1-\Delta_{us}| \in[0.984,1.017]$ from $K_{\mu 2}/\pi_{\mu 2}$ and
$|1-\Delta_{us}| \in[0.965,1.025]$ from the ratio $\tau\to \nu K/\pi$. The real solutions are then,
$\varsigma_l^*\varsigma_d /M_{H^\pm}^2 \in[-0.07,0.07]~\mathrm{GeV}^{-2}$ or
$[8.14,8.28]~\mathrm{GeV}^{-2}$. The larger real solution is already excluded by the $B\to\tau\nu$ data.

\subsection{Semileptonic decays of pseudoscalar mesons}
Semileptonic decays receive contributions from a charged scalar as well, but in this case the leading SM amplitude is not helicity suppressed, therefore the relative influence is smaller.
In addition, there are momentum-dependent form factors involved. The decay amplitude $M\rightarrow M' l\bar\nu_l$ is characterized by two form factors, $f_+(t)$ and $f_0(t)$ associated with the P-wave and S-wave projections of the crossed-channel matrix element
$\langle 0|\bar u_i\gamma^\mu d_j|M \bar M'\rangle$. The scalar-exchange amplitude only contributes to the scalar form factor; it amounts to a multiplicative correction
\begin{equation}\label{eq:sFFmod}
\tilde f_0(t)\, = \, f_0(t)\;\left( 1 + \delta_{ij}\, t\right)\, ,
\end{equation}
where
\begin{equation}
\delta_{ij}\,\equiv \, -\frac{\varsigma_l^*}{M_{H^{\pm}}^2}\,
\frac{\varsigma_um_{u_i}-\varsigma_dm_{d_j}}{m_{u_i}-m_{d_j}}\,.
\end{equation}

The determination of the CKM matrix element $|V_{ij}|$ is not contaminated by the new-physics contribution, because it is governed by the vector form factor. One measures the electron mode $M\rightarrow M' e\bar\nu_e$, where the scalar contribution is heavily suppressed by the electron mass, determining the product $|V_{ij}|\, |f_+(t_0)|$, with $t_0=0$ for light-quark transitions and $ t_0 = (m_M-m_{M'})^2$ for heavy quarks. A theoretical calculation of $|f_+(t_0)|$ is then needed to extract $|V_{ij}|$. The sensitivity to the scalar contribution can only be achieved in semileptonic decays into heavier leptons. Whenever available, one can make use of the differential decay distribution to separate the scalar and vector amplitudes. In any case, theoretical determinations of the scalar and vector form factors are needed to extract information on $\delta_{ij}$.

\subsubsection{$B\to D\tau\nu_\tau$}

To reduce the uncertainty from the vector form factor, let us consider the ratio
\begin{equation}
\frac{\Br (B\rightarrow D \tau \nu_{\tau})}{\Br (B \rightarrow D e\nu_e)}\, =\,
a_0 + a_1 \left(m_B^2-m_D^2\right) \mathrm{Re}(\delta_{cb}) + a_2 \left(m_B^2-m_D^2\right)^2 |\delta_{cb}|^2 \; .
\end{equation}
The coefficients $a_i$, which contain the dependence on the strong-interaction dynamics, have been studied recently and parametrized
in terms of the vector form-factor slope $\rho^2$ and the scalar density $\Delta(v_B\cdot v_D)\equiv\Delta$, assumed to be constant
\cite{Deschamps:2009rh,Kamenik:2008tj}. We make use of these parametrizations, taking for the two parameters the values indicated in table~\ref{tab::hadronic}.
The function $\Delta(v_B\cdot v_D)\propto f_0(t)/f_+(t)$ has been studied in the lattice, in the range
$v_B\cdot v_D = 1$--1.2, and found to be consistent with a constant value $\Delta = 0.46\pm 0.02$, very close to its static-limit approximation $(m_B-m_D)/(m_B+m_D)$ \cite{deDivitiis:2007uk}.

\begin{figure}[tbh]
\begin{center}
\begin{tabular}{c c}
\includegraphics[width=6cm]{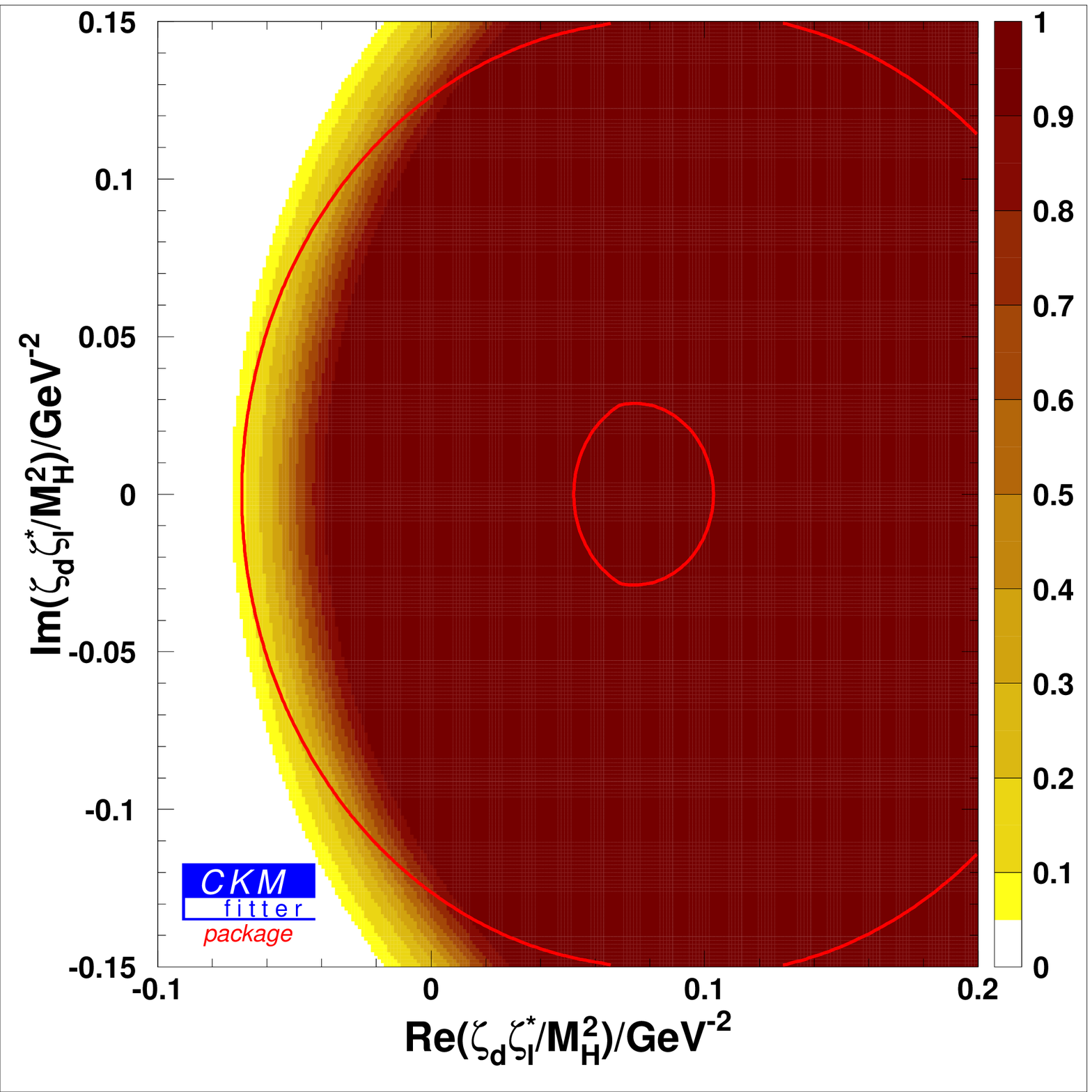} &  \hskip .5cm
\includegraphics[width=6cm]{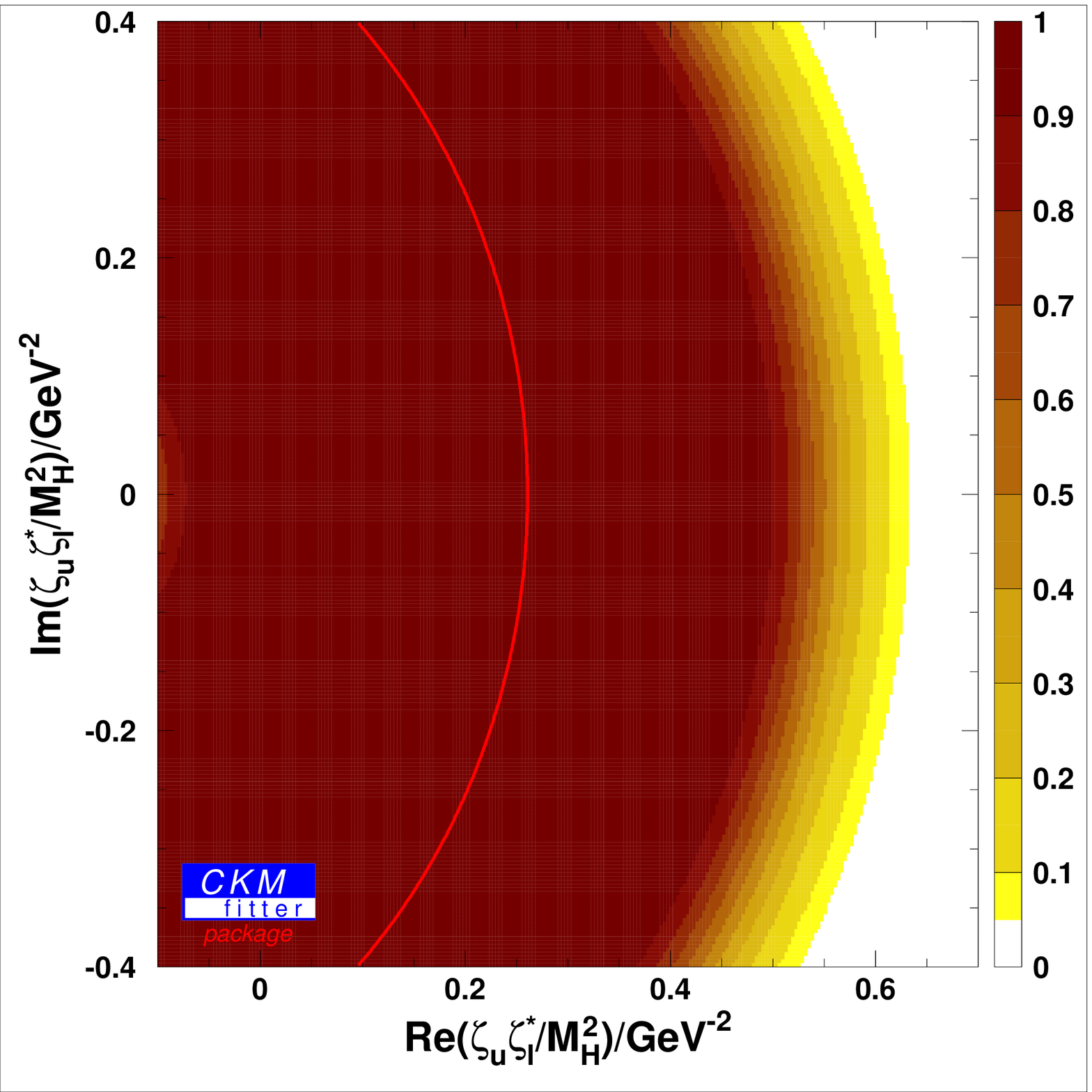}
\end{tabular}
\caption{\it Constraints from $B\to D\tau\nu_\tau$, in units of $GeV^{-2}$, plotted in the complex plane for $\varsigma_l^*\varsigma_{d}/M_{H^{\pm}}^2$ (left) and $\varsigma_l^*\varsigma_{u}/M_{H^{\pm}}^2$ (right), using $D\to\mu\nu$ and $B\to\tau\nu$ to constrain the combination not shown, respectively. The colours indicate $1-CL$, the red lines the constraint (95\% CL) for $\varsigma_l^*\varsigma_{u,d}/M_{H^\pm}^2\to0$.\label{fig::bdlnu}}
\end{center}
\end{figure}

We obtain once more a correlation between
$\varsigma_l^*\varsigma_u/M_{H^{\pm}}^2$ and $\varsigma_l^*\varsigma_d/M_{H^{\pm}}^2$, where the term proportional to the charm quark mass is in general potentially more important than in the type II model.
The results are shown in figure~\ref{fig::bdlnu} for both parameter combinations. As can be seen there, the constraint on $\varsigma_d\varsigma_l^*/M_{H^{\pm}}^2$ is consistent with the information %
coming from $B\to\tau\nu$ and the leptonic decays of light mesons, but does not constrain this combination further
as long as only the information of $B\to D\tau\nu_\tau$ is used. The red lines indicating the constraint for $\varsigma_l^*\varsigma_u\to0$, however, show that the semileptonic decay can exclude a small region around $(0.08,0)$, once that combination is bound to be small. We will use this to exclude the second real solution for $\varsigma_l^*\varsigma_d/M_{H^\pm}^2$ with aid of the processes $\epsilon_K,Z\to b\bar{b}$ and $\tau\to\mu\nu\nu$ (see figure~\ref{fig::globalfit}).
Also, when plotted in the complex $\varsigma_l^*\varsigma_{u}/M_{H^{\pm}}^2$ plane, it becomes apparent that this constraint is important to exclude the second real solution allowed by $D_{(s)}\to\ell\nu$ decays, already using only the information from leptonic decays in addition (see again figure~\ref{fig::globalfit}).

\begin{figure}[tbh]
\begin{center}
\includegraphics[width=8cm]{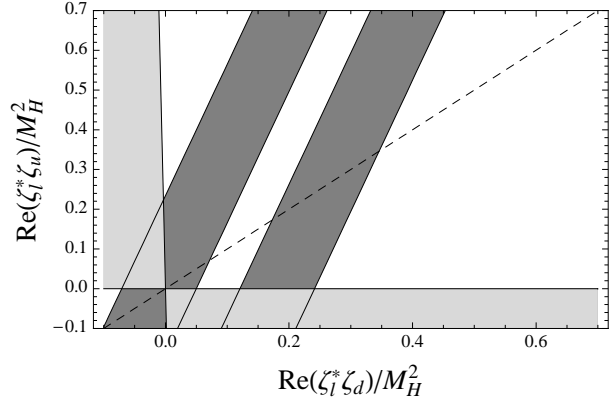}
\caption{\it Allowed regions for $Re(\varsigma_l^*\varsigma_d)/M_{H^{\pm}}^2$ and $Re(\varsigma_l^*\varsigma_u)/M_{H^{\pm}}^2$ from the process $B\to D\tau\nu$ at $95\%$ CL (grey), in units of $GeV^{-2}$, assuming that their imaginary parts are zero. The projections for the \thdmws s of types I/X (dashed line) and II (lighter grey area, $\tan{\beta}\in [0.1,60]$) are also shown.\label{fig::bdlnureal}}
\end{center}
\end{figure}

Considering the limit of real $\varsigma_f$'s, the correlation between the real parts is visualized in figure~\ref{fig::bdlnureal}, together with the cuts corresponding to the different models with $\mathcal{Z}_2$ symmetries. The plot shows that the $m_b$ and $m_c$ terms have potentially similar influence in this case.

It has been pointed out in \cite{Nierste:2008qe} that measuring the spectrum instead of just the branching ratio will increase the sensitivity of this channel. This, however, has not been done up to now, due to lack of statistics.

\subsubsection{$K\to\pi\ell\nu$}

In semileptonic kaon decays the Callan-Treiman theorem \cite{Callan:1966hu,Dashen:1969bh} allows to relate the scalar form factor at the kinematic point $t_{_{\mathrm{CT}}}=m_K^2-m_\pi^2$ to the ratio of kaon and pion decay constants:
$C\equiv f_0(t_{_{\mathrm{CT}}})/f_+(0) =\frac{f_K}{f_\pi}\frac{1}{f_+(0)}+\Delta_{_{\mathrm{CT}}}$,
where $\Delta_{_{\mathrm{CT}}} = (-3.5\pm 8)\cdot 10^{-3}$ is a small $\chi$PT correction of
$O[m_\pi^2/(4\pi f_\pi)^2]$ \cite{Gasser:1984ux,Kastner:2008ch,Passemar:2010cc}. Using a twice-subtracted dispersion relation for $f_0(t)$ \cite{Bernard:2006gy}, the constant $C$ has been determined from the $K_{\mu 3}$ data by KLOE \cite{:2007yza}, KTeV \cite{Abouzaid:2009ry} and NA48 \cite{Lai:2007dx}. In the average quoted in table~\ref{tab::measurements} the NA48 result has been excluded because it disagrees with the other two measurements by more than $2\sigma$.

\begin{figure}[hbt]
\begin{center}
\includegraphics[width=5cm]{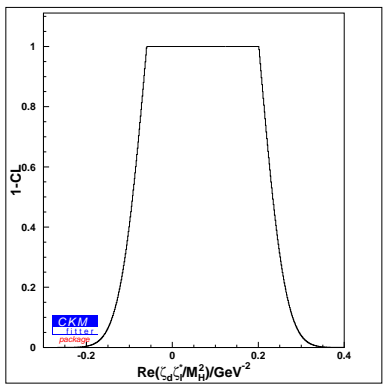}
\caption{\it Constraint from the direct measurement of $\log C$, in units of $GeV^{-2}$. \label{fig::kl3_logC}}
\end{center}
\end{figure}

In the presence of charged-scalar contributions, the scalar form factor gets modified as indicated in Eq.~(\ref{eq:sFFmod}), inducing a corresponding change in $C$. Taking into account that the analyzed experimental distribution is only sensitive to $|\tilde f_0(t)|^2$,
to first order in the new-physics correction $\delta_{us}$, the measured value of $C$ corresponds to
\begin{equation}
\log C=\log\left(\frac{f_K}{f_\pi}\frac{1}{f_+(0)}+\Delta_{CT}\right)
+ \mbox{Re}\left[ \delta_{us} (m_K^2-m_\pi^2)\right] \, .
\end{equation}
The resulting constraint on the real part of $\varsigma_d\varsigma_l^*/M_{H^{\pm}}^2$ is shown in figure~\ref{fig::kl3_logC}, leading to
\begin{equation}
\mathrm{Re}\left(\frac{\varsigma_l^*\varsigma_d}{M_{H^{\pm}}^2}\right)\;\in\; \mathrm{[-0.16,0.30]}~\mbox{GeV}^{-2} \quad (95\%~\mathrm{CL})\, ,
\end{equation}
which is in agreement with the previous constraints, but with larger uncertainties. This might change in the near future, due to improved lattice determinations of $f_+(0)$ and $f_K/f_\pi$, as well as improved experimental precision, e.g. from NA62 or KLOE-2.

\subsection{Global fit to leptonic and semileptonic decays}
\begin{figure}[tbh]
\centering{
\includegraphics[width=60mm]{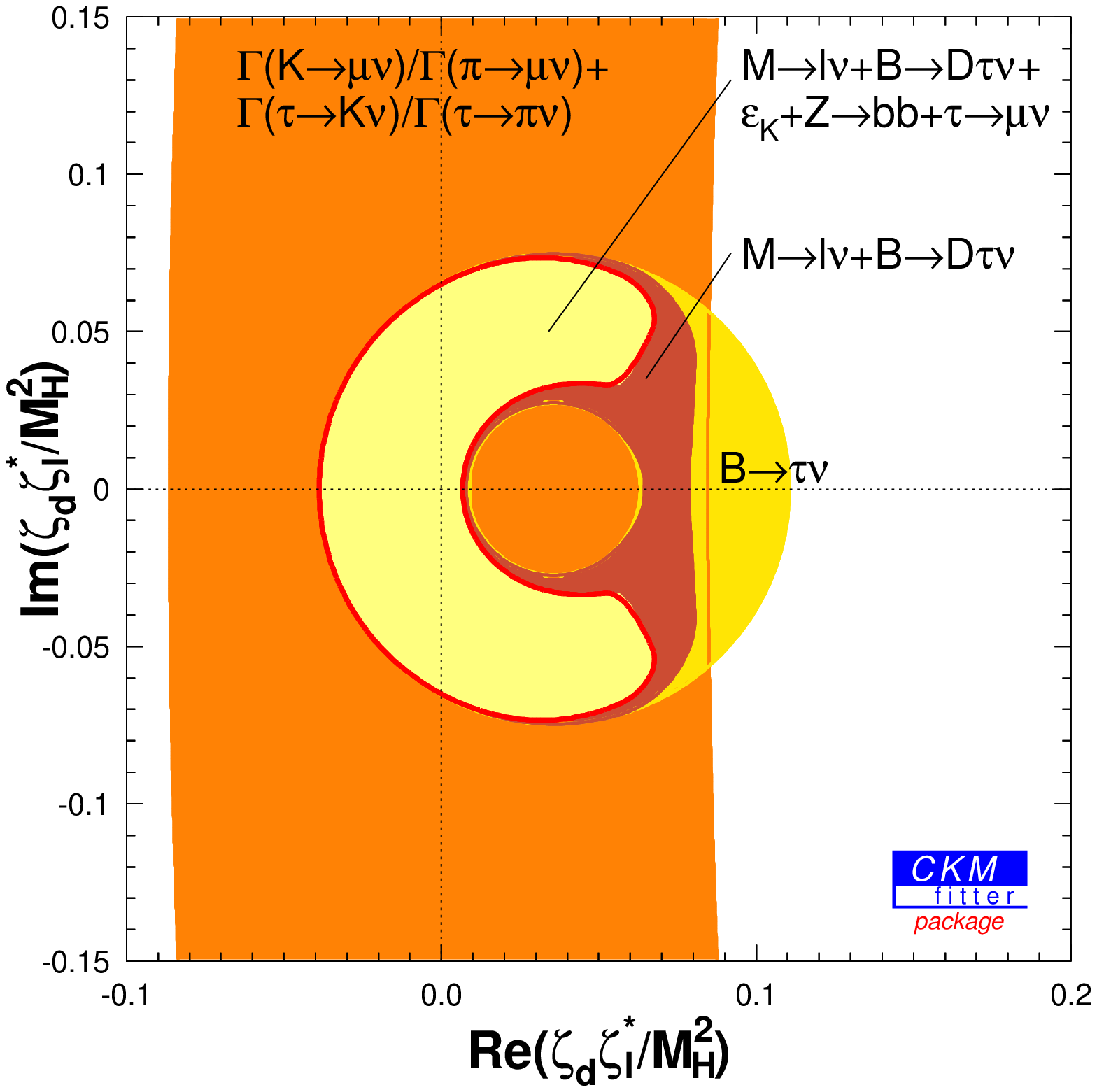}\quad\quad\includegraphics[width=60mm]{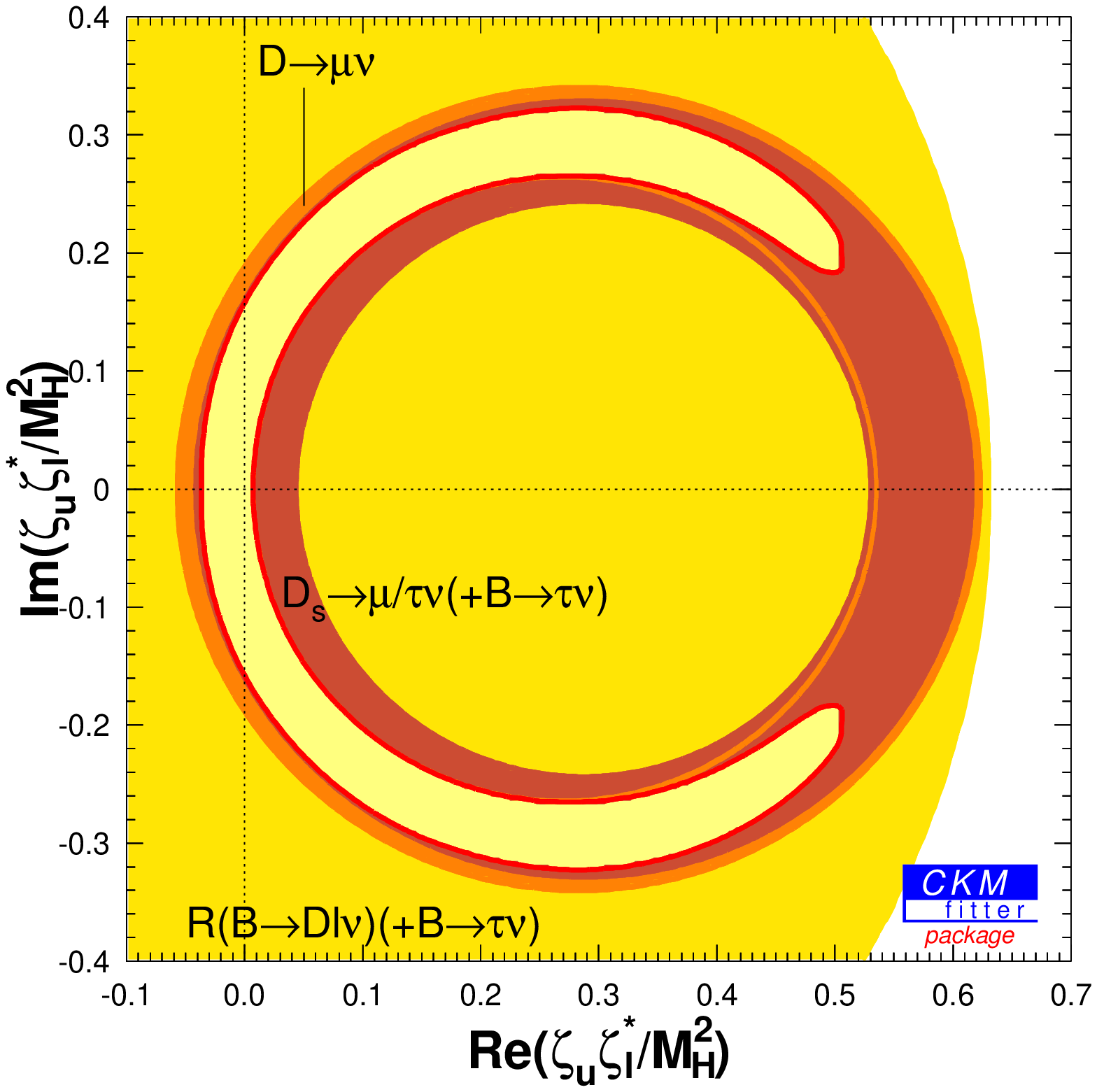}
\caption{\it $\varsigma_d\varsigma_l^*/M_{H^{\pm}}^2$ (left) and $\varsigma_u\varsigma_l^*/M_{H^{\pm}}^2$ (right) in the complex plane, in units of $GeV^{-2}$, constrained by leptonic and semileptonic decays. The inner yellow area shows the allowed region at 95\% CL, in the case of $\varsigma_d\varsigma_l^*/M_{H^{\pm}}^2$ using additional information (see text).
\label{fig::globalfit}} }
\end{figure}

Combining the information from all leptonic and semileptonic decays discussed before, one gets the constraints shown in figure~\ref{fig::globalfit}.
$|\varsigma_d\varsigma_l^*/M_{H^{\pm}}^2|$ is bounded to be smaller than $\sim0.1~{\rm GeV}^{-2}$ (95\% CL) from these decays alone, while for $\varsigma_u\varsigma_l^*/M_{H^{\pm}}^2$ the constraints are relatively weak,
due to the similar masses of the mesons in the leptonic decays.
Note that in both cases there are two real solutions.
For the combination $\varsigma_u\varsigma_l^*/M_{H^{\pm}}^2$, one real solution is excluded in the global fit at 95\%~CL, while the other, including the SM point of vanishing couplings remains allowed. As mentioned before, this exclusion is due to $B\to D\tau\nu$ in combination with the constraint on $\varsigma_d\varsigma_l^*/M_{H^{\pm}}^2$. For the latter, the situation is more complicated. The second solution remains allowed, due to the overlapping of the two main constraints in both regions and the weak constraint on $\varsigma_u\varsigma_l^*/M_{H^{\pm}}^2$ derived from semileptonic decays. However, using in addition the information coming from leptonic $\tau$ decays in (\ref{eq:sigma_l}), the lower Higgs mass bound from LEP and the constraint from $\epsilon_K,Z\to\bar{b}b$ (see section \ref{Ztobb}) in a conservative way, $|\varsigma_u\varsigma_l^*|/M_{H^\pm}^2\lesssim0.01~\mbox{GeV}^{-2}$,
the second real solution for $\varsigma_d\varsigma_l^*/M_{H^{\pm}}^2$ is excluded as well by $B\to D\tau\nu$.

\section{Loop-induced processes}\label{sec:loops}

For processes where new-physics contributions appear only through quantum loop effects,
the situation becomes obviously more difficult, regarding not only the calculation but also the interpretation of the results. If the SM amplitude is also mediated only by loops,
the relative importance of the charged-scalar contributions is expected to be higher, but this implies also a higher sensitivity to the framework in which the A2HDM is eventually to be embedded in. In the following we make the assumption that for the observables under discussion the dominant new-physics corrections are those generated by the charged scalar. Moreover, since no significant signal for new physics has been found up to now in flavour observables, we assume these effects to be subleading with respect to the SM contribution.

\subsection[$Z\to b\bar{b}$]{$\mathbf{Z\to b\bar{b}}$\label{Ztobb}}

The high-precision data collected at LEP and SLD has made it possible to accurately test the SM electroweak loop corrections at the
$Z$ scale, providing information on the Higgs mass and useful constraints on many new-physics scenarios. While most $Z$-peak observables are only sensitive to the gauge-boson selfenergies, the decay $Z\to b\bar{b}$ provides valuable information on fermionic vertex corrections induced by charged-current exchanges. Since $V_{tb}\approx 1$, those loop diagrams involving virtual top quarks generate quantum corrections to the $Zb\bar b$ vertex, which are absent in the $Zd\bar d$ and $Z s\bar s$ vertices. These corrections are enhanced by a factor $m_t^2$, allowing for a quite accurate determination of the top quark mass \cite{Bernabeu:1987me,Bernabeu:1990ws}. The same arguments apply to the charged-scalar contributions present in the \athdmws, providing a sensitive probe of the corresponding $H^+\bar t b$ coupling.
For very large values of $|\varsigma_d|$ this decay would also be sensitive to contributions from neutral scalars \cite{Haber:1999zh};
we don't consider this possibility here. However, given a not too small value for $\varsigma_l$, (semi-)leptonic decays can be used to exclude that possibility.

Therefore, we assume the dominance of charged-scalar effects in the following, allowing only for $|\varsigma_d|\leq50$.
We disregard the information coming from the forward-backward polarization asymmetry $A_b$, because the scalar-exchange contributions to $A_b$ are small compared to the present uncertainties.

It is convenient to normalize the $Z\to b\bar{b}$ decay width to the total hadronic width of the $Z$, because many QCD and electroweak corrections cancel in the ratio, amplifying the sensitivity to the wanted vertex contribution \cite{Bernabeu:1990ws}.
Within the \athdmws, this ratio can be written as \cite{Haber:1999zh,Field:1997gz,Degrassi:2010ne}
\begin{equation}\label{Eq::Ztobb1}
R_b \equiv \frac{\Gamma(Z\to\bar{b}b)}{\Gamma(Z\to \mbox{hadrons})}\, =\,
\left[1+\frac{S_b}{s_b}\, C_b^{\rm QCD}\right]^{-1}\, ,
\end{equation}
where
\begin{equation}\label{Eq::Ztobb2}
s_q = \left[(\bar{g}_b^L-\bar{g}_b^R)^2 + (\bar{g}_b^L+\bar{g}_b^R)^2\right]
\,\left( 1 +\frac{3\alpha}{4\pi}\, Q_q^2\right)\, ,
\qquad\qquad
S_b \equiv\sum_{q\not= b,t} s_q\, ,
\end{equation}
with
$C_b^{\rm QCD} = 1.0086$ being a factor including QCD
corrections \cite{Chetyrkin:1996ia}.
The \athdm contributions are encoded through the effective left- and right-handed $Zb\bar b$-couplings:
\begin{eqnarray}
\bar{g}_b^L&=&\bar{g}_{b,SM}^L\, +\, \frac{\sqrt{2}\, G_F M_W^2}{16\pi^2}\; \frac{m_t^2}{M_W^2}\, |\varsigma_u|^2\;
\left[ f_1(t_h) + \frac{\alpha_s}{3\pi}\, f_2(t_h)\right]\,,\\
\bar{g}_b^R&=&\bar{g}_{b,SM}^R\, -\, \frac{\sqrt{2}\, G_F M_W^2}{16\pi^2}\;\frac{m_b^2}{M_W^2}\, |\varsigma_d|^2\;
\left[ f_1(t_h) + \frac{\alpha_s}{3\pi}\, f_2(t_h)\right]\,,
\end{eqnarray}
where $t_h\equiv m_t^2/M_{H^\pm}^2$,
$f_1(t_h) = [t_h^2-t_h-t_h\log{t_h}]/(1-t_h)^2$ and the function $f_2(t_h)$ governing the NLO correction is given in \cite{Degrassi:2010ne}. If running quark masses $\bar{m}_t(M_Z)$ and $\bar{m}_b(M_Z)$ are used, this NLO QCD correction is small.
The light-quark coupling contribution $S_b = 1.3214$ \cite{LEPZresonance,Degrassi:2010ne} is not sensitive to the new-physics effects.
The SM values of the couplings $\bar{g}_{b,SM}^{L,R}$, given in table~\ref{tab::hadronic}, have been computed removing the $Z\to b\bar b$
information from the standard electroweak fit \cite{LEPZresonance,Degrassi:2010ne}.

\begin{figure}[tb]
\begin{center}
\includegraphics[width=70mm]{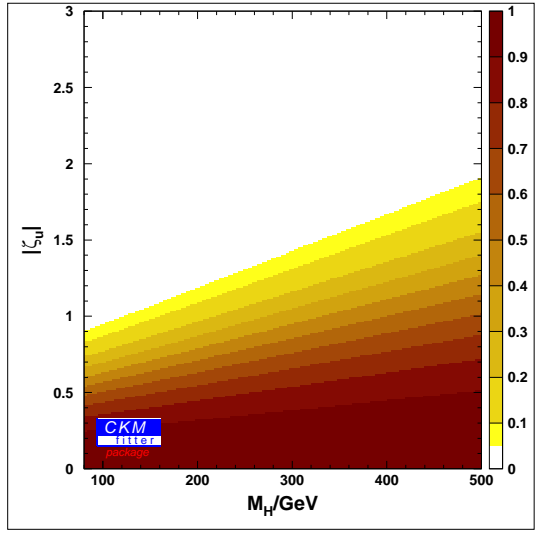}
\caption{\it \label{fig:Rb} Constraint from $R_b$ in the $|\varsigma_u|-M_{H^\pm}$ plane ($M_{H^\pm}$ in GeV units), allowing for $|\varsigma_d|\leq50$.}
\end{center}
\end{figure}

In contrast to the leptonic and semileptonic constraints discussed before, here the parameters $|\varsigma_{u,d}|$ enter directly, allowing to bound them without information on $|\varsigma_l|$. The constraint resulting from the input values in tables~\ref{tab::hadronic} and~\ref{tab::measurements} is shown in figure~\ref{fig:Rb}.
The constraint is plotted in the $|\varsigma_u|-M_{H^\pm}$ plane, as obviously it is much weaker for $|\varsigma_d|$, due to the relative factor $m_b/m_t$. For large scalar masses, the constraint weakens as the effects start to decouple, reflected in $\lim_{t_h\to0}f_{1,2}(t_h)=0$. In the range of scalar masses considered, it leads to a 95\%~CL upper bound 
$|\varsigma_u|\leq 0.91~(1.91)$,
for $M_{H^\pm}= 80\; (500)$~GeV. The upper bound increases linearly with $M_{H^\pm}$, implying
\begin{equation}\label{eq:zeta_u_limit}
\frac{|\varsigma_u|}{M_{H^\pm}}\; <\; 0.0024~\mathrm{GeV}^{-1} + \frac{0.72}{M_{H^\pm}} < 0.011~\mathrm{GeV}^{-1}\, ,
\end{equation}
where we have used the lower bound on the charged-scalar mass from LEP searches, $M_{H^\pm}> 78.6$~GeV (95\% CL) \cite{Pich:2009sp,:2001xy}.
Combined with the limit on $|\varsigma_l/M_{H^\pm}|$ from leptonic $\tau$ decays, this already constrains the combination $|\varsigma_u\varsigma_l^*|/M_{H^\pm}^2$ much stronger than the global fit to (semi)leptonic decays, leading to
\begin{equation}\label{eq:zeta_ul_limit}
\frac{|\varsigma_u\varsigma_l^*|}{M_{H^\pm}^2}\; <\; 0.005~\mathrm{GeV}^{-2}\,,
\end{equation}
however only with the additional assumptions of $|\varsigma_d|\leq50$ and charged-scalar effects dominating the new-physics contributions to $R_b$.
The range allowed for $|\varsigma_d|$ in the fit does not influence the upper bound on $|\varsigma_u|$, apart from the exclusion of neutral-scalar effects, since both contributions can only lower the value for $R_b$ and both are allowed to vanish in the fit. Therefore the upper limit stems from points with $|\varsigma_d|=0$.

\subsection[$B^0$-$\bar B^0$ mixing]{$\mathbf{B^0}$-$\mathbf{\bar B^0}$ mixing}
\label{subsec:Bmixing}

The mixing of neutral $B$ mesons is very sensitive to charged-scalar effects, as the leading contribution stems from top-quark loops, rendering the new-physics and SM contributions comparable. Besides the high precision of the measurement for the mass difference $\Delta m_{B^0}$, the $B^0_s$ mixing is especially interesting due to the observed tension in its phase \cite{Abazov:2010hv,Barberio:2008fa}. In the usual \thdmws s with a $\mathcal{Z}_2$ symmetry the scalar couplings are necessarily real, leading to a vanishing contribution to this phase. However, the complex Yukawa couplings $\varsigma_{u,d}$ of the \athdm provide a potential new-physics contribution, which could account for the experimentally observed phase.

In the SM, the calculation is simplified by the fact that only one operator contributes, denoted $\mathcal{O}^{\rm VLL}$ below. In the presence of a charged scalar, an enlarged effective Hamiltonian
\begin{equation}
\mathcal{H}^{\Delta B=2}_{\rm eff}=\frac{G_F^2 M_W^2}{16\pi^2}\; \left( V_{td}^* V_{tb}^{\phantom{*}}\right)^2\;\sum_{i}
C_i(\mu)\,\mathcal{O}_i
\end{equation}
has to be considered, involving a basis of eight operators \cite{Gerard:1984bg,Gabbiani:1996hi,Buras:2001ra,Becirevic:2001jj}:
\begin{eqnarray}\label{eq:DB=2operators}
\mathcal{O}^{\rm VLL,VRR}&=&\left(\bar{d}^\alpha\gamma_\mu \cP_{L,R} b^\alpha\right)\left(\bar{d}^\beta\gamma^\mu \cP_{L,R} b^\beta\right)\,,\nonumber\\
\mathcal{O}_1^{\rm LR}&=&\left(\bar{d}^\alpha\gamma_\mu \cP_L b^\alpha\right)\left(\bar{d}^\beta\gamma^\mu \cP_R b^\beta\right)\,,\nonumber\\
\mathcal{O}_2^{\rm LR}&=&\left(\bar{d}^\alpha \cP_L b^\alpha\right)\left(\bar{d}^\beta \cP_R b^\beta\right)\,,\\
\mathcal{O}_1^{\rm SLL,SRR}&=&\left(\bar{d}^\alpha \cP_{L,R} b^\alpha\right)\left(\bar{d}^\beta \cP_{L,R} b^\beta\right)\, ,\nonumber\\
\mathcal{O}_2^{\rm SLL,SRR}&=&\left(\bar{d}^\alpha \sigma_{\mu\nu}\cP_{L,R} b^\alpha\right)\left(\bar{d}^\beta \sigma^{\mu\nu}\cP_{L,R} b^\beta\right)\, ,\nonumber
\end{eqnarray}
with $\alpha,\beta$ being colour indices and $\sigma^{\mu\nu}=\frac{1}{2}[\gamma^\mu,\gamma^\nu]$. We have written the effective Hamiltonian relevant for $B^0_d$-$\bar B^0_d$ mixing; the mixing of $B_s^0$ mesons is described by the analogous expression, changing the label $d$ to $s$ everywhere.

We have performed the matching of the underlying \athdm and the low-energy effective Hamiltonian at the scale $\mu_{tW}\sim M_W,m_t$.
The resulting Wilson coefficients, given in the appendix, reproduce the SM result as well as the matching for the \thdm in the limit $m_d\to0$, given in \cite{Urban:1997gw}.
As noted above, the contribution of the \athdm to $C_{\rm VLL}(\mu_{tW})$ is an $\mathcal{O}(1)$ effect. For that reason, we calculate this contribution at NLO, implementing the results of \cite{Urban:1997gw} within the \athdmws .\footnote{Note, that there are several smaller errors in that paper, most of which have been pointed out in \cite{WahabElKaffas:2007xd}.}
Owing to their chirality structure, the remaining Wilson coefficients are all suppressed by powers of the light-quark mass $m_d$ ($m_s$ in the $B^0_s$ case), except $C^1_{\rm SRR}$ which is proportional to $m_b^2$. Restricting the parameter ranges to $|\varsigma_u|\in[0,5]$ and $|\varsigma_d|\in[0,50]$, the ratio $|C_i(\mu_{tW})/C_{\rm VLL}(\mu_{tW})|$ is then below two percent for all operators apart from $\mathcal{O}_1^{\rm SRR}$. Since the matrix elements for the $B^0$ mixing do not contain the large (chiral) enhancement factors
present in the kaon system,
this allows us to restrict ourselves to two operators only. Moreover, the ratio $C^1_{\rm SRR}/C_{\rm VLL}$ is a small quantity
($10\%$ at most for $|\varsigma_d|\leq25$, still below $40\%$ for $|\varsigma_d|=50$) and therefore a leading-order estimate of the
 $\mathcal{O}_1^{\rm SRR}$ contribution is enough for our purposes, while the dominant $\mathcal{O}^{\rm VLL}$ contribution is included at NLO.

The strong $(m_s-m_d)/M_W$ suppression of SU(3)-breaking effects implies that, for the parameter ranges considered, the ratio $\Delta m_{B^0_s}/\Delta m_{B^0_d}$ is unaffected by charged-scalar contributions and can be used in the CKM fit. Note, however, that in the limit $|\varsigma_d|\gg50,\,|\varsigma_u|\ll1$, which corresponds to the large--$\tan\beta$ scenario in the type II model, the contribution from $\mathcal{O}^{\rm VRR}$ might become the dominant new-physics correction to $B_s^0$ mixing, but remains small compared to the SM one.

We use the ratio $\Delta m_{B^0_s}/\Delta m_{B^0_d}$ to determine the apex $(\bar\rho,\bar\eta)$ of the unitarity triangle, and bound the charged-scalar parameters with the $B^0_s$ mixing information.
The resulting constraint from $\Delta m_{B^0_s}$ in the $M_{H^\pm}$-- $|\varsigma_u|$ plane is shown in figure~{\ref{fig::deltams}, using the scales $\mu_{tW}=m_t$ and $\mu_b=4.2$~GeV. The error includes the variations in the CKM parameters, $f_{B^0_s}$, $\hat B_{B^0_s}$ and the experimental uncertainty. The leading $\mathcal{O}^{\rm VLL}$ contribution depends on $|\varsigma_u|^2$ only, while $C^1_{\rm SRR}$ is proportional to $\varsigma_u^*\varsigma_d^{\phantom{*}} = |\varsigma_u| |\varsigma_d|  \e^{i\varphi}$, $\varphi$ being the relative phase between the two Yukawa couplings. To determine the allowed region shown in figure~{\ref{fig::deltams}, we have varied $\varsigma_d$ in the range $|\varsigma_d| < 50$ and $\varphi\in [0,2\pi]$.

\begin{figure}[tb]
\centering{
\includegraphics[width=70mm]{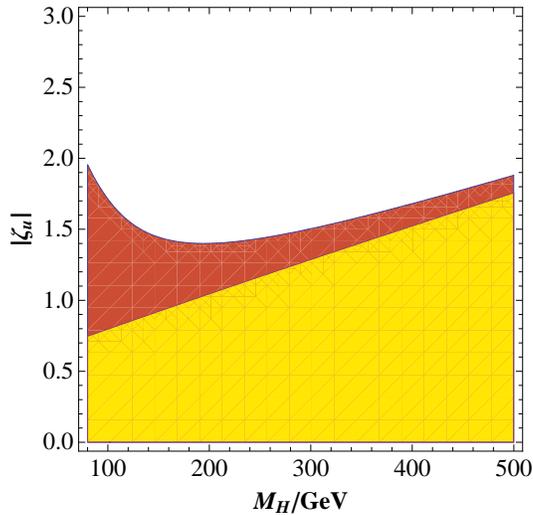}
}
\caption{\it The $95\%~CL$ constraint coming from $\Delta m_{B^0_s}$ in the $M_{H^\pm}$-- $|\varsigma_u|$ plane for $|\varsigma_d|\in[0,50]$, varying in addition the relative phase $\varphi$ in $[0,2\pi]$. The excluded area lies above the dark (red) region only. In yellow the allowed area for $\varsigma_d=0$ is shown.\label{fig::deltams}}
\end{figure}

Interestingly, the dominant contribution to a possible phase shift in the mixing is also the one from $\mathcal{O}_1^{\rm SRR}$. The factor $M_W^4D_0(m_t,M_{H^\pm})$ (see appendix) varies between zero and $\sim-3\%$ for scalar masses between 50 and 500~GeV, while $4m_b^2 m_t^4/M_W^6\sim10\%$. For relatively large values of the product $|\varsigma_u^*\varsigma_d^{\phantom{*}}|$ ($\gtrsim20$) this factor can contribute sizeably to the $B^0$ mixing phase, as long as $M_{H^\pm}$ is relatively small. The sign of the shift is obviously not fixed, but depends on the sign of the relative Yukawa phase $\varphi$. As long as $|\varsigma_d|$ is not too large, the effect is the same in $B^0_d$ and $B^0_s$.

The D0 experiment has measured very recently \cite{Abazov:2010hv} a like-sign dimuon charge asymmetry leading to $A_{sl}^b=-0.00957\pm0.00251\pm0.00146$, which differs by over three standard deviations from the SM prediction \cite{Lenz:2006hd,Ciuchini:2003ww}. The measurement includes contributions from $B^0_d$ and $B^0_s$ mesons, corresponding to
$A_{sl}^b=(0.506\pm0.043)\, a_{sl}^d+(0.494\pm 0.043)\, a_{sl}^s$, with ($q=d,s$)
\begin{equation}
a_{sl}^{q} \, =\, \mathrm{Im}\left(\frac{\Gamma_{12}^q}{M_{12}^q}\right)\, =\,
\frac{|\Gamma_{12}^q|}{|M_{12}^q|}\,\sin{\phi_q}\, =\,
\frac{\Delta\Gamma_{B^0_q}}{\Delta M_{B^0_q}}\,\tan{\phi_q}\, ,
\end{equation}
where $M_{12}^q -\frac{i}{2}\, \Gamma_{12}^q\equiv \langle B^0_q|\mathcal{H}^{\Delta B=2}_{\mathrm{eff}}|\bar{B}_q^0\rangle$.
While this result needs certainly confirmation, we will explore some of its consequences for the parameters of the \athdm  in the following.
Using the current experimental value for the asymmetry in the $B^0_d$ system, $a_{sl}^d = -0.0047\pm 0.0046$ \cite{Barberio:2008fa}, the measured value of $\Delta M_{B^0_s}$ and the SM prediction for $\Delta\Gamma_{B^0_s}$, the D0 asymmetry implies
$\sin\phi_s=-2.7\pm1.4\pm1.6$, showing that the central value of this measurement is incompatible with the assumption of negligible influence of new physics on $\Gamma_s^{12}$, while the uncertainties are large enough to allow every value for the mixing phase at $2\sigma$.
Using in addition the direct measurement of $a_{sl}^s$ through
$B^0_s\to\mu^+D_s^-X$ decays by D0 \cite{Abazov:2009wg}, $a_{sl}^s=-0.0017\pm0.0091$, results in
$\sin\phi_s=-1.7\pm1.1\pm1.0$. 
Note that (part of) the observed deviation may also be due to the possibility of bad convergence of the operator product expansion (OPE) \cite{Grinstein:2001nu,Grinstein:2001zq}, related to the relatively low effective energy scale $m_b-2m_c$. However, no signs for a breakdown were found in the above calculation. Note also that in \cite{Berger:2010wt} it has been argued that such a large value violates a ``coherence bound'' derived by demanding monotonicity of the Stokes vector in the $B_s$ system. The possibility of NP influence on the rate as an explanation for this measurement has recently been discussed in \cite{Bauer:2010dg,Dighe:2010nj,Deshpande:2010hy}. The authors of \cite{Bauer:2010dg} conclude, that most of the possible operators are strongly constrained by other processes (including the one discussed in \cite{Dighe:2010nj}), leaving little space for an $\mathcal{O}(1)$ contribution to $\Gamma_{12}^s$.

Hints of a large $\phi_s$ value have been also obtained previously
from $B_s^0\to J/\psi\phi$ decays \cite{Abazov:2007tx,Aaltonen:2007he,:2008fj}, where the extraction of the phase might however be influenced by contributions to the decay amplitude:
in the SM, one of the reasons why this decay is ``golden'' is the fact, that the potentially relatively large penguin contributions have the same phase as the leading (colour-suppressed) tree amplitude, and therefore do not spoil the extraction of the mixing phase from the time-dependent $CP$ asymmetry. However, this is no longer true in the \athdmws: the charged-scalar penguin contributions include terms similar to their leading SM counterparts, with an additional factor of $\varsigma_u^*\varsigma_d m_bm_t/M_{H^\pm}^2\sim\mathcal{O}(1)$, thereby providing a second weak phase in the decay amplitude. Quantitatively assessing the influence of these contributions would require a reliable calculation of the corresponding matrix elements, which is however not available; we are thus left with the possibility of a semi-quantitive analysis only, e.g. along the lines of \cite{Feldmann:2008fb}, which we however do not consider 
here.

The SM predicts a very small positive value for $\phi_s$ and a much larger and negative result for
$\phi_d$. The theoretical values quoted in \cite{Lenz:2006hd} are
$\phi_s= 0.24^\circ\pm 0.08^\circ$ and $\phi_d= -5.2^\circ\, {}^{+\, 1.5^\circ}_{-\, 2.1^\circ}$.

\begin{figure}[hbt]
\begin{center}
\begin{tabular}{c c}
\includegraphics[width=7cm]{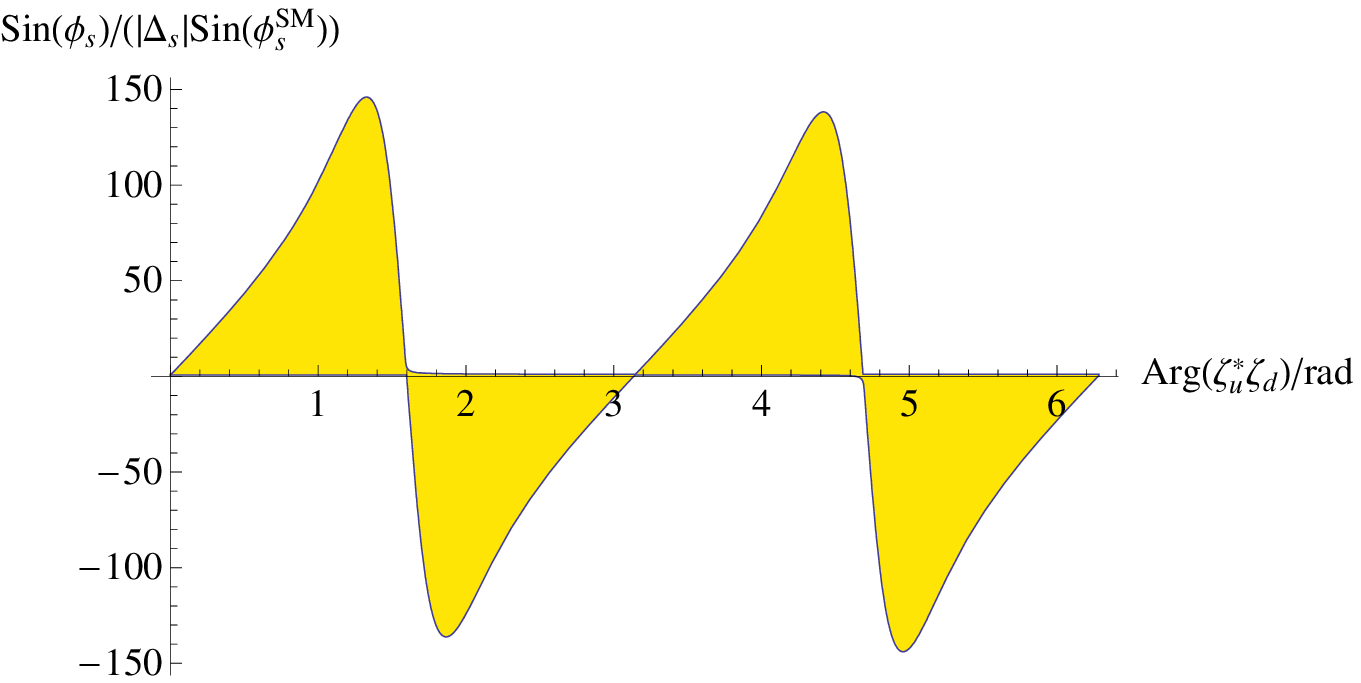} &  \hskip .5cm
\includegraphics[width=7cm]{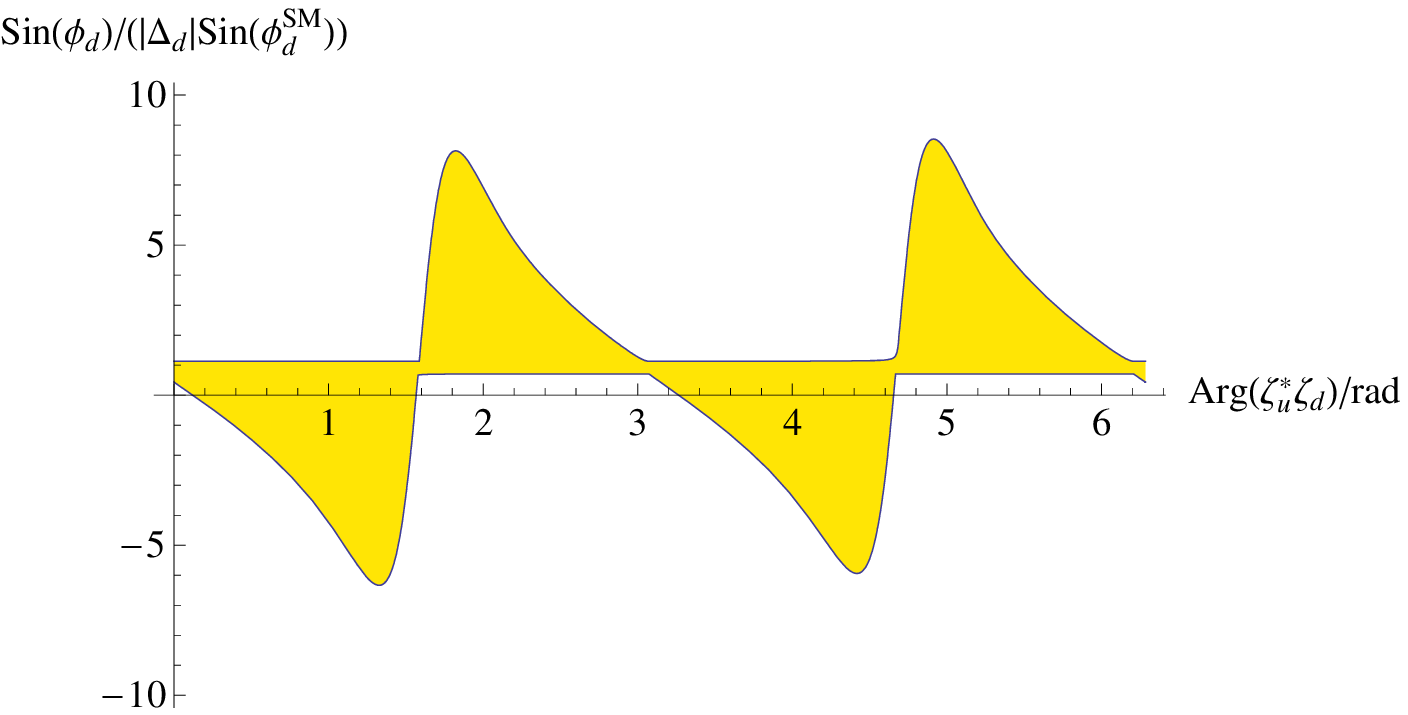}
\end{tabular}
\caption{\it Dependence of $\sin{\phi_q}/(|\Delta_q|\sin{\phi_q^{\mathrm{SM}}})$ on
$\varphi\equiv\arg{(\varsigma_u^*\varsigma_d^{\protect\phantom{*}})}$. \label{fig:Phis_correlation}}
\end{center}\end{figure}

Assuming that the charged-scalar contributions are the only relevant new-physics effects, we
can analyze the possibility to accommodate a large $\phi_s$ phase within the \athdmws.
In figure~\ref{fig:Phis_correlation} we plot the allowed range for
$\sin{\phi_q}/(|\Delta_q|\sin{\phi_q^{\mathrm{SM}}})$, where
$\Delta_q\equiv M_{12}^q/M_{12}^{q,\mathrm{SM}}$, as a function of the relative Yukawa phase
$\varphi\equiv\arg{(\varsigma_u^*\varsigma_d^{\protect\phantom{*}})}$.
The other scalar parameters have been varied in the ranges
$|\varsigma_d|\in[0,50]$, $M_{H^\pm}\in [80,500]$~GeV, and $|\varsigma_u|$ according to the allowed range from $\epsilon_K,Z\to\bar{b}b$, which includes only values for $(|\varsigma_u|,M_{H^\pm})$ which lead to acceptable values for $\Delta m_{s,d}$. While it is indeed possible
to obtain a large value of $\phi_s$, the predicted equality of $\Delta_s$ and $\Delta_d$
implies a strong anti-correlation of $\sin{\phi_d}/(|\Delta_d|\sin{\phi_d^{\mathrm{SM}}})$
and $\sin{\phi_s}/(|\Delta_s|\sin{\phi_s^{\mathrm{SM}}})$, due to the different sign (and size)
of $\phi_d^{\mathrm{SM}}$ and $\phi_s^{\mathrm{SM}}$.
This leads to a prediction for the sign of $a_{sl}^d$, which could be verified/falsified, once higher experimental precision is achieved.
As can be seen, the preferred negative sign for the $a_{sl}^s$ asymmetry implies $\varphi\in[\pi/2,\pi],[3\pi/2,2\pi]$, and for possible large values the Yukawa phase should not be close to $0,\pi$ (obviously).

Figure \ref{fig:Phis} shows the dependence of $\sin{\phi_s}/(|\Delta_s|\sin{\phi_s^{\mathrm{SM}}})$ with
$|\varsigma_d|$ (left) and $M_{H^\pm}$ (right), varying the remaining parameters within their allowed ranges. If large values for the $a_{sl}^s$ asymmetry are confirmed (within the physical range $|\sin{\phi_s}|\leq 1$), this would point towards large values of $|\varsigma_d|$ and small charged scalar masses.
\begin{figure}[hbt]
\begin{center}
\begin{tabular}{c c}
\includegraphics[height=4cm]{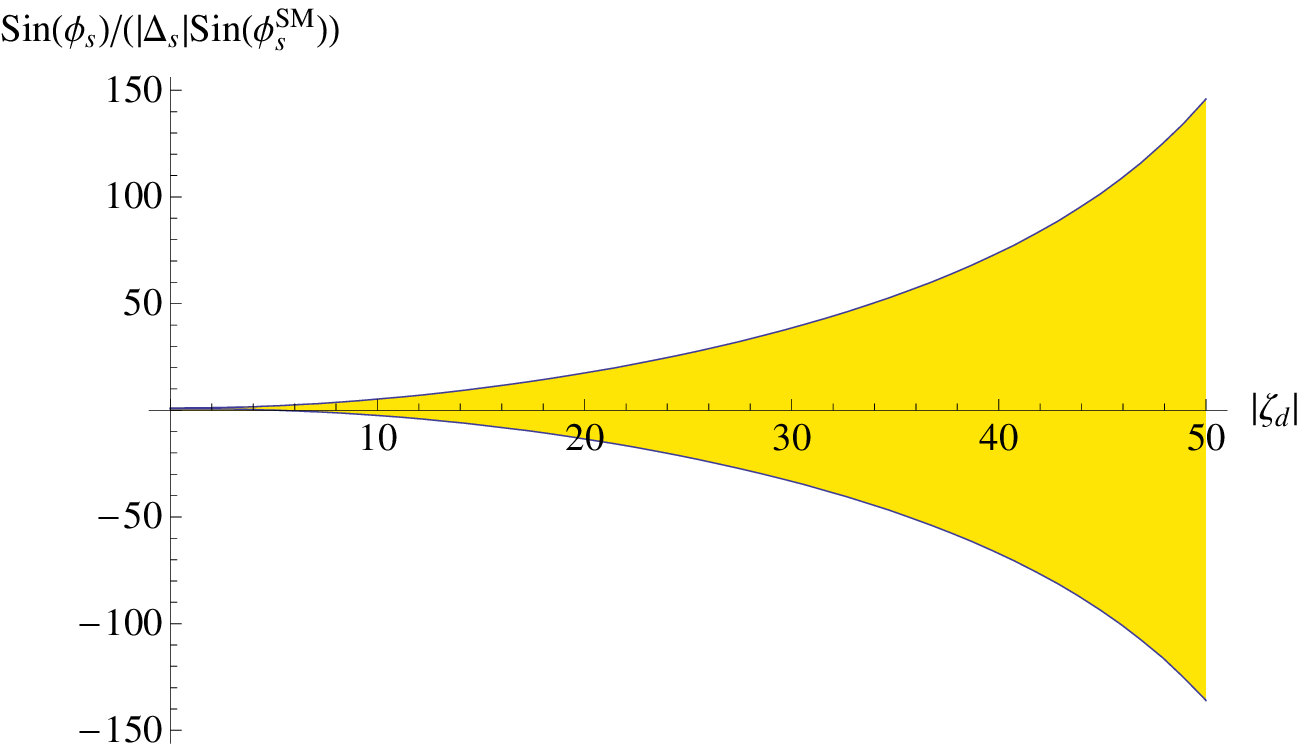} &  \hskip .5cm
\includegraphics[height=4cm]{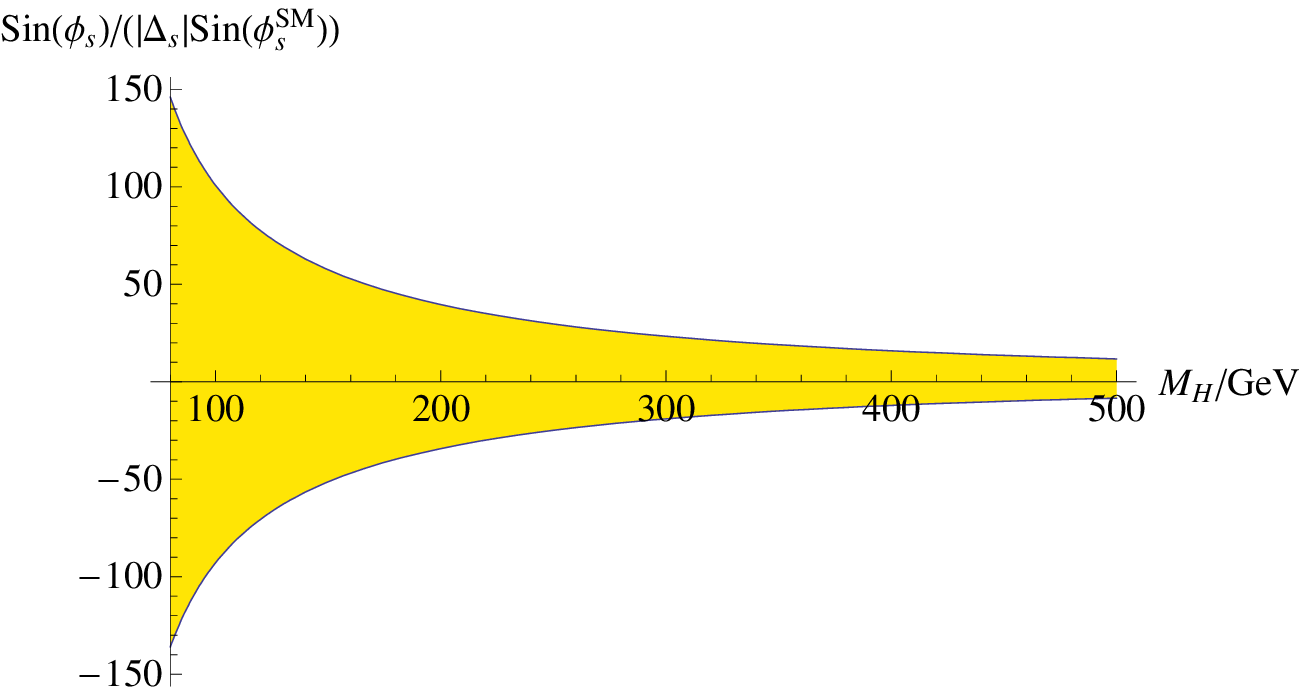}
\end{tabular}
\caption{\it Dependence of $\sin{\phi_s}/(|\Delta_s|\sin{\phi_s^{\mathrm{SM}}})$ on
$|\varsigma_d|$ (left) and $M_{H^\pm}$ (right). \label{fig:Phis}}
\end{center}\end{figure}
Finally we show in figure~\ref{fig::Phis_correlation_btos} the plots from figure~\ref{fig:Phis_correlation} again, restricting the product $|\varsigma_u\varsigma_d^*|\leq20$ (see section~\ref{sec::btos}). The corresponding maximal asymmetry is correspondingly smaller, but still relative factors up to $\sim 60$ are allowed for $B_s$ with respect to the SM.

\begin{figure}[!hbt]
\begin{center}
\begin{tabular}{c c}
\includegraphics[width=7cm]{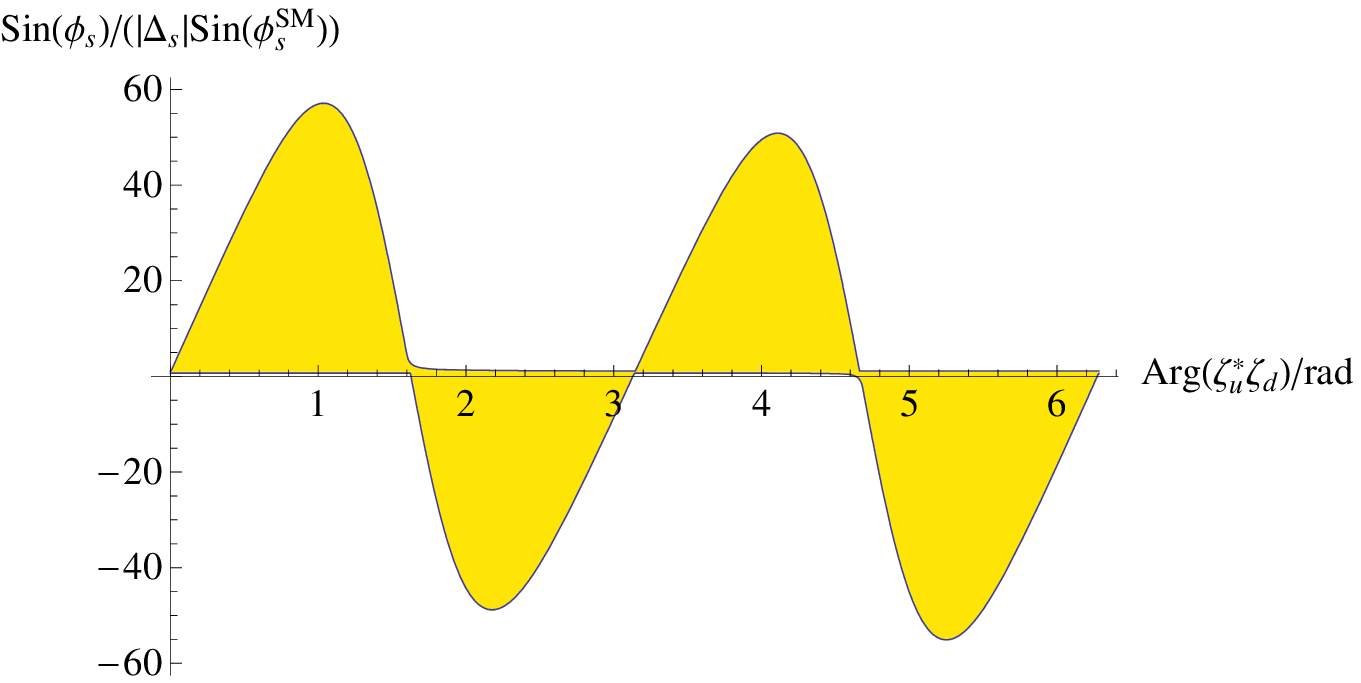} &  \hskip .5cm
\includegraphics[width=7cm]{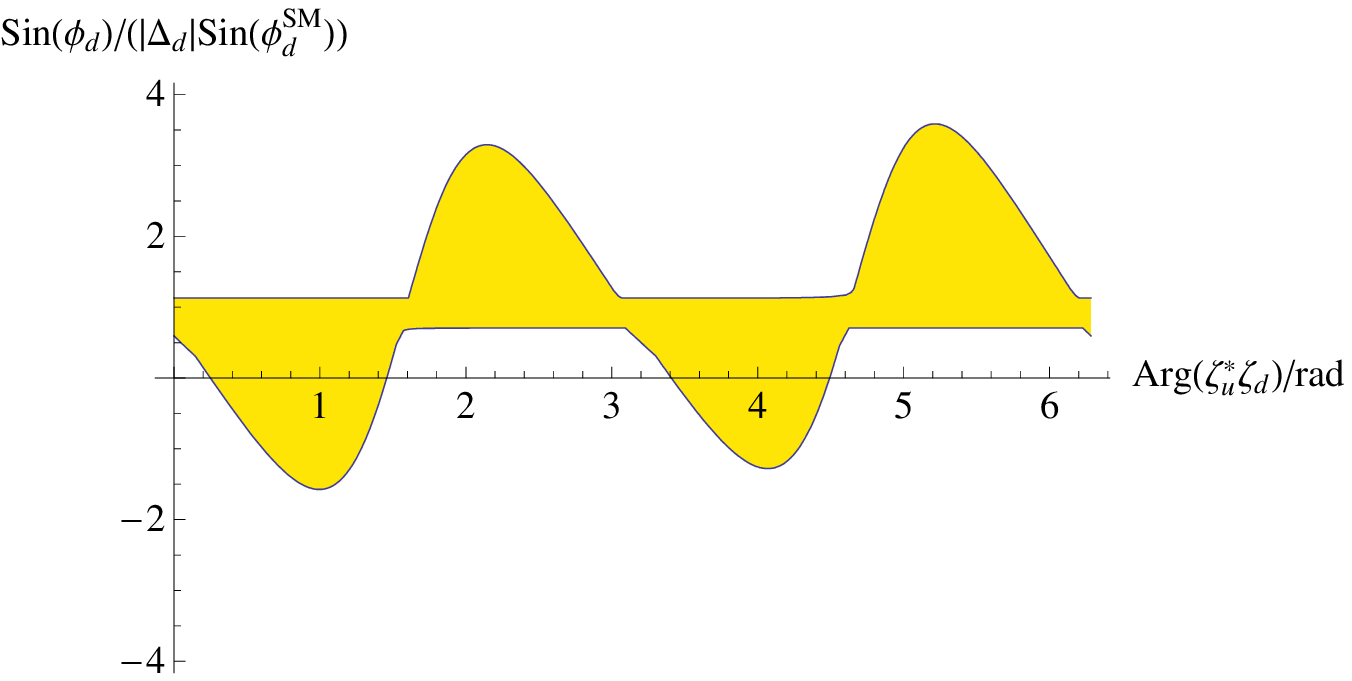}
\end{tabular}
\caption{\it Dependence of $\sin{\phi_q}/(|\Delta_q|\sin{\phi_q^{\mathrm{SM}}})$ on
$\varphi\equiv\arg{(\varsigma_u^*\varsigma_d^{\protect\phantom{*}})}$, constraining $|\varsigma_u\varsigma_d^*|\leq20$. \label{fig::Phis_correlation_btos}}
\end{center}\end{figure}

Additional contributions to $\phi_s$ could be induced by neutral scalar exchanges, through the effective FCNC operator in Eq.(\ref{eq:FCNCop}) appearing at the one-loop level. Also, a sizable Yukawa phase $\varphi\equiv\arg(\varsigma_u^*\varsigma_d^{\phantom{*}})$ could generate observable signals in other $CP$-violating observables not yet included in our analysis. A detailed discussion of these effects and their corresponding constraints on the model parameters is postponed to future work.


\subsection[$K^0$-$\bar K^0$ mixing: $\epsilon_K$]{$\mathbf{K^0}$-$\mathbf{\bar K^0}$ mixing: $\mathbf{\epsilon_K}$}
\label{subsec:Kmixing}

The $\Delta S=2$ effective Hamiltonian is described by the same basis of four-quark operators given in (\ref{eq:DB=2operators}), changing the flavour $b$ to $s$ everywhere. However, the small light-quark masses $m_d$ and $m_s$ suppress now the contributions from all operators except
$\cO^{\mathrm{VLL}}$. Another difference with respect to $B^0$ mixing is that, owing to the different CKM factors, one needs to consider the virtual contributions from top and charm quark exchanges within the box diagrams:
\begin{equation}
\mathcal{H}^{\Delta S=2}_{\rm eff}\, =\,\frac{G_F^2 M_W^2}{16\pi^2}\; \left\{
\lambda_t^2\, C_{\mathrm{VLL}}^{tt}(\mu) +
\lambda_c^2\, C_{\mathrm{VLL}}^{cc}(\mu) +
2 \lambda_t\lambda_c\, C_{\mathrm{VLL}}^{ct}(\mu) \right\}\,
\left(\bar{d}\gamma_\mu \cP_{L} s\right)\left(\bar{d}\gamma^\mu \cP_{L} s\right)\, .
\end{equation}
Since $\lambda_t\equiv V_{td}^* V_{ts}^{\phantom{*}}\sim A^2\lambda^5$ while
$\lambda_t\equiv V_{cd}^* V_{cs}^{\phantom{*}}\sim\lambda$, in spite of the $m_c^2/m_t^2$
relative suppression, the charm loop gives the dominant short-distance contribution to
$\Delta m_K$. There are in addition large corrections from long-distance physics, which make it difficult to extract from $\Delta m_K$ useful constraints on the new-physics amplitude.

More interesting is the $CP$-violating parameter $\epsilon_K$, which can be written in the form
\begin{equation}
\epsilon_K\, =\, \frac{\kappa_\epsilon\;\e^{i\phi_\epsilon}}{\sqrt{2}}\;
\frac{\mathrm{Im} (M_{12})}{\Delta m_K}\, ,
\end{equation}
where $\kappa_\epsilon = 0.94\pm 0.02$ takes into account small long-distance corrections
\cite{Buras:2008nn,Buras:2010pz}. The top and charm contributions are now weighted by
less hierarchical
CKM factors
$\mathrm{Im}(\lambda_t^2)\sim \lambda^4\mathrm{Im}(\lambda_c\lambda_t)\sim\lambda^4\mathrm{Im}(\lambda_c^2)
$; the mass hierarchy compensates for this, implying that the top quark gives the most important contribution to $\epsilon_K$.

The relevant Wilson coefficients $C_{\mathrm{VLL}}^{qq'}$, containing the SM and new-physics contributions, are given in the appendix. The corrections induced by the charged scalar are proportional to $|\varsigma_u|^2$ and $|\varsigma_u|^4$. All contributions from the coupling $\varsigma_d$ are absent in the limit $m_{d,s}=0$.
\begin{figure}[!hbt]
\centering{
\includegraphics[width=70mm]{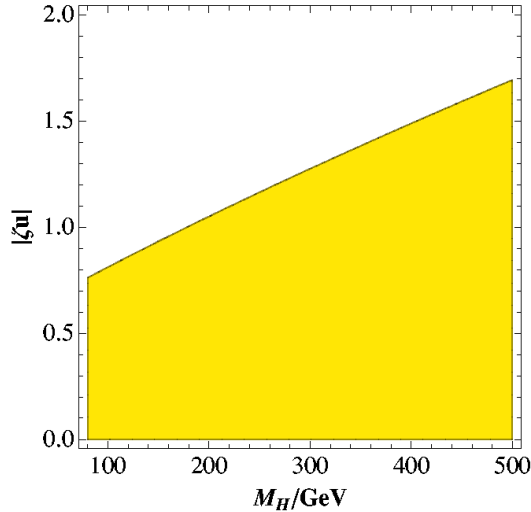}
\caption{\it 95\% CL constraints from $\epsilon_K$. \label{fig::EpsK}}}
\end{figure}
The matrix element $\langle K^0|\mathcal{H}^{\Delta S=2}_{\mathrm{eff}}|\bar{K}^0\rangle$
is parametrized through the hadronic quantity $f_K^2 \hat{B}_K$. We use the numerical values of $f_K/f_\pi$ and $\hat{B}_K$, given in table~\ref{tab::hadronic}, together with the phenomenological
determination of the pion decay constant from $\Gamma(\pi^+\to\mu^+\nu_\mu)$,
$f_\pi = 130.4\pm 0.04\pm 0.2$~MeV \cite{Amsler:2008zzb}.
Figure \ref{fig::EpsK} shows the constraint obtained from $\epsilon_K$ in the plane
$M_{H^\pm}$ -- $|\varsigma_u|$. It is very similar to the one extracted from $Z\to b\bar b$, and even slightly stronger.


\subsection[$\bar{B} \rightarrow X_s \gamma$]{$\mathbf{\bar{B} \rightarrow X_s \boldsymbol{\gamma}}$\label{sec::btos}}

The radiative decay $\bar{B}\rightarrow X_s \gamma$ has been calculated at NNLO in the SM,
leading to the prediction $\Br (\bar{B} \rightarrow X_s \gamma)_{\mathrm{SM}}=(3.15\pm0.23)\times10^{-4}$ \cite{Misiak:2006zs}.
In the \thdm the decay amplitude is known at NLO
\cite{Ciuchini:1997xe,Borzumati:1998tg,Degrassi:2010ne,Ciafaloni:1997un}. Following the steps given in \cite{Misiak:2006ab}, one can
express the branching ratio as
\begin{equation}
\Br (\bar{B} \rightarrow X_s \gamma)_{E_{\gamma}>E_0}\, =\, \Br (\bar{B} \rightarrow X_c e\bar{\nu})_{\mathrm{exp}} \left|  \frac{V_{ts}^*V_{tb}}{V_{cb}}   \right|^2 \frac{6\alpha}{\pi C_B}\; [P(E_0)+N(E_0)] \; ,
\end{equation}
where the phase-space factor
$C_B = |V_{ub}/V_{cb}|^2 \Gamma(\bar B\to X_c e\bar\nu)/\Gamma(\bar B\to X_u e\bar\nu) = 0.580 \pm 0.016$ \cite{Bauer:2004ve}
accounts for the $m_c$ dependence of $\Br (\bar{B} \rightarrow X_c e\bar{\nu})$. Normalizing the result with the $\bar{B} \rightarrow X_c e\bar{\nu}$ transition, cancels the leading non-perturbative corrections of order $\Lambda^2/m_b^2$ and minimizes many sources of uncertainties, such as those generated by the CKM quark-mixing factors, the dependence on $m_b^5$ and the sensitivity to $m_c$.
The subleading non-perturbative contributions are contained in $N(E_0)$, 
which includes corrections of $\cO(\Lambda^2/m_c^2)$
\cite{Gambino:2001ew}, $\cO(\Lambda^3/m_b^3)$, $\cO(\Lambda^3/m_b m_c^2)$ \cite{Bauer:1997fe} and $\cO(\alpha_s\Lambda^2/(m_b-2E_0)^2)$ \cite{Neubert:2004dd}.
The relevant combination of CKM factors is given by
\begin{equation}
\left|\frac{V_{ts}^*V_{tb}}{V_{cb}}\right|^2\; =\; 1 + \lambda^2 (2\bar\rho-1) + \lambda^4 (\bar\rho^2 + \bar\eta^2 -A^2) +\cO(\lambda^6)
\; =\; 0.963\pm 0.002\pm 0.005\, ,
\end{equation}
where the sensitivity to the apex $(\bar\rho,\bar\eta)$ of the unitarity triangle is suppressed by two powers of $\lambda$.

For $m_s=0$ the effective low-energy operator basis remains the same as in the SM. The modifications induced by new-physics contributions
appear only in the Wilson coefficients, which are included in the perturbative part $P(E_0)$:
\begin{equation}\label{eq:Ci-bsg}
C_i^{\mathrm{eff}}(\mu_W)=C_{i,SM}+|\varsigma_u|^2\; C_{i,uu}-(\varsigma_u^*\varsigma_d{\phantom{*}})\; C_{i,ud} \; ,
\end{equation}
where $\varsigma_u^*\varsigma_d{\phantom{*}}=|\varsigma_u||\varsigma_d| e^{i\varphi}$, $\varphi$ being the relative phase. The virtual top-quark contributions dominate the coefficients $C_{i,uu}$ and $C_{i,ud}$; their explicit expressions as a function of $m_t$ can be found in \cite{Degrassi:2010ne}. Depending on the value of the phase $\varphi$, the combined effect of the two terms $C_{i,uu}$ and $C_{i,ud}$ can be rather different. For instance, these two terms tend to cancel each other in the type I model where $\varphi=0$, while in the type II version with $\varphi =\pi$ they add constructively.

Since the new-physics contribution is only calculated up to NLO, terms in the branching ratio of $\mathcal O(\alpha_s^2)$ coming from the square of the \thdm amplitude are neglected consistently. In some regions of the parameter space, leading to large new-physics effects of opposite sign to the SM amplitude, the cancellations between the two contributions enhance the sensitivity to higher-order QCD corrections, generating in some cases unphysical results (for instance in the type I model at small values of $\tan{\beta}$) \cite{Borzumati:1998tg}.
Fortunately, the most problematic region  (large values of $|\varsigma_u|$) is already excluded by the constraints from $Z\to\bar{b}b$ and $\Delta m_{B^0_s}$. The inclusion of the SM NNLO contributions substantially improves the reliability of the theoretical predictions.

\begin{figure}[!hbt]
\begin{center}
\begin{tabular}{cc}
\includegraphics[width=7cm]{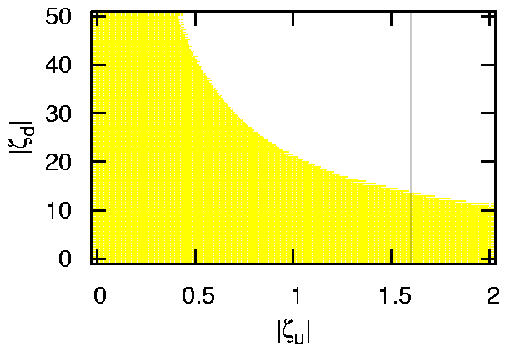} &
\includegraphics[width=7.1cm]{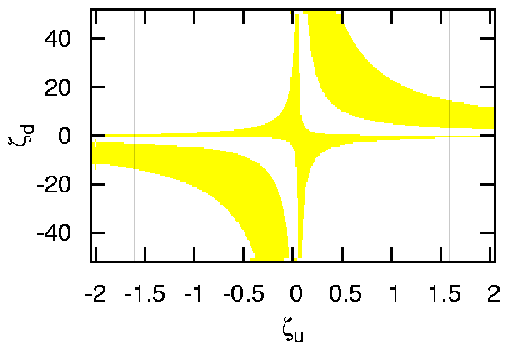} \\
\end{tabular}
\caption{\label{ud} \it
Constraints on $\varsigma_u$ and $\varsigma_d$ from $\bar{B} \rightarrow X_s \gamma$, taking $M_{H^\pm}\in[80,500]$~GeV. The white areas are excluded at $95\%$ CL. The black line corresponds to the upper limit from $\epsilon_K, Z\to\bar{b}b$ on $|\varsigma_u|$.
In the left panel, the relative phase has been varied in the range $\varphi \in [0,2\pi]$. The right panel assumes real couplings.}
\end{center}
\end{figure}

To extract the information on the \athdm couplings, we take into account the latest experimental values, given in table \ref{tab::measurements},
and use the same renormalization scales as in \cite{Misiak:2006ab}
($\mu_0=160$ GeV, $\mu_b=2.5$ GeV and $\mu_c=1.5$ GeV as central values and the same ranges of variation). We follow again the RFit approach, adding the theoretical uncertainty linearly to the systematic error.
The resulting constraints on $|\varsigma_u|$ and $|\varsigma_d|$ are shown in figure~\ref{ud}, varying the charged-scalar mass in the range $M_{H^\pm}\in[80,500]$~GeV. The white areas are excluded at 95\% CL. In the left plot, the phase $\varphi$ has been scanned in the whole range from $0$ to $2\pi$; the resulting constraints are not very strong because a destructive interference between the two terms in (\ref{eq:Ci-bsg}) can be adjusted through the relative phase. In the range $|\varsigma_u|<2$, one finds roughly $|\varsigma_d| |\varsigma_u|<20$ (95\% CL).
More stringent bounds are obtained at fixed values of the relative phase. This is shown in the right plot, where $\varsigma_u$ and $\varsigma_d$ have been assumed to be real (i.e. $\varphi=0$ or $\pi$). In that case, couplings of different sign are excluded, except at very small values, while a broad region of large equal-sign couplings is allowed, reflecting again the possibility of a destructive interference.

The sensitivity to the charged-scalar mass is illustrated in figure~\ref{zetad}, which shows the constraints on $|\varsigma_d|$ versus
$M_{H^\pm}$ for fixed values of $\varsigma_u=0.5$ (left) and $\varsigma_u=1.5$ (right).
Again, in the upper plots the relative phase has been varied in the whole range $\varphi \in [0,2\pi]$, while the lower plots assume real couplings.
\begin{figure}[!hbt]
\begin{center}
\begin{tabular}{cc}
\includegraphics[width=7cm]{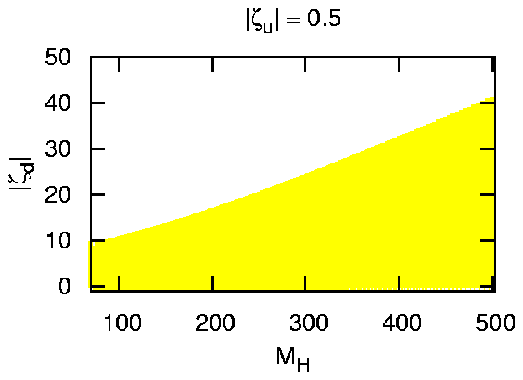} & \includegraphics[width=7cm]{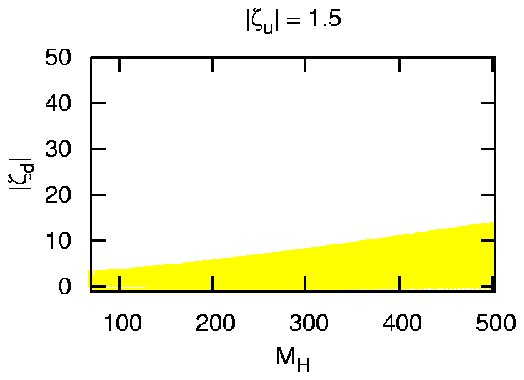} \\[15pt]
\includegraphics[width=7cm]{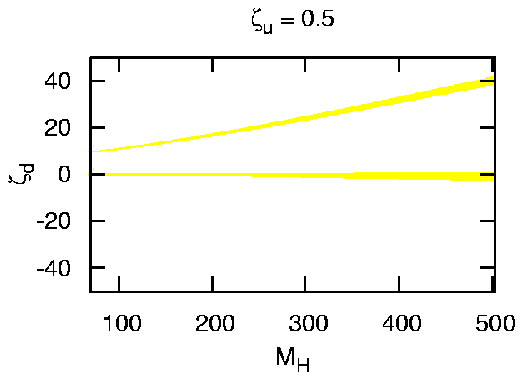} & \includegraphics[width=7cm]{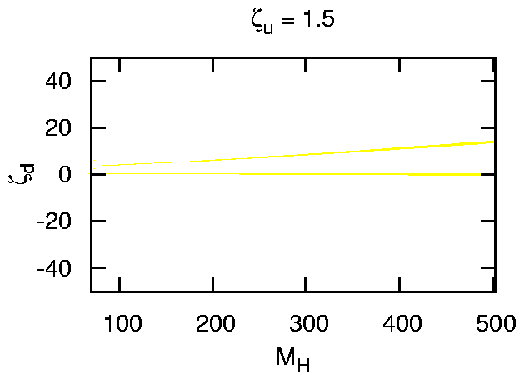}
\end{tabular}
\end{center}
\caption{\label{zetad} \it Constraints on $|\varsigma_d|$ versus $M_{H^\pm}$ (in GeV) from $\bar{B} \rightarrow X_s \gamma$, for $\varsigma_u=0.5$ (left) and $\varsigma_u=1.5$ (right). The white areas are excluded at $95\%$ CL.
In the upper panels, the phase has been varied in the range $\varphi \in [0,2\pi]$. The lower panels assume real couplings.}
\end{figure}
Figure~\ref{zetau} shows the constraints on the $|\varsigma_{u}|-M_{H^\pm}$ plane, for $\varsigma_d=0$. Finally, in figure~\ref{Mahmoudiplots} we show the constraints obtained for fixed values of the charged-scalar mass, assuming $\varsigma_u$ and $\varsigma_d$ to be real. We reproduce in this case the qualitative behaviour obtained in \cite{Mahmoudi:2009zx}.
\begin{figure}[!hbt]
\begin{center}
\includegraphics[width=7cm]{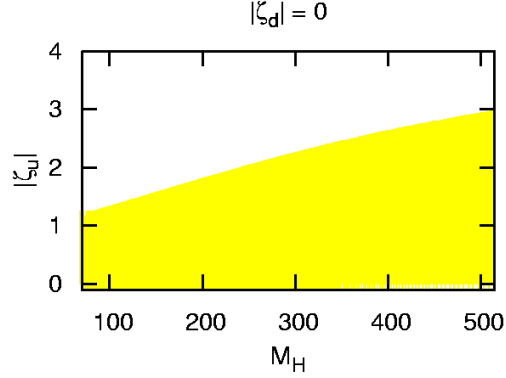}
\caption{\label{zetau} \it Constraints on $|\varsigma_u|$ versus $M_{H^\pm}$ (in GeV) from $\bar{B} \rightarrow X_s \gamma$, for $\varsigma_d=0$.
The white area is excluded at $95\%$~CL.}
\end{center}
\end{figure}
\begin{figure}[!hbt]
\begin{center}
\begin{tabular}{cc}
\includegraphics[width=7cm]{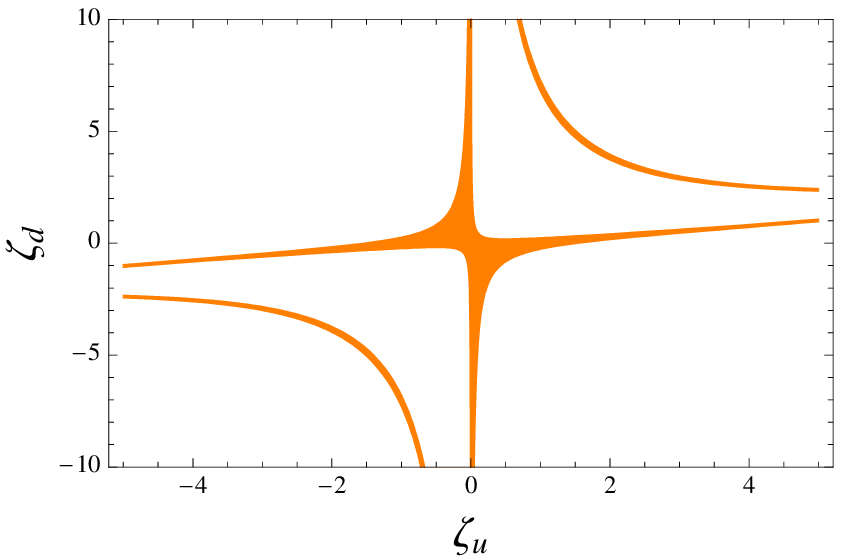} &
\includegraphics[width=7cm]{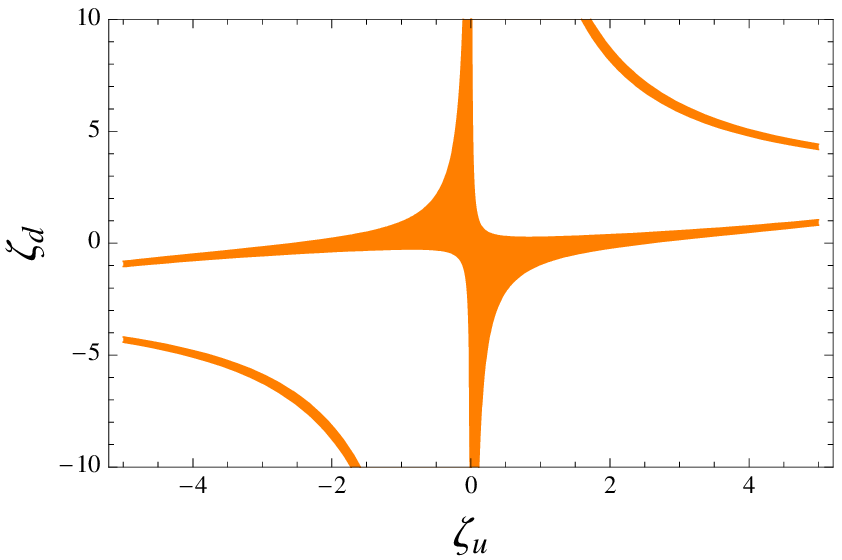} \\
\end{tabular}
\caption{\it Constraints on $\varsigma_d$ versus $\varsigma_u$ ($95\%$~CL) from $\bar{B} \rightarrow X_s \gamma$, assuming real couplings and taking $M_{H^\pm}=150$~GeV (left) and $M_{H^\pm}=400$~GeV (right).
\label{Mahmoudiplots}}
\end{center}
\end{figure}

We observe that for small values of $|\varsigma_u|$ no constraint on $\varsigma_d$ is obtained, because in the limit $|\varsigma_u|\to 0$ the SM is recovered, which is compatible with the data. With growing $|\varsigma_u|$ a bound on $|\varsigma_d|$ emerges, corresponding to $|\varsigma_u\varsigma_d|\lesssim 20$. For $\varsigma_d=0$ on the other hand, a limit of $|\varsigma_u|\lesssim 3$ can be observed for large scalar masses around 500 GeV, strengthening to $|\varsigma_u|\lesssim 1.3$ for smaller values of $M_{H^\pm}$. The overall constraint is relatively weak compared to the strong bound on $M_{H^\pm}$ obtained in the type II \thdmws,
due to the correlation $\varsigma_u\varsigma_d=-1$.
However, it can be seen from the plots with vanishing phase and/or a fixed value for $|\varsigma_{d,u}|$ that this strength is recovered, once some parameters are constrained independently. Comparing the plots with complex input parameters to their real counterparts, we observe that the effect of the relative phase is mainly to 
extend the allowed bands in a way that the excluded space between them is rendered allowed, too.

We have also analyzed the $CP$ rate asymmetry, defined as
\begin{equation}
a_{CP}= \frac{BR(\bar{B}\to X_{s}\gamma)-BR(B\to X_{\bar{s}}\gamma)}{BR(\bar{B}\to X_s\gamma)+BR(B\to X_{\bar{s}}\gamma)} \,,
\end{equation}
which is predicted to be tiny in the SM.
Once the constraints from the branching ratio are implemented in the A2HDM, the predicted asymmetry is smaller than the present experimental bounds. Thus, one does not obtain further constraints on the model parameters.
A sizable Yukawa phase $\varphi$ could generate values of the $CP$-asymmetry large enough to be relevant for future high-precision experimental analyses. However, a NNLO analysis of the theoretical prediction appears to be needed to reduce the presently large theoretical uncertainties and fully exploit such a measurement.


\section{Discussion}\label{summary}

Imposing natural flavour conservation through discrete $\mathcal{Z}_2$ symmetries, one finds that the CKM phase is the only source of $CP$ violation
in the resulting \thdmws s. During the last thirty years, it has been common lore to assume that this is a more general fact, i.e. that the absence of tree-level FCNCs implies the absence of additional phases beyond the CKM one. The \athdm provides an explicit counter-example, where FCNC couplings are absent at the Lagrangian level, while additional unconstrained complex phases generate new sources of $CP$ violation. Since all Yukawa couplings are proportional to fermion masses, the \athdm gives rise to an interesting hierarchy of FCNC effects, avoiding the stringent experimental constraints for light-quark systems and predicting at the same time interesting signals in heavy-quark transitions. The flavour-blind phases present in the model open a very interesting phenomenology which is worth to be investigated. The built-in flavour symmetries protect very efficiently the \athdm from unwanted FCNC effects generated through quantum corrections. At the one-loop level the only allowed FCNC local structures are the two operators in 
(\ref{eq:FCNCop}), which could have very interesting (and computable) implications for $B_s^0$ mixing.

Besides the fermion masses and mixings, the charged-scalar couplings of the \athdm are fully characterized by three complex parameters $\varsigma_f$. In the previous sections, we have analyzed the impact of the $H^\pm$ contribution to different observables, where it is expected to be the dominant new-physics effect. Using conservatively estimated hadronic parameters and up-to-date data, we have inferred the present constraints on the new-physics parameters involved in these processes.

Leptonic tau decays provide a direct bound on the leptonic Yukawa coupling:
$|\varsigma_l|/M_{H^\pm}\le 0.40~\mbox{GeV}^{-1}$ (95\% CL).
From semileptonic processes constraints on the products $\varsigma_l^*\varsigma_u/M_{H^\pm}^2$ and $\varsigma_l^*\varsigma_d/M_{H^\pm}^2$ are derived.
The leptonic decays of heavy-light mesons allow us to disentangle the effects from $\varsigma_u$ and $\varsigma_d$.
Thus, from $B\to\tau\nu$ we derive an annular constraint in the complex plane $\varsigma_l^*\varsigma_d/M_{H^\pm}^2$ (figure~\ref{fig::btaudmu}a), implying the absolute bound
$|\varsigma_l^*\varsigma_d/M_{H^{\pm}}^2| < 0.108~\mathrm{GeV}^{-2}$ (95\% CL).
For real Yukawa couplings there is a two-fold sign ambiguity generating two possible solutions,
the expected one around $\Delta_{ij}=0$ (the SM amplitude dominates) and its mirror around
$\Delta_{ij}=2$, corresponding to a new-physics contribution twice as large as the SM one and of opposite sign. The real solutions are $\varsigma_l^*\varsigma_d/M_{H^{\pm}}^2 \in[-0.036,0.008]~\mathrm{GeV}^{-2}$ and $[0.065,0.108]~\mathrm{GeV}^{-2}$.

Similar, but slightly weaker constraints on $\varsigma_l^*\varsigma_u/M_{H^\pm}^2$ are obtained
from the decays $D\to\mu\nu$ (figure~\ref{fig::btaudmu}b) and $D_s\to (\tau,\mu)\nu$
(figure~\ref{fig::dslnu}); in the last case the bounds from $B\to\tau\nu$ are used to get rid of the small $\varsigma_d$ contamination proportional to the strange quark mass.
The resulting absolute bound
$|\varsigma_l^*\varsigma_u/M_{H^{\pm}}^2| < 0.6~\mathrm{GeV}^{-2}$ (95\% CL) is rather weak, but the upper limit corresponds to a new-physics contribution twice as large as the SM one, a very unlikely situation. The annular form of these constraints results in much stronger limits, once this possibility is excluded by other processes.
For real Yukawa couplings, one finds $\varsigma_l^*\varsigma_u/M_{H^{\pm}}^2 \in[-0.005,0.037]~\mathrm{GeV}^{-2}$ or $[0.511,0.535]~\mathrm{GeV}^{-2}$, at 95\% CL.

Owing to the quark-mass suppression, the absolute constraints obtained from leptonic decays of light mesons (figure \ref{fig::kopilnu}) are obviously much weaker. However, the excellent experimental precision achieved in $\pi$ and $K$ decays implies a narrow allowed annular region. For real Yukawa couplings this translates into quite stringent bounds:
$\varsigma_l^*\varsigma_d /M_{H^\pm}^2 \in[-0.07,0.07]~\mathrm{GeV}^{-2}$ or
$[8.14,8.28]~\mathrm{GeV}^{-2}$ (95\% CL). The uncertainties are dominated by the present theoretical knowledge of the ratio $f_K/f_\pi$.

Independent information is obtained from the semileptonic decays of pseudoscalar mesons, through the scalar form-factor contribution. One needs, however, to disentangle the dominant vector form-factor amplitude, which does not contain any charged-scalar effect and is correlated with the usual measurement of the corresponding CKM mixing factor. The present constraints from the ratio $\Br (B\rightarrow D \tau \nu_{\tau})/\Br (B \rightarrow D e\nu_e)$, shown in figures \ref{fig::bdlnu} and \ref{fig::bdlnureal}, are not very strong
by themselves, but allow in combination with other processes the exclusion of the second real solutions in the $\varsigma_{u,d}\varsigma_l^*/M_{H^\pm}^2$ planes.
A future measurement of the differential distribution in $B\rightarrow D \tau \nu_{\tau}$ would obviously increase the sensitivity to the scalar contribution. In spite of the strange-mass suppression, the much higher experimental accuracy achieved in the analysis of $K\to\pi l\nu$ decays allows to derive the bound
$\mathrm{Re}(\varsigma_l^*\varsigma_d/M_{H^{\pm}}^2)\in[-0.16,0.30]~\mbox{GeV}^{-2}$ (95\%~CL).
This already excludes the second real solution (a scalar amplitude larger than the SM one) obtained
from $K_{\mu 2}/\pi_{\mu 2}$.

Combining the information from all leptonic and semileptonic decays analyzed, one gets the constraints shown in figure \ref{fig::globalfit}.

The flavour-conserving decay $Z\to b\bar b$  provides a very stringent constraint on $|\varsigma_u|$. Since $V_{tb}\approx 1$, the one-loop contributions involving virtual top quarks completely dominate both the SM ($W^\pm$) and the new-physics ($H^\pm$) radiative corrections.
In contrast to leptonic and semileptonic processes, where the charged-scalar effects are necessarily proportional to $\varsigma_l$, the $Z\to b\bar b$ amplitude gives direct access to
$\varsigma_u$ and $\varsigma_d$. Owing to the relative factor $m_b/m_t$ which suppresses the
$\varsigma_d$ contribution, one gets finally the constraints on $|\varsigma_u|$ shown in figure \ref{fig:Rb}  (assuming $|\varsigma_d|\leq50$). At 95\% CL, we obtain
$|\varsigma_u|< 0.91\; (1.91)$, for $M_{H^\pm}= 80\; (500)$~GeV. The upper bound increases linearly with $M_{H^\pm}$, implying $|\varsigma_u|/M_{H^\pm}< 0.0024~\mathrm{GeV}^{-1} + \frac{0.72}{M_{H^\pm}} < 0.011~\mathrm{GeV}^{-1}$, where we have used the LEP lower bound on the charged-scalar mass $M_{H^\pm}> 78.6$~GeV (95\% CL) \cite{Pich:2009sp,:2001xy}.
Together with the tau-decay constraint on $|\varsigma_l| /M_{H^\pm}$, this gives the limit
$|\varsigma_u\varsigma_l^*|/M_{H^\pm}^2<0.005~\mathrm{GeV}^{-2}$, which is much stronger than the information extracted from the global fit to leptonic and semileptonic decays.

Quite similar information can be extracted from $B^0$ mixing, which is also dominated by one-loop contributions involving virtual top quarks. The smallness of the $m_s/M_W$ ratio
implies that SU(3)-breaking corrections are negligible; therefore, the charged-scalar contributions cancel in the ratio $\Delta m_{B^0_s}/\Delta m_{B^0_d}$, which can be used in the CKM fit. Only two $\Delta B=2$ four-quark operators are numerically relevant; the one generating the leading SM amplitude gets new-physics contributions proportional to $|\varsigma_u|^{2,4}$, while the other operator generates subleading corrections proportional to $(\varsigma_u^*\varsigma_d^{\phantom{*}})^{1,2} m_b^2/M_W^2$. Scanning the parameter ranges $|\varsigma_d|<50$ and $\varphi\in [0,2\pi]$, where $\varphi$ is the relative phase between $\varsigma_u$ and $\varsigma_d$, the measured $B^0_s$ mixing amplitude implies the constraints
shown in figure~{\ref{fig::deltams}, in the plane $M_{H^\pm}$-- $|\varsigma_u|$. At 95\% CL,
one gets 
$|\varsigma_u|< 0.00279\, M_{H^\pm} + 0.27 + 117/M_{H^\pm}$, for $M_{H^\pm}\in [80, 500]$, in GeV units. 

The charged-scalar contribution could accommodate a large $B^0_s$ mixing phase, without spoiling the agreement in the $B_d$ system, although it is not possible to reach a value as large as hinted at by the present D0 central value, which is however at odds with the rate difference being unaffected by new physics (unless the calculation of the rate difference is affected by problems regarding the OPE). If confirmed, a large phase $\phi_s$ would point towards large values of $|\varsigma_d|$, small charged-scalar masses and a sizable Yukawa phase $\varphi$. The preferred negative sign for the $a^s_{sl}$ asymmetry would require $\varphi\in[\pi/2,\pi],[3\pi/2,2\pi]$. Additional contributions to $\phi_s$ could be induced by neutral scalar exchanges, through the effective FCNC operator in Eq.(\ref{eq:FCNCop}) appearing at the one-loop level. Large Yukawa phases could be constrained by other $CP$-violating observables not yet included in our analysis. A detailed discussion of these effects is postponed to future work.

The observable $\epsilon_K$ leads again to a similar constraint, even slightly more restrictive than the ones from $B^0$ mixing and $Z\to\bar{b}b$. Although $CP$ violating, this observable is insensitive to the new-physics phases, as the relevant contributions involve $|\varsigma_u|$, only. We obtain at $95\%$~CL $|\varsigma_u|\leq 0.560+2.647\,10^{-3}M_{H^\pm}-1.049\,10^{-6}M_{H^\pm}^2+6.153\,10^{-10}M_{H^\pm}^3$ in units of GeV.

The radiative decay $\bar B\to X_s\gamma$ provides another important source of information. There are two different charged-scalar contributions, proportional again to  $|\varsigma_u|^{2}$ and $\varsigma_u^*\varsigma_d^{\phantom{*}}$, but in this case the two have similar sizes. Their combined effect can be quite different depending on the value of the relative phase $\varphi$.
This results in rather weak limits because a destructive interference can be adjusted through this phase. The resulting constraints on $|\varsigma_u|$ and $|\varsigma_d|$ are shown in figure~\ref{ud}, varying the charged-scalar mass in the range $M_{H^\pm}\in[80,500]$~GeV.
Scanning the phase $\varphi$ in the whole range from $0$ to $2\pi$, and imposing $|\varsigma_u|<3$, one finds roughly $|\varsigma_d| |\varsigma_u|<20$ (95\% CL).
Much stronger bounds are obtained at fixed values of the relative phase.
Assuming real values of $\varsigma_u$ and $\varsigma_d$ (i.e. $\varphi=0$ or $\pi$), one finds that couplings of different sign are excluded, except at very small values, while a broad region of large equal-sign couplings is allowed, reflecting again the possibility of a destructive interference.
Figures~\ref{zetad}, \ref{zetau} and \ref{Mahmoudiplots} show the sensitivity of the $\bar B\to X_s\gamma$ constraints to the different unknown parameters: $M_{H^\pm}$, $|\varsigma_u|$, $|\varsigma_d|$ and $\varphi$.

\begin{figure}[tbh]
\begin{center}
\begin{tabular}{cc}
\includegraphics[width=7cm]{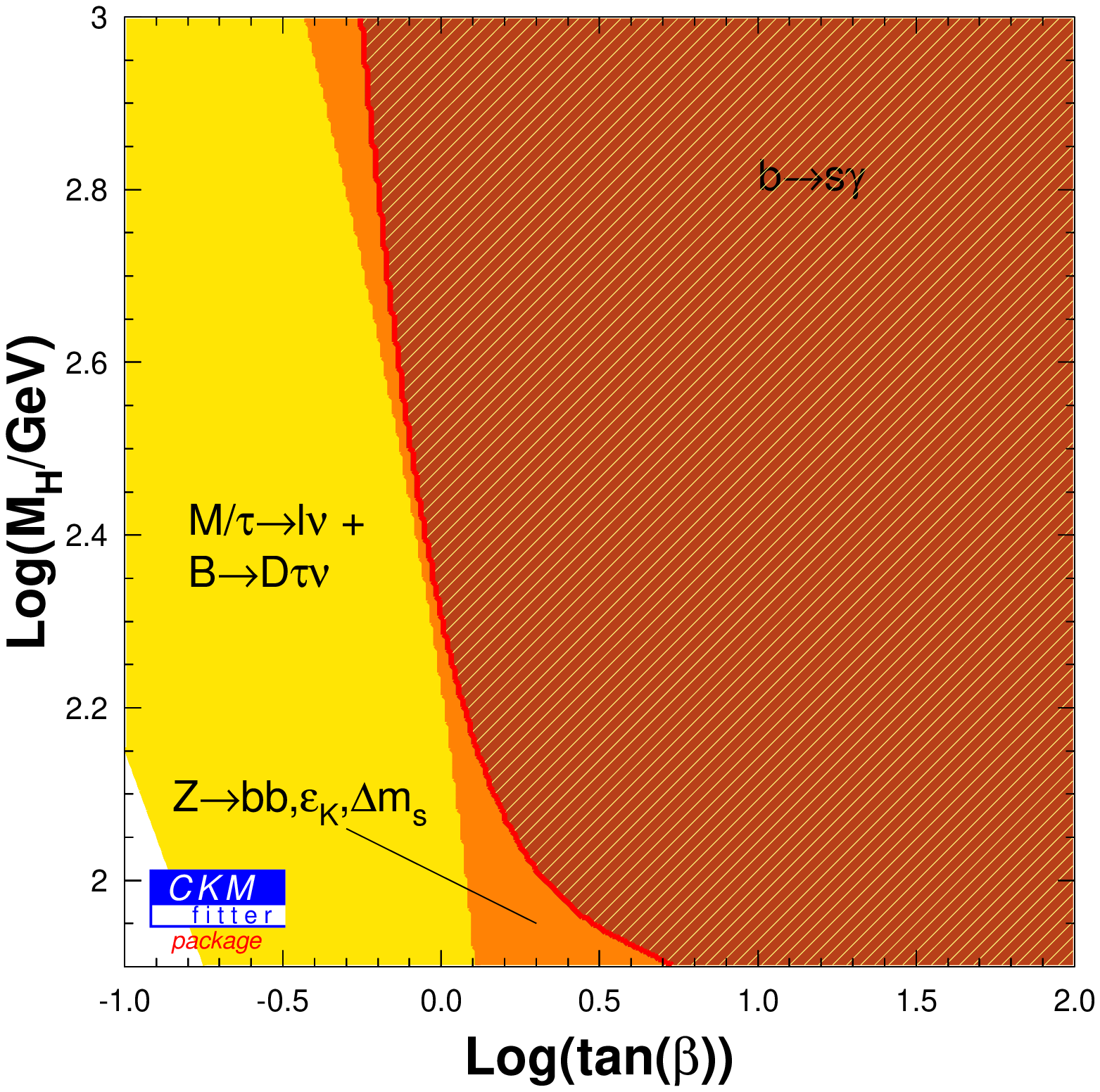} & \includegraphics[width=7cm]{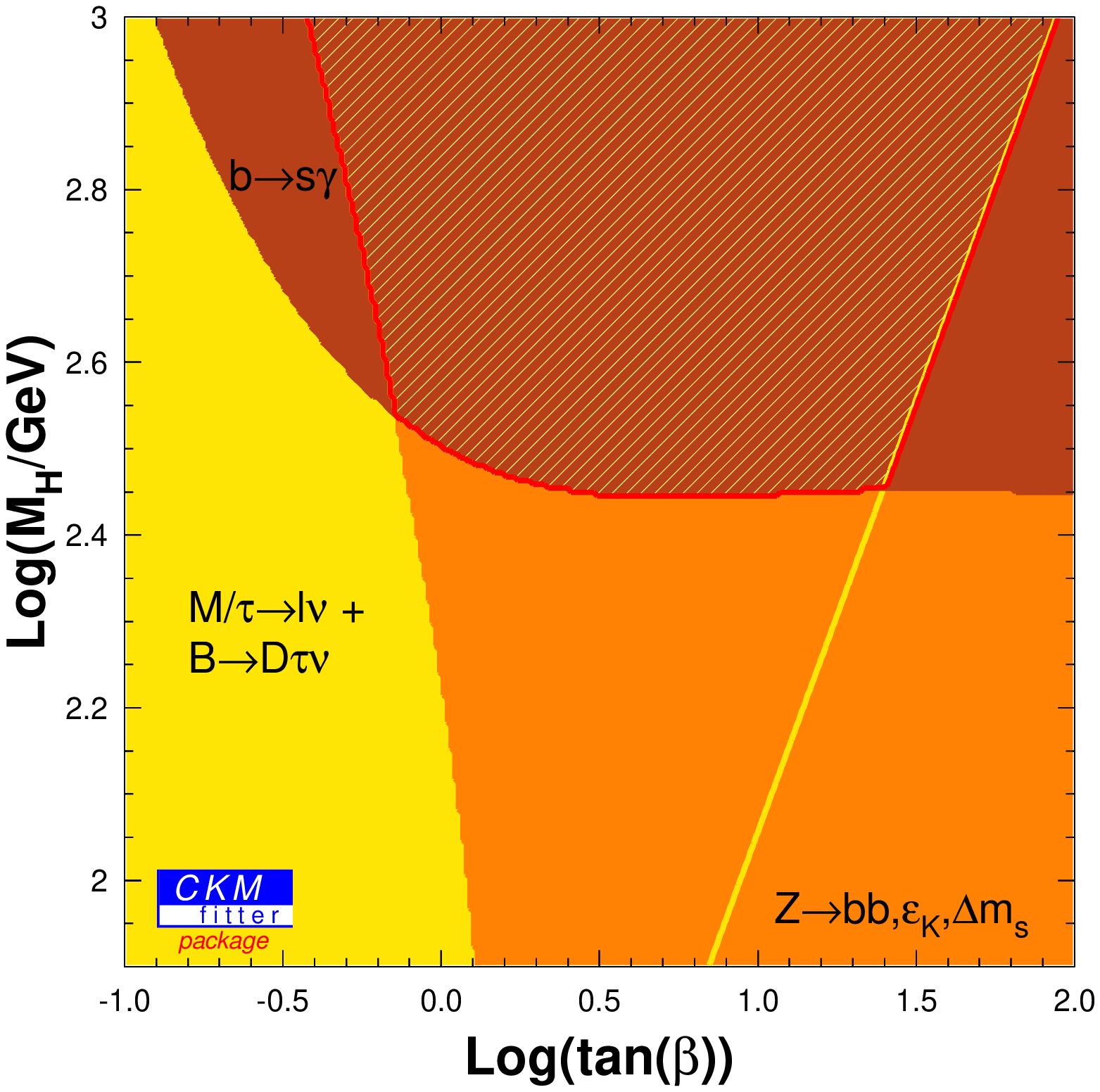}\\
\includegraphics[width=7cm]{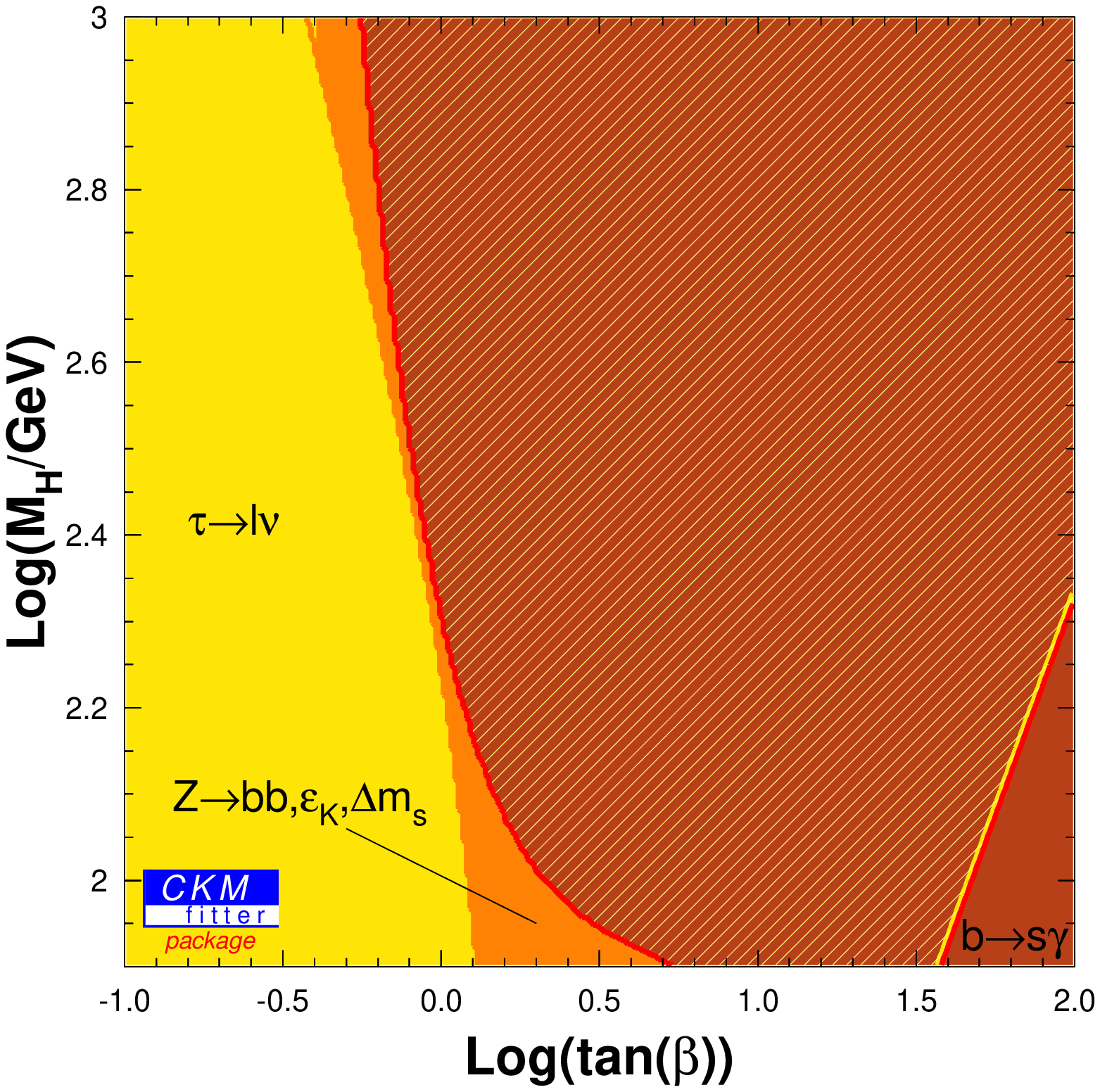} & \includegraphics[width=7cm]{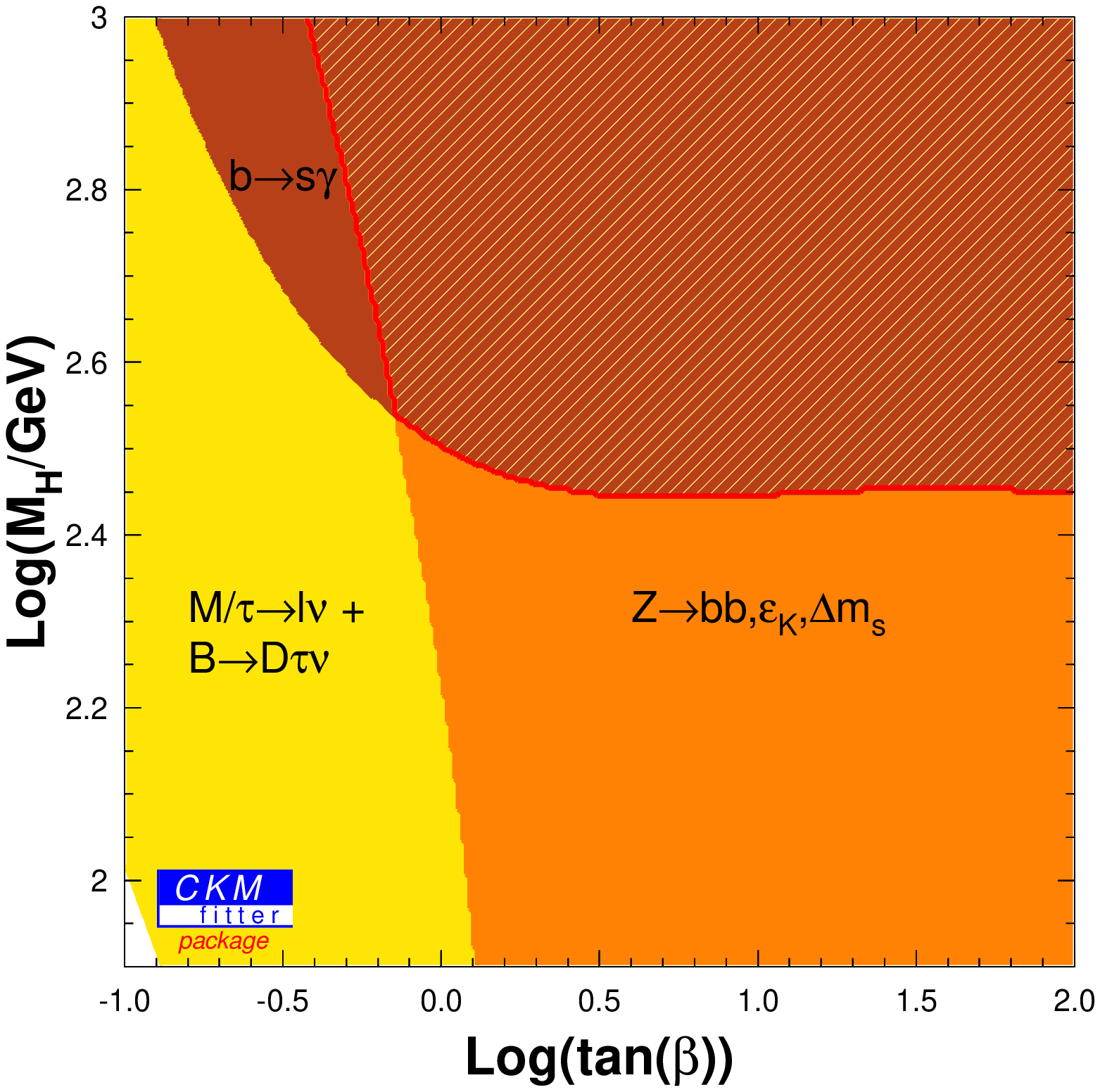}\\
\end{tabular}
\caption{\it Constraints on $M_{H^\pm}$ (in GeV) versus $\tan{\beta}$ ($95\%$~CL),
in the \thdm models of types I (upper-left), II (upper-right), X (lower-left) and Y (lower-right).
\label{fig:usualbounds}}
\end{center}
\end{figure}

The constraints discussed so far apply to the general \athdm framework, with three arbitrary complex parameters $\varsigma_f$. The limits become of course much stronger in particular models where these parameters are correlated. Figures~\ref{fig:usualbounds} show the combined constraints on the $\tan{\beta}$--$M_{H^\pm}$ plane for the different $\mathcal{Z}_2$ models. The bounds from
$Z\to b\bar b$, $\epsilon_K$, $\Delta m_{B^0_s}$ and $\bar{B}\to X_s\gamma$ are obviously identical for the models of type I and X and also for type II and Y. In the type I/X case,
$\varsigma_u^2 = \varsigma_d^2=\varsigma_u\varsigma_d = \cot^2{\beta}$ and the scalar amplitudes grow for decreasing values of $\tan{\beta}$. For type II/Y, this behaviour is only observed in the $\varsigma_u^2$ term, while $\varsigma_d^2=\tan^2{\beta}$ and $\varsigma_u\varsigma_d=-1$;
the decay $\bar{B}\to X_s\gamma$ provides then a very strong lower bound on the scalar mass, $M_{H^\pm}> 277~\mathrm{GeV}$ (95\% CL), due to the constructive interference of the two contributing amplitudes. The $\varsigma_l$ coupling gives rise to different constraints from leptonic and semileptonic decays in each of the four models. Our results agree with the
qualitative behaviour found in previous analyses \cite{Aoki:2009ha,WahabElKaffas:2007xd,Deschamps:2009rh,Flacher:2008zq,Bona:2009cj,Mahmoudi:2009zx,Misiak:2006zs,Logan:2009uf,Logan:2010ag,Akeroyd:2008ac,Akeroyd:2009tn,Barenboim:2007sk,Gupta:2009wn}, the small differences originating from the slightly different inputs adopted.

The \athdm is not the most general version of a \thdm without tree-level FCNCs. To avoid the unwanted FCNCs one just needs diagonal Yukawa matrices $Y_f$ in the fermion mass-eigenstate basis,
i.e. $Y_d = \mathrm{diag}(y_d, y_s, y_b)$, $Y_u = \mathrm{diag}(y_u, y_c, y_t)$ and $Y_l = \mathrm{diag}(y_e, y_\mu, y_\tau)$, with arbitrary parameters $y_i$. This more general scenario
can be formally described by the Lagrangian (\ref{lagrangian}) with the substitution $\varsigma_f M_f\to Y_f$. One could still use nine dimensionless parameters $\varsigma_f\equiv y_f/m_f$, one for each charged fermion \cite{Ahn:2010zza}, but in this case this is just a redefinition of the Yukawa couplings $y_f$ because a priori nothing relates them to the fermion masses \cite{Joshipura:2010tz}. The hierarchy of couplings characteristic of the \athdm ansatz is lost and one can no longer justify that the leading charged-scalar effects originate in the heavier fermion couplings (it becomes an assumption). With this caveat in mind, our results can still be applied in this case, but most correlations among different processes disappear because the associated constraints correspond now to different $\varsigma_f$ parameters. 
For instance, the constraint in (\ref{eq:sigma_l}) refers to 
$\sqrt{|\varsigma_\tau\varsigma_\mu^*|}$ 
and figures \ref{fig::btaudmu} to $\varsigma_\tau^*\varsigma_b$ (left) and
$\varsigma_\mu^*\varsigma_c$ (right).

The \athdm provides a general setting to discuss the phenomenology of \thdmws s, satisfying in a natural way the requirement of very suppressed FCNC effects. The alignment conditions imply Yukawa couplings proportional to the corresponding fermion masses, which is  supported by the data (bounds of order 1 for the $\varsigma_f$ parameters). While including as
limiting cases all $\mathcal Z_2$ models, the \athdm incorporates possible new sources of $CP$ violation through the $\varsigma_f$ phases. The additional freedom introduced by these phases makes easier to avoid some low-energy constraints, resulting in weaker limits than in
the usual scenarios with discrete $\mathcal Z_2$ symmetries. A detailed analysis of $CP$-violating observables is clearly needed to investigate the allowed ranges for these phases and their potential phenomenological relevance \cite{JPT:2010}.

At the moment, the data does not show any clear deviation from the SM. Therefore, we have derived upper limits on the Yukawa parameters. Nevertheless, we have already pointed out that
the \athdm could account for a sizeable $B^0_s$ mixing phase, as suggested by the present
$B_s\to J/\psi \phi$ and like-sign dimuon data. Our bounds could be made stronger, adopting more aggressive estimates for the hadronic parameters entering the analysis, but we have preferred to be on the conservative side and infer solid limits for later use. Improvements are to be expected on one hand from better theoretical determinations of the hadronic inputs, and on the other hand from more accurate measurements at NA62 (kaons), LHCb ($\Delta m_{d,s}, B_s\to J/\psi\phi$), a future Super-B factory ($\tau$, $b\to s\gamma, \Delta m_d, B\to\ell\nu, B\to D\ell\nu$), or a linear collider with Giga-Z option ($R_b$).
The agreement of the different bounds in the vicinity of zero is trivial, when the SM agrees with the data. If signals for new-physics are found at LHC, the analysis presented here will be capable of quantifying the agreement (or disagreement) of the data with the \athdmws, and with the different implementations of the \thdm based on $\mathcal{Z}_2$ symmetries, in one step.

\section*{Note added}
After this work was finished, two relevant papers have been posted in the archives. In Ref.~\cite{Braeuninger:2010td} an approximate solution to the renormalization-group equations of the \athdm is analyzed and the generated FCNC terms are studied numerically; the results presented there agree with our FCNC operator (\ref{eq:FCNCop}) and it is concluded that the induced FCNC effects are well below the present experimental bounds.
Ref.~\cite{Buras:2010mh} analyzes the strength of FCNC effects mediated by neutral scalars
in a minimal-flavour-violating framework containing two Higgs doublets, assuming a perturbative expansion around the type II model.
The tree-level alignment conditions of Ref.~\cite{Pich:2009sp} are reproduced, the one-loop FCNC structures in (\ref{eq:FCNCop}) are discussed and their coefficients are estimated at large $\tan{\beta}$ in the decoupling limit. The phenomenological analysis of Ref.~\cite{Buras:2010mh} emphasizes the potential relevance of the flavour-blind phases present in the \athdm to accommodate the recent hints of a large $B^0_s$ mixing phase through neutral-Higgs exchange.

\acknowledgments{
The authors would like to thank Mikolaj Misiak and Nazila Mahmoudi for helpful discussions, and Guiseppe Degrassi for clarifying comments on Eqs.~(\ref{Eq::Ztobb1}) and (\ref{Eq::Ztobb2}).
This work has been supported in part by the EU MRTN network FLAVIAnet [Contract No. MRTN-CT-2006-035482], by MICINN, Spain 
[Grants FPA2007-60323 and Consolider-Ingenio 2010 Program CSD2007-00042 --CPAN--] and by Generalitat Valenciana [Prometeo/2008/069]. The work of P.T. is funded through an FPU Grant (MICINN, Spain).}

\appendix

\section{$\mathbf{\Delta F=2}$ effective Hamiltonian}\label{appendix}

\subsection[$\Delta B=2$]{$\mathbf{\Delta B=2}$}

At lowest order, the $\Delta F=2$ transitions are mediated by box diagrams with exchanges of $W^\pm$ and/or $H^\pm$ propagators.
Performing the matching between the \athdm amplitude and the low-energy effective Hamiltonian ${\cal H}_{\mathrm{eff}}^{\Delta F=2}$,
at the scale $\mu_{tW}\sim M_W, m_t$, one obtains the Wilson coefficients $C_i(\mu)$. We have derived the LO results given in table~\ref{tab::matching}, where $x_W\equiv m_t^2/M_W^2$ and $x_H\equiv m_t^2/M_{H^\pm}^2$. They can be expressed in terms of the two four-point functions
\cite{Buras:2001mb}:
\begin{eqnarray}
D_0(m_1,m_2,M_1,M_2) &\equiv&
\frac{m_2^2\;\log{\left(\frac{m_2^2}{m_1^2}\right)}}{(m_2^2-m_1^2)(m_2^2-M_1^2)(m_2^2-M_2^2)}+
\nonumber\\ 
&& + \frac{M_1^2\;\log{\left(\frac{M_1^2}{m_1^2}\right)}}{(M_1^2-m_1^2)(M_1^2-m_2^2)(M_1^2-M_2^2)}+\nonumber\\
&&+ \frac{M_2^2\;\log{\left(\frac{M_2^2}{m_1^2}\right)}}{(M_2^2-m_1^2)(M_2^2-m_2^2)(M_2^2-M_1^2)}\; ,
\end{eqnarray}
\begin{eqnarray}
D_2(m_1,m_2,M_1,M_2) &\equiv&
\frac{m_2^4\;\log{\left(\frac{m_2^2}{m_1^2}\right)}}{(m_2^2-m_1^2)(m_2^2-M_1^2)(m_2^2-M_2^2)}+\nonumber\\ 
&& + \,\frac{M_1^4\;\log{\left(\frac{M_1^2}{m_1^2}\right)}}{(M_1^2-m_1^2)(M_1^2-m_2^2)(M_1^2-M_2^2)}+\nonumber\\
&& + \,\frac{M_2^4\;\log{\left(\frac{M_2^2}{m_1^2}\right)}}{(M_2^2-m_1^2)(M_2^2-m_2^2)(M_2^2-M_1^2)}\; ,
\end{eqnarray}
through ($i=0,2$)
\begin{eqnarray}
D_i(m,M_1,M_2) &\equiv &\lim_{m_2\to m}\; D_i(m,m_2,M_1,M_2)\, ,
\\
D_i(m,M)&\equiv & \lim_{M_2\to M}\; D_i(m,M,M_2)\, ,
\\
\overline{D}_2(m,M_1,M_2) &\equiv & D_2(m,M_1,M_2) - D_2(0,M_1,M_2)\, .
\end{eqnarray}

These one-loop contributions involve virtual propagators of up-type quarks ($u,c,t$). Once the GIM cancellation is taken into account, the up and charm contributions vanish in the limit $m_{u,c}\to 0$, which we have adopted. Thus, the $B$ meson mixing is completely dominated by the top-quark contributions (the different CKM factors have all a similar size for $B^0_d$ mixing,
$V_{ud}^* V^{\phantom{*}}_{ub}\sim V_{cd}^* V^{\phantom{*}}_{cb}\sim V_{td}^* V^{\phantom{*}}_{tb}\sim A \lambda^3$, while in the $B^0_s$ case $V_{us}^* V^{\phantom{*}}_{ub}\sim A \lambda^4$ and $V_{cs}^* V^{\phantom{*}}_{cb}\sim V_{ts}^* V^{\phantom{*}}_{tb}\sim A \lambda^2$).
Since the scalar couplings are proportional to quark masses, we have maintained the masses of the external down-type quarks.
In the limit $m_d\to 0$, we reproduce the results given in \cite{Urban:1997gw}. The only Wilson coefficients which are not suppressed by
powers of $m_d$ are $C_{\rm VLL}$ and $C^1_{\rm SRR}$. Therefore, for all practical purposes, one can neglect the remaining operators.

\mytab[tab::matching]{|l|l|}{\hline
$\mathcal{O}_i$           & $C_i(\mu_{tW})$ \\[3pt]\hline
$\mathcal{O}^{\rm VLL}$   & $(4x_W+x_W^2)M_W^2D_2(m_t,M_W)-8x_W^2M_W^4D_0(m_t,M_W)$+\\
                          & $\mbox{}+2|\varsigma_u|^2x_W^2\left[M_W^2D_2(m_t,M_W,M_{H^\pm})-4M_W^4D_0(m_t,M_W,M_{H^\pm})\right]+$\\
			  & $\mbox{}+|\varsigma_u|^4x_W^2M_W^2D_2(m_t,M_{H^\pm})$\\[5pt]
$\mathcal{O}^{\rm VRR}$   & $\frac{m_d^2m_b^2}{M_W^4}\left[ |\varsigma_d|^4x_HM_W^2D_2(m_t,M_{H^\pm})+|\varsigma_d|^2 M_W^2 \overline{D}_2(m_t,M_W,M_{H^\pm})\right]$\\[5pt]
$\mathcal{O}_1^{\rm LR}$  & $2\frac{m_dm_b}{M_W^2}x_W\left[|\varsigma_d|^2|\varsigma_u|^2M_W^2D_2(m_t,M_{H^\pm})+2\, {\rm Re}(\varsigma_d^*\varsigma_u)M_W^2D_2(m_t,M_W,M_{H^\pm})\right]$\\[5pt]
$\mathcal{O}_2^{\rm LR}$  & $2\frac{m_dm_b}{M_W^2}\left[4|\varsigma_d|^2|\varsigma_u|^2x_WM_W^4D_0(m_t,M_{H^\pm})
-4|\varsigma_d|^2 M_W^2 \overline{D}_2(m_t,M_W,M_{H^\pm})+\right.$\\
                          & $\left.\mbox{}\qquad\quad\,+(|\varsigma_d|^2+|\varsigma_u|^2)x_WM_W^2D_2(m_t,M_W,M_{H^\pm})\right]$\\[5pt]
$\mathcal{O}_1^{\rm SLL}$ & $4\frac{m_d^2}{M_W^2}x_W^2\left[(\varsigma_u\varsigma_d^*)^2M_W^4D_0(m_t,M_{H^\pm})+2\varsigma_u\varsigma_d^*M_W^4D_0(m_t,M_W,M_{H^\pm})\right]$\\[5pt]
$\mathcal{O}_2^{\rm SLL}$ & 0\\[5pt]
$\mathcal{O}_1^{\rm SRR}$ & $4\frac{m_b^2}{M_W^2}x_W^2\left[(\varsigma_d\varsigma_u^*)^2M_W^4D_0(m_t,M_{H^\pm})+2\varsigma_d\varsigma_u^*M_W^4D_0(m_t,M_W,M_{H^\pm})\right]$\\[5pt]
$\mathcal{O}_2^{\rm SRR}$ & 0\\\hline
}
{Leading-order Wilson coefficients for the $\Delta B = 2$ operators 
given above.
The quark masses from the scalar couplings are to be taken at the matching scale $\mu_{tW}$.
}

The running for $\mathcal{O}_1^{SRR}$ is performed using the results of \cite{Buras:2001ra},
\begin{equation}
\left(\begin{array}{c}C^1_{\rm SRR}(\mu_b)\\C^2_{\rm SRR}(\mu_b)\end{array}\right)=
\left(\begin{array}{c c}
[\eta_{11}(\mu_b)]_{\rm SRR} & [\eta_{12}(\mu_b)]_{\rm SRR}\\{}
[\eta_{21}(\mu_b)]_{\rm SRR} & [\eta_{22}(\mu_b)]_{\rm SRR}
\end{array}\right)
\left(
\begin{array}{c}
C^1_{\rm SRR}(\mu_{tW})\\
C^2_{\rm SRR}(\mu_{tW})
\end{array}\right)\,,
\end{equation}
with
\begin{eqnarray}
[\eta_{11}(\mu_b)]_{\rm SRR} &=& 1.0153\,\eta_5^{-0.6315}-0.0153\,\eta_5^{0.7184}\,,\\{}
[\eta_{12}(\mu_b)]_{\rm SRR} &=& 1.9325\, (\eta_5^{-0.6315}-\eta_5^{0.7184})\,,\\{}
[\eta_{21}(\mu_b)]_{\rm SRR} &=& 0.0081\, (\eta_5^{0.7184}-\eta_5^{-0.6315})\,,\\{}
[\eta_{22}(\mu_b)]_{\rm SRR} &=& 1.0153\,\eta_5^{0.7184}-0.0153\,\eta_5^{-0.6315}\,.
\end{eqnarray}
These are leading-order expressions, but they have been evaluated with the two-loop expression for $\alpha_s$ in $\eta_5=\frac{\alpha_s^{(5)}(\mu_{tW})}{\alpha_s^{(5)}(\mu_{b})}\sim0.7$.

The corresponding matrix elements are given by
\begin{eqnarray}
\langle\mathcal{O}^{\rm VLL}\rangle(\mu) &=& \phantom{-}\frac{1}{3}m_{B^0_d}f_{B^0_d}^2B^{\rm VLL}(\mu)\,,\\
\langle\mathcal{O}_1^{\rm SRR}\rangle(\mu) &=& -\frac{5}{24}\left(\frac{m_{B^0_d}}{m_b(\mu)+m_d(\mu)}\right)^2m_{B^0_d}f_{B^0_d}^2B_1^{\rm SRR}(\mu)\, ,\\
\langle\mathcal{O}_2^{\rm SRR}\rangle(\mu) &=& -\frac{1}{2}\left(\frac{m_{B^0_d}}{m_b(\mu)+m_d(\mu)}\right)^2m_{B^0_d}f_{B^0_d}^2B_2^{\rm SRR}(\mu)\,,
\end{eqnarray}
the $B_i(\mu)$ parametrizing the deviation from the naive factorization limit. These $B_i(\mu)$ factors have been evaluated in the quenched approximation on the lattice in \cite{Becirevic:2001xt}, using a different operator basis. The connection reads (see again \cite{Buras:2001ra}, given here with both operators in the same scheme)
\begin{equation}
B^{\rm SRR}_1(\mu) \; =\; B_2(\mu)\, , \qquad\qquad
B^{\rm SRR}_2(\mu) \; =\; \frac{5}{3}B_2(\mu)-\frac{2}{3}B_3(\mu)\,.
\end{equation}
From \cite{Becirevic:2001xt} we arrive at the values given in table~\ref{tab::Bfactors} by adding again all systematic uncertainties linearly.

\mytab[tab::Bfactors]{|l|c|c|}{\hline
                           & $B^0_d$                  & $B^0_s$\\\hline
$B_2^{\overline{MS}}(m_b)$ & $0.83\pm0.03\pm0.06$ & $0.84\pm0.02\pm0.06$\\
$B_3^{\overline{MS}}(m_b)$ & $0.90\pm0.06\pm0.12$ & $0.91\pm0.03\pm0.12$\\\hline}
{$B$-parameters for $B^0_{d,s}$ mixing from \cite{Becirevic:2001xt}. Systematic errors added linearly.}

The wanted $B^0_d$-$\bar B^0_d$ mixing amplitude is given by
\begin{eqnarray}
\langle B^0|\mathcal{H}^{\Delta B=2}_{\mathrm{eff}}|\bar{B}^0\rangle &=& \frac{G_F^2M_W^2}{16\pi^2}(V_{td}^*V_{tb}^{\phantom{*}})^2f_{B^0_d}^2M_{B^0_d}^2\times\nonumber\\
&&\times\left[\frac{2}{3}\,\hat{B}_{B^0_d}\,\eta_B(x_W,x_H)\, C_{\rm VLL}(\mu_{tW})+\right.\\
&&\left. \mbox{}\quad\, + \frac{m_{B^0_D}^2}{(m_b(\mu_b)+m_d(\mu_b))^2}\left[\eta_{\rm SRR}(\mu_b,\mu_{tW})\,\mathbf{C}_{\rm SRR}(\mu_{tW})\right]^T\mathbf{B}_{\rm SRR}(\mu_b) \right]\,,\nonumber
\end{eqnarray}
with
\begin{equation}
\mathbf{B}_{\rm SRR}(\mu_b)=\left(\begin{array}{c}-\frac{5}{12}B_{2,d}(\mu_b)\\\frac{2}{3}B_{3,d}(\mu_b)-\frac{5}{3}B_{2,d}(\mu_b)\end{array}\right)\,.
\end{equation}
From this, we get the relevant observables as
\begin{eqnarray}
\Delta m_{B^0_d} &=& \frac{1}{m_{B^0_d}}|\langle B^0_d|\mathcal{H}^{\Delta B=2}_{\mathrm{eff}}|\bar{B}^0_d\rangle|\, ,\\
\phi_{B^0_d} &=& -\mbox{Arg}\left[\langle B^0_d|\mathcal{H}^{\Delta B=2}_{\mathrm{eff}}|\bar{B}^0_d\rangle\right]\,.
\end{eqnarray}

The analogous expressions for $B^0_s$-$\bar B^0_s$ mixing are trivially obtained changing the label $d$ to $s$ everywhere.

\subsection[$\Delta S=2$]{$\mathbf{\Delta S=2}$}
For the Kaon mixing amplitude, we have calculated the LO matching coefficients completely anologous to the $\Delta B=2$ coefficients, keeping the charm mass finite. Due to the strong suppression of all other operators by light quark masses we can choose the LO matching coefficients to be
\begin{eqnarray}
C_{\mathcal{O}_{\rm VLL}}^{cc}&=& (4x_W^{cc}+x_W^{cc\,2})M_W^2D_2(m_c,M_W)-8x_W^{cc\,2}M_W^4D_0(m_c,M_W)\,,\nonumber\\
C_{\mathcal{O}_{\rm VLL}}^{ct}&=& (4x_W^{ct}+x_W^{ct\,2})M_W^2D_2(m_c,m_t,M_W)-8x_W^{ct\,2}M_W^4D_0(m_c,m_t,M_W)+\nonumber\\
                              & & +2|\varsigma_u|^2x_W^{ct\,2}\left[M_W^2D_2(m_c,m_t,M_W,M_{H^\pm})-4M_W^4D_0(m_c,m_t,M_W,M_{H^\pm})\right]+\nonumber\\
			      & & +|\varsigma_u|^4x_W^{ct\,2}M_W^2D_2(m_c,m_t,M_{H^\pm})\,,\\
C_{\mathcal{O}_{\rm VLL}}^{tt}&=& (4x_W+x_W^2)M_W^2D_2(m_t,M_W)-8x_W^2M_W^4D_0(m_t,M_W)+\nonumber\\
                              & & +2|\varsigma_u|^2x_W^2\left[M_W^2D_2(m_t,M_W,M_{H^\pm})-4M_W^4D_0(m_t,M_W,M_{H^\pm})\right]+\nonumber\\
			      & & +|\varsigma_u|^4x_W^2M_W^2D_2(m_t,M_{H^\pm})\,,\nonumber\\
C_{\mathcal{O}_{i}}&=& 0 \quad(i\neq \rm VLL)\,,\nonumber
\end{eqnarray}
where the loop functions $D_{0,2}$ have been defined in appendix A, and $x_W^{ct}=m_cm_t/M_W^2$. In the calculation, we use the NLO results for the SM which have been calculated in \cite{Buras:1990fn,Herrlich:1996vf}, while the NLO charged scalar contributions to the top contribution are again taken from \cite{Urban:1997gw}, corrected and applied to our scenario.

\bibliography{bibliography}

\end{document}